\documentclass{aa}  
\usepackage{graphicx}
\usepackage{amsmath}
\usepackage{txfonts}
\usepackage{multirow}
\usepackage{booktabs}
\usepackage{subcaption}
\def\ms{\hbox{\,m\,s$^{-1}$}}         %m.s -1
\def\msq{\hbox{\,m$^{2}$\,s$^{-2}$}}  
\def\kms{\hbox{\,km\,s$^{-1}$}}
\newcommand{\dstuc}{DS\,Tuc}
\newcommand{\xmm}{{\em XMM-Newton}}

\newcommand{\lxu}{{erg s$^{-1}$}}
\newcommand{\vsini}{{$v \sin i$}}
\begin{document}

   \title{Constraints on the mass and atmospheric composition and evolution of the low-density young planet DS\,Tuc A\,b
\thanks{Based on observations made with ESO Telescopes at the La Silla Paranal Observatory under programme ID 
0103.C-0759(A), 	% HARPS P103
0104.C-0798(A). % ESPRESSO P104
%Cheops GO programme ID 0020 
}
\thanks{Based on observations obtained with \xmm, an ESA science mission with instruments and contributions directly funded by ESA Member States and NASA.}} 

\author{S. Benatti\inst{1},
M. Damasso\inst{2},
F. Borsa\inst{3},
D. Locci\inst{1},
I. Pillitteri\inst{1},
S. Desidera\inst{4},
A. Maggio\inst{1},
G. Micela\inst{1},
S. Wolk\inst{5},
R.~Claudi\inst{4}, 
L. Malavolta\inst{6},
D. Modirrousta-Galian\inst{1}
}

\institute{INAF -- Osservatorio Astronomico di Palermo, Piazza del Parlamento, 1, I-90134, Palermo, Italy \\
  \email{serena.benatti@inaf.it}
\and INAF -- Osservatorio Astrofisico di Torino, Via Osservatorio 20, I-10025, Pino Torinese (TO), Italy 
\and  INAF -- Osservatorio Astronomico di Brera, Via E. Bianchi 46, I-23807 Merate (LC), Italy 
\and INAF -- Osservatorio Astronomico di Padova, Vicolo dell'Osservatorio 5, I-35122, Padova, Italy  
\and Smithsonian Astrophysical Observatory, MS 4, 60 Garden St., Cambridge, MA 02138 
\and Dipartimento di Fisica e Astronomia -- Universit\`a di Padova, Vicolo dell'Osservatorio 3, I-35122 Padova 
}

   \date{Received ; accepted }

% \abstract{}{}{}{}{} 
% 5 {} token are mandatory
 
  \abstract
{The observations of young close-in exoplanets are providing first indications of the characteristics of the population and, in turn, clues on the early stages of their evolution. Transiting planets at young ages are also key benchmarks for our understanding of planetary evolution through the verification of atmospheric escape models. %
}
{We performed a radial velocity (RV) monitoring of the 40 Myr old star DS\,Tuc\,A with HARPS at the ESO-3.6m to determine the planetary mass of its 8.14-days planet, first revealed by the NASA \textit{TESS} satellite. We also observed two planetary transits with HARPS and ESPRESSO at ESO-VLT, to measure the Rossiter-McLaughlin (RM) effect and characterise the planetary atmosphere. We measured the high-energy emission of the host with \xmm~observations to investigate models for atmospheric evaporation. 
}
{We employed Gaussian Processes (GP) regression to model the high level of the stellar activity, which is more than 40 times larger than the expected RV planetary signal. GPs were also used to correct the stellar contribution to the RV signal of the RM effect. We extracted the transmission spectrum of DS\,Tuc\,A\,b from the ESPRESSO data and searched for atmospheric elements/molecules either by single-line retrieval and by performing cross-correlation with a set of theoretical templates. Through a set of simulations, we evaluated different scenarios for the atmospheric photo-evaporation of the planet induced by the strong XUV stellar irradiation.% from the young host.
}
{While the stellar activity prevented us from obtaining a clear detection of the planetary signal from the RVs, we set a robust mass upper limit of 14.4 M$_{\oplus}$ for DS\,Tuc\,A\,b. 
We also confirm that the planetary system is almost (but not perfectly) aligned. The strong level of stellar activity hampers the detection of any atmospheric compounds, in line with other studies presented in the literature.  
The expected evolution of DS\,Tuc\,A\,b from our grid of models indicates that the planetary radius after the photo-evaporation phase will be 1.8-2.0 R$_{\oplus}$, falling within the Fulton gap. }
{The comparison of the available parameters of known young transiting planets with the distribution of their mature counterpart confirms that the former are characterised by a low density, with DS\,Tuc\,A\,b being one of the less dense. A clear determination of their distribution is still affected by the lack of a robust mass measurement, in particular for planets younger than $\sim$100 Myr. }


\titlerunning{DS\,Tuc}
\authorrunning{S. Benatti et al.}
   \maketitle
%
%-------------------------------------------------------------------

\section{Introduction}
\label{sec:intro}
In the last few years, the initial observations and studies of young close-in exoplanets have been carried out (e.g. \citealt{2017AJ....154..224R,2018ApJ...855..115B,2020A&A...638A...5C}). A crucial role in this field is played by space-based surveys for transiting planets, such as those performed with the NASA \textit{Kepler/K2} \citep{2003SPIE.4854..129B,2014PASP..126..398H} and the Transiting Exoplanet Survey Satellite (\textit{TESS}, \citealt{2015JATIS...1a4003R}) missions. The latter is providing a significant number of close-in planet candidates around young stars \citep[e.g.][]{2020MNRAS.495.4924N,2020MNRAS.496.1197B,2020MNRAS.498.5972N}, thanks to  systematic observations of about 85\% of the sky. 
Early discoveries of hot Jupiters around very young stars \citep{donati2016,yu2017,rizzuto2020} suggested a larger frequency of close-in massive planets with respect to older stars. The discrepancy between the two age regimes was interpreted with different scenarios, considering an intense atmospheric escape due to the strong XUV irradiation from the young host or planet engulfment in early evolutionary stages \citep{2017MNRAS.465.3343D}.

The apparent high frequency of young hot Jupiters has been questioned from the recent results by \cite{2018A&A...613A..50C}, \cite{2020MNRAS.491.5660D} and \cite{2020A&A...642A.133D}, that put doubts on the detection of such type of companions. Moreover, the first results of \textit{TESS} observations suggest that the typical radius of close-in planets with age < 100 Myr is lower than one Jupiter radius (e.g. \citealt{2019A&A...630A..81B,2020Natur.582..497P,2021A&A...645A..71C}). 
The on-going contraction and the effect of the intense high-energy irradiation from their host star suggest the possibility that many of them could actually be highly inflated lower mass planets, as first showed by \cite{2015ApJ...807....3K} and \cite{2016AJ....152...61M}, and confirmed by theoretical predictions as in \cite{2019A&A...623A..85L}.
This working hypothesis implies:
\begin{itemize}
 \item A dependency of the planetary mass-radius relation with the age, which is well-constrained for the older population (e.g. \citealt{2017ApJ...834...17C,2017A&A...604A..83B}), while it is still unexplored for young planets ranging from a few Myr up to hundreds of Myr.
  \item  The key role of the photo-evaporation of the primordial hydrogen envelope in shaping the exoplanet population throughout the system lifetime. Super Earths and mini Neptunes (radii between 1-4 R$_{\oplus}$ and masses lower than 20 M$_{\oplus}$) at short orbital distances are more susceptible to lose part or all of their gaseous envelope (see e.g. \citealt{2012ApJ...761...59L}), creating a gap between smaller super-Earth planets that have lost all the hydrogen/helium envelope, and the larger sub-Neptune planets that kept some of it (e.g. \citealt{2013ApJ...775..105O}).
 \item A difficult radial velocity (RV) characterization, since the host stars show an intense level of the stellar activity able to bury the small planetary signal (up to tens of times lower than the typical RV dispersion). The typically large rotation velocity of the star also degrades the quality of the spectrum in terms of RV precision and the use of very dense time sampling and sophisticated tools, necessary to model the stellar activity, is required (e.g. \citealt{2020A&A...642A.133D}). 
 \item A very low density and large scale height that make young planets, in principle, excellent targets for atmospheric characterization; however, studies suggest the presence of high-altitude clouds, hazes and dust grains that increase atmospheric opacity, possibly leading to a null detection of atmospheric features \citep{2019ApJ...873L...1W,2020ApJ...890...93G}.
\end{itemize}

\cite{2019A&A...630A..81B} and \cite{2019ApJ...880L..17N} identified the first young planet detected by \textit{TESS} in its first month of observations, around the main component of the DS Tuc binary system in the Tuc-Hor association (G6 spectral type, visual magnitude = 8.2, estimated age $\sim 40$ Myr, TESS Object of Interest: TOI-200, \textit{Gaia} DR3 source ID: 6387058411482257536). The planetary orbital period from the transit fit is 8.14 days and the planetary radius is about 0.5 R$_{\rm J}$ (5.6 R$_{\oplus}$). The upper limit on the planetary mass is set to 1.3 M$_{\rm J}$ from \cite{2019A&A...630A..81B} by using a few high-resolution archive spectra, and 0.09 M$_{\rm J}$(28 M$_{\oplus}$) from \cite{2019ApJ...880L..17N} according to the mass-radius relation \citep{2017ApJ...834...17C}.
\cite{2020AJ....159..112M} and \cite{2020ApJ...892L..21Z}  found the planetary orbit well aligned with the equatorial plane of the host through the observation of the Rossiter-McLaughlin (RM) effect \citep{1924ApJ....60...15R,1924ApJ....60...22M} during the transit of DS\,Tuc\,A\,b.

Within the framework of coordinated observing programs with the High Accuracy Radial velocity Planet Searcher (HARPS, \citealt{2003Msngr.114...20M}) spectrograph at the 3.6m-ESO telescope (La Silla Observatory, Chile), the Echelle SPectrograph for Rocky Exoplanets and Stable Spectroscopic Observations (ESPRESSO, \citealt{2021A&A...645A..96P}) at VLT (Paranal Observatory, Chile), and \xmm\ we now aim to measure the mass of DS\,Tuc\,A\,b, to confirm the orbital inclination, and to characterise its atmosphere.
We will show that the impact of the stellar activity on our data is particularly significant, but nevertheless, we were able to reduce the current value of the mass upper limit for the planet, and to investigate the scenario of atmospheric escape by using mass and radius evolutionary models.

This paper is organised as follows: in Sect. \ref{sec:data} we describe the HARPS and ESPRESSO datasets; the RV modelling with the GP regression  are provided in Sect. \ref{sec:rv}. The analysis of the RM effect from ESPRESSO data is reported in Sect. \ref{sec:rml}, while the results of the atmospheric characterization are presented in Sect. \ref{sec:atmo}. We present the results from \xmm~ observations in Sect. \ref{sec:x} and describe our photo-evaporation models in Sect. \ref{sec:models}. We  provide a discussion in Sect. \ref{sec:discussion} and draw our conclusions in Sect. \ref{sec:concl}.

%--------------------------------------------------------------------------------------

\section{Datasets}\label{sec:data}
\subsection{HARPS - RV monitoring}\label{sec:data_harps_rv}
We monitored DS\,Tuc\,A with HARPS at the 3.6m telescope at the ESO-La Silla Observatory (Chile), within the framework of the HARPS time-sharing collaboration (programme ID 0103.C-0759(A), PI Benatti). Thanks to this strategy, we benefited from the time shared among several observing programs and gathered a time series composed by 76 spectra spanning from June to September 2019.
As previously mentioned, DS\,Tuc\,A\,belongs to a visual binary system, with a projected separation from the stellar companion (K3 spectral type) of about 5\arcsec. The presence of the companion does not affect our RV measurement, since the aperture of the HARPS fibre correspond to 1\arcsec~ on the sky.
When possible, during the maximum visibility period of our target, we obtained two spectra in the same night with a few hours of separation. Doing so, we managed to collect nine couples of data points for a better knowledge of the intra-night RV scatter. The mean signal-to-noise ratio (S/N) of the spectra at 5500 \AA  ~is 108, with typical exposure time of 900 s.
 
\begin{figure}
    \centering
    \includegraphics[trim={1cm 3cm 2cm 2cm},clip,width=0.35\textwidth, angle=270]{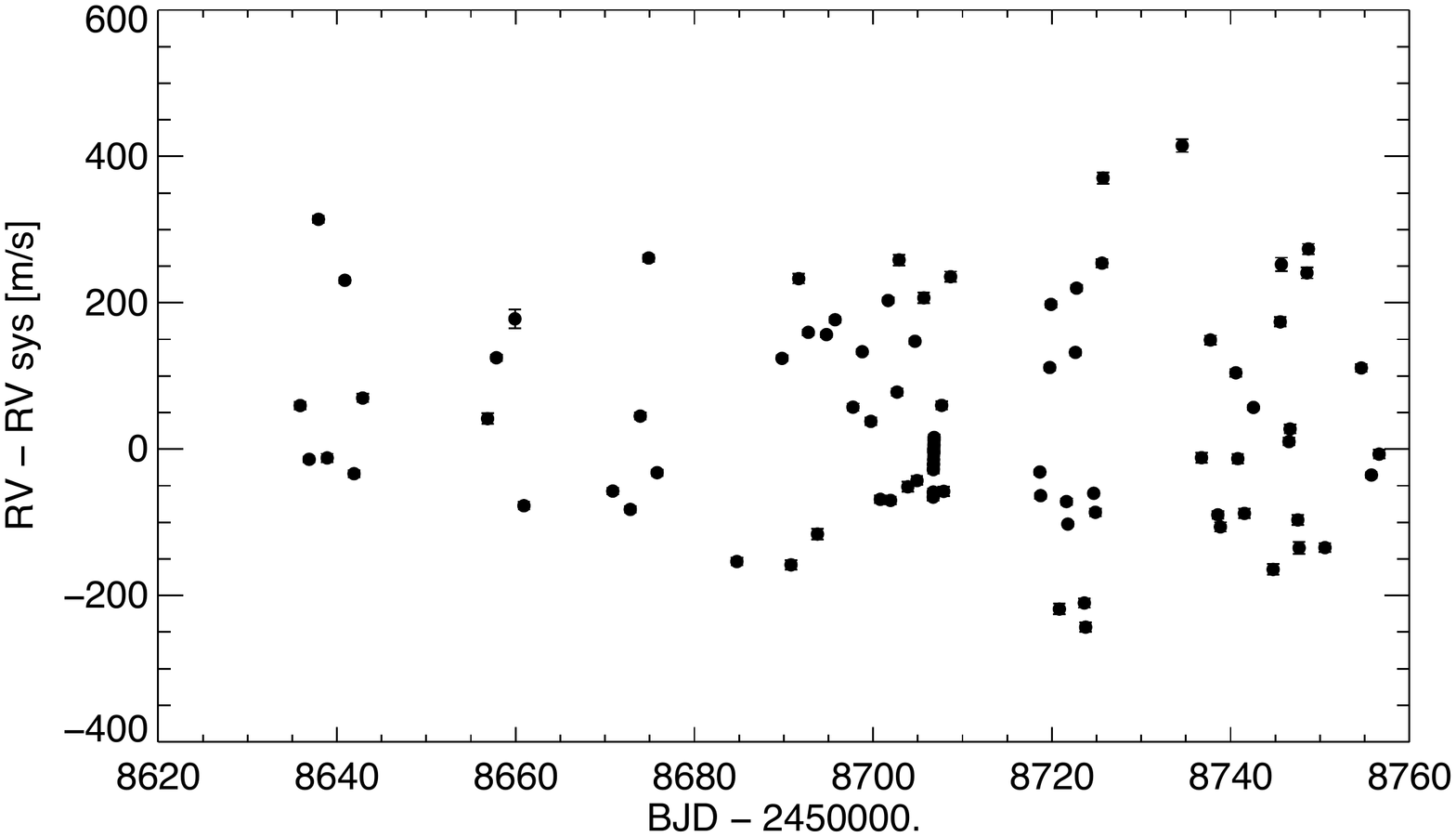}
    \caption{HARPS RV time series of DS\,Tuc\,A as extracted from the TERRA pipeline. A bulk of data at JD = 8706 corresponds to the RVs collected during  the planetary transit. }
    \label{fig:rv}
\end{figure}

The HARPS spectra were reduced through the Data Reduction Software (DRS), which estimates the RVs by computing the Cross Correlation function (CCF) between the spectra and a digital mask depicting the features of a G2 star (see \citealt{2002A&A...388..632P} and references therein). In particular, since DS\,Tuc\,A is a moderately fast rotator (\vsini$ = 15.5 \pm 1.5$ \kms, \citealt{2019A&A...630A..81B}) the CCF profile is broadened, so we adapted the width of the window for the CCF evaluation by using the offline version of the DRS.
The RV dispersion of the whole time series and the mean RV uncertainty obtained with the HARPS DRS are 163 \ms~and 5.2 \ms, respectively. 
We also obtained an alternative processing of the RVs by using the pipeline TERRA \citep{2012ApJS..200...15A}. In case of active stars, TERRA is more efficient in evaluating the RVs (see \citealt{2020A&A...638A...5C,2020A&A...642A.133D}), since it employs one of the observed spectra as a template of the line profiles, that is more suitable to describe potential distortions and asymmetries. For the TERRA dataset, the RV dispersion and the mean RV uncertainty are 138 \ms~and 5.4 \ms, respectively. 
The following analysis is based on the RV time series obtained with TERRA, reported in Table \ref{tab:harps} with the flag ``RV'' and displayed in Fig. \ref{fig:rv}. 

Finally, we extracted the time series of several activity indicators such as the Ca II H+K lines and H$\alpha$ by using the ACTIN code \citep{2018JOSS....3..667G}, while the HARPS DRS provides the bisector span (BIS) of the CCF, useful to trace the distortion of the line profiles produced by the stellar activity. These time series are reported in Table \ref{tab:harps}.

\longtab{
\begin{longtable}{ccccccccc}
\caption{\label{tab:harps} Time series of the HARPS spectra for DS\,Tuc\,A: BJD$_{\rm TDB}$, radial velocity (RV) and radial velocity error (RV$_{\rm err}$) obtained with TERRA, CaII H\&K and H$\alpha$ with uncertainties as obtained with ACTIN, and the bisector span as obtained with HARPS DRS. The ``RV'' and ``T'' flags indicate that the corresponding spectrum was collected within the regular RV monitoring or during the transit sequence on the 2019-08-10 night.} \\
\hline\hline
BJD$_{\rm TDB}$--2\,450\,000 & RV & RV$_{\rm Err}$ & CaII & CaII$_{\rm Err}$ & H$\alpha$ & H$\alpha _{\rm Err}$ & BIS & Flag\\
 & [\ms] & [\ms]  & & & & & [km\,s$^{-1}$] & \\
\hline\hline
8635.89130668  &  59.50  &  4.91  &  0.444  &  0.002  &  0.3871  &  0.0005  &  -0.4464  & RV\\
8636.89799636  &  -13.97  &  3.86  &  0.411  &  0.001  &  0.3451  &  0.0004  &  -0.1248  & RV\\
8637.93607589  &  314.00  &  5.00  &  0.415  &  0.001  &  0.3524  &  0.0005  &  -0.5868  & RV\\
8638.91698802  &  -12.08  &  5.86  &  0.438  &  0.002  &  0.3780  &  0.0006  &  -0.3072  & RV\\
8640.88845254  &  230.68  &  3.85  &  0.421  &  0.001  &  0.3498  &  0.0004  &  -0.3919  & RV\\
8641.91842789  &  -33.60  &  4.74  &  0.422  &  0.001  &  0.3607  &  0.0005  &  -0.1449  & RV\\
8642.88840014  &  69.76  &  5.62  &  0.415  &  0.001  &  0.3525  &  0.0004  &  -0.1545  & RV\\
8656.85954333  &  41.67  &  7.27  &  0.463  &  0.002  &  0.3693  &  0.0006  &  -0.1417  & RV\\
8657.84593114  &  124.86  &  4.49  &  0.438  &  0.001  &  0.3640  &  0.0005  &  -0.1844  & RV\\
8659.91816404  &  178.01  &  13.16  &  0.413  &  0.005  &  0.3616  &  0.0011  &  -0.5279  & RV\\
8660.92880982  &  -77.38  &  5.21  &  0.454  &  0.002  &  0.3651  &  0.0006  &  0.0397  & RV\\
8670.86816236  &  -57.37  &  4.47  &  0.409  &  0.001  &  0.3505  &  0.0005  &  -0.1168 & RV \\
8672.82861343  &  -82.44  &  4.33  &  0.435  &  0.001  &  0.3635  &  0.0005  &  0.0845  & RV\\
8673.94434424  &  45.12  &  5.00  &  0.408  &  0.002  &  0.3470  &  0.0006  &  -0.1414  & RV\\
8674.88487363  &  261.09  &  4.71  &  0.452  &  0.001  &  0.3651  &  0.0004  &  -0.5998  & RV\\
8675.83004360  &  -32.19  &  4.38  &  0.412  &  0.001  &  0.3547  &  0.0005  &  -0.1436  & RV\\
8684.75359416  &  -153.67  &  5.06  &  0.413  &  0.001  &  0.3427  &  0.0005  &  0.1608  & RV\\
8689.80413404  &  123.97  &  3.97  &  0.451  &  0.001  &  0.3673  &  0.0004  &  -0.4472  & RV\\
8690.81137592  &  -158.19  &  6.49  &  0.440  &  0.002  &  0.3740  &  0.0006  &  0.2255  & RV\\
8691.67147470  &  232.93  &  6.48  &  0.418  &  0.001  &  0.3397  &  0.0005  &  -0.1061  & RV\\
8692.74643523  &  159.58  &  4.23  &  0.453  &  0.001  &  0.3730  &  0.0005  &  -0.4451  & RV\\
8693.76675327  &  -116.11  &  7.33  &  0.419  &  0.002  &  0.3495  &  0.0007  &  0.2205  & RV\\
8694.77703562  &  156.49  &  3.33  &  0.422  &  0.001  &  0.3480  &  0.0003  &  -0.3006  & RV\\
8695.74383313  &  176.78  &  4.21  &  0.446  &  0.001  &  0.3723  &  0.0004  &  -0.4815  & RV\\
8697.74233316  &  57.29  &  5.15  &  0.431  &  0.001  &  0.3558  &  0.0005  &  0.0487  & RV\\
8698.78033186  &  133.05  &  3.60  &  0.468  &  0.001  &  0.3780  &  0.0004  &  -0.6300 & RV \\
8699.74559814  &  37.90  &  5.21  &  0.402  &  0.001  &  0.3376  &  0.0004  &  -0.3187  & RV\\
8700.80316672  &  -68.66  &  5.26  &  0.423  &  0.001  &  0.3605  &  0.0004  &  0.2708 &  RV\\
8701.67682266  &  203.12  &  3.87  &  0.448  &  0.001  &  0.3719  &  0.0004  &  -0.7354  & RV\\
8701.94247023  &  -70.11  &  5.30  &  0.442  &  0.002  &  0.3635  &  0.0005  &  -0.2413  & RV\\
8702.67621831  &  77.91  &  5.17  &  0.404  &  0.001  &  0.3360  &  0.0004  &  -0.3471  & RV\\
8702.90324307  &  258.49  &  7.30  &  0.401  &  0.001  &  0.3235  &  0.0004  &  -0.4011  & RV\\
8703.87843634  &  -51.55  &  6.98  &  0.423  &  0.002  &  0.3485  &  0.0006  &  0.0896  & RV\\
8704.67943978  &  147.36  &  3.85  &  0.451  &  0.002  &  0.3692  &  0.0006  &  -0.6723 &  RV\\
8704.90640627  &  -43.03  &  6.04  &  0.454  &  0.002  &  0.3677  &  0.0005  &  -0.2653  & RV\\
8705.67400640  &  206.66  &  7.61  &  0.410  &  0.001  &  0.3401  &  0.0005  &  -0.4021  & RV\\
8706.61070215  &  -38.60  &  5.36  &  0.434  &  0.002  &  0.3572  &  0.0006  &  -0.4464  & T\\
8706.61835256  &  -22.48  &  5.56  &  0.436  &  0.002  &  0.3577  &  0.0006  &  -0.1248  & T\\
8706.62516964  &  -43.32  &  9.20  &  0.418  &  0.004  &  0.3601  &  0.0012  &  -0.5868  & T\\
8706.63309783  &  -4.30  &  4.58  &  0.439  &  0.002  &  0.3613  &  0.0006  &  -0.3072  & T\\
8706.64088714  &  -26.58  &  8.02  &  0.436  &  0.003  &  0.3594  &  0.0008  &  -0.3919  & T\\
8706.64777367  &  -24.28  &  4.04  &  0.426  &  0.001  &  0.3561  &  0.0005  &  -0.1449  & T\\
8706.65459075  &  -27.85  &  4.02  &  0.431  &  0.001  &  0.3548  &  0.0005  &  -0.1545  & T\\
8706.66265783  &  -30.70  &  3.91  &  0.430  &  0.001  &  0.3549  &  0.0005  &  -0.1417  & T\\
8706.66933603  &  -42.81  &  3.65  &  0.426  &  0.001  &  0.3528  &  0.0004  &  -0.1844  & T\\
8706.67587534  &  -59.20  &  3.51  &  0.428  &  0.002  &  0.3531  &  0.0006  &  -0.5279  & T\\
8706.68449796  &  -58.07  &  3.98  &  0.430  &  0.001  &  0.3532  &  0.0005  &  0.0397  & T\\
8706.69125718  &  -83.17  &  3.53  &  0.425  &  0.001  &  0.3524  &  0.0005  &  -0.1168  & T\\
8706.69847936  &  -76.66  &  4.24  &  0.422  &  0.002  &  0.3510  &  0.0005  &  0.0845  & T\\
8706.70585199  &  -80.41  &  3.71  &  0.421  &  0.001  &  0.3496  &  0.0005  &  -0.1414  & T\\
8706.71343296  &  -77.40  &  3.61  &  0.421  &  0.002  &  0.3505  &  0.0005  &  -0.5998  & T\\
8706.72046995  &  -65.66  &  4.41  &  0.420  &  0.001  &  0.3492  &  0.0004  &  0.2547  & T\\
8706.72762268  &  -58.68  &  5.17  &  0.419  &  0.002  &  0.3486  &  0.0005  &  0.2333  & T\\
8706.73520365  &  -28.03  &  5.44  &  0.421  &  0.002  &  0.3477  &  0.0005  &  0.3352  & T\\
8706.74229851  &  -27.05  &  4.96  &  0.420  &  0.002  &  0.3476  &  0.0005  &  0.1241  & T\\
8706.74996050  &  -20.94  &  5.22  &  0.416  &  0.002  &  0.3477  &  0.0005  &  0.1167  & T\\
8706.75725212  &  -20.40  &  4.78  &  0.423  &  0.001  &  0.3485  &  0.0005  &  0.0901  & T\\
8706.76427754  &  -14.31  &  4.43  &  0.422  &  0.001  &  0.3478  &  0.0004  &  0.1128  & T\\
8706.77151129  &  0.00  &  4.49  &  0.415  &  0.001  &  0.3464  &  0.0005  &  0.1294  & T\\
8706.77902281  &  -3.49  &  4.18  &  0.414  &  0.001  &  0.3466  &  0.0005  &  0.1199  & T\\
8706.78619869  &  -3.75  &  4.15  &  0.416  &  0.001  &  0.3441  &  0.0005  &  0.1547  & T\\
8706.79364077  &  -4.42  &  4.96  &  0.412  &  0.002  &  0.3434  &  0.0005  &  0.1692  & T\\
8706.80073563  &  5.70  &  4.45  &  0.416  &  0.002  &  0.3432  &  0.0005  &  0.1532  & T\\
8706.80824715  &  5.27  &  4.23  &  0.414  &  0.002  &  0.3443  &  0.0005  &  0.1712  & T\\
8706.81548090  &  12.46  &  3.95  &  0.412  &  0.001  &  0.3437  &  0.0005  &  0.1268  & T\\
8706.82271465  &  15.91  &  4.32  &  0.410  &  0.001  &  0.3436  &  0.0005  &  0.0760 & T \\
8707.66553283  &  59.72  &  5.47  &  0.438  &  0.002  &  0.3625  &  0.0006  &  0.0458 &  RV\\
8707.89608667  &  -57.76  &  6.66  &  0.436  &  0.002  &  0.3668  &  0.0007  &  -0.0361 &  RV\\
8708.66817548  &  235.59  &  7.20  &  0.417  &  0.001  &  0.3464  &  0.0004  &  -0.0152  & RV\\
8718.65239025  &  -31.28  &  3.36  &  0.420  &  0.001  &  0.3505  &  0.0004  &  -0.0044  & RV\\
8718.73854620  &  -63.61  &  3.77  &  0.428  &  0.001  &  0.3517  &  0.0004  &  -0.0462  & RV\\
8719.75900301  &  111.50  &  4.24  &  0.420  &  0.001  &  0.3497  &  0.0004  &  -0.0401  & RV\\
8719.89868802  &  197.73  &  4.42  &  0.426  &  0.001  &  0.3532  &  0.0004  &  -0.0897  & RV\\
8720.83499031  &  -218.60  &  7.36  &  0.454  &  0.002  &  0.3667  &  0.0004  &  -0.0781 & RV \\
8721.62961638  &  -71.67  &  4.48  &  0.423  &  0.001  &  0.3475  &  0.0005  &  -0.1045 &  RV\\
8721.78080557  &  -102.51  &  3.39  &  0.427  &  0.001  &  0.3585  &  0.0004  &  -0.1119 & RV \\
8722.63039518  &  132.16  &  3.68  &  0.426  &  0.001  &  0.3510  &  0.0004  &  -0.1392  & RV\\
8722.75914257  &  219.94  &  4.31  &  0.413  &  0.001  &  0.3455  &  0.0004  &  -0.1343  & RV\\
8723.63079120  &  -210.32  &  6.32  &  0.450  &  0.001  &  0.3460  &  0.0003  &  -0.1693 & RV \\
8723.77537148  &  -243.33  &  6.56  &  0.471  &  0.001  &  0.3674  &  0.0004  &  -0.1318  & RV\\
8724.67972700  &  -60.40  &  2.80  &  0.418  &  0.001  &  0.3458  &  0.0003  &  -0.1631  & RV\\
8724.86420211  &  -86.37  &  5.26  &  0.414  &  0.002  &  0.3429  &  0.0005  &  -0.7668  & RV\\
8725.60413925  &  254.11  &  5.40  &  0.408  &  0.001  &  0.3457  &  0.0004  &  -0.0965  & RV\\
8725.72294442  &  370.55  &  7.79  &  0.413  &  0.001  &  0.3428  &  0.0004  &  -0.2182  & RV\\
8734.57673936  &  414.74  &  8.79  &  0.447  &  0.002  &  0.3650  &  0.0007  &  -0.0498  & RV\\
8736.73823590  &  -11.74  &  6.65  &  0.404  &  0.002  &  0.3438  &  0.0006  &  0.0396  & RV\\
8737.73487100  &  149.12  &  6.05  &  0.457  &  0.002  &  0.3683  &  0.0007  &  -0.2909  & RV\\
8738.56543443  &  -89.84  &  5.03  &  0.439  &  0.002  &  0.3601  &  0.0006  &  -0.2747  & RV\\
8738.85498514  &  -106.33  &  6.46  &  0.434  &  0.002  &  0.3541  &  0.0007  &  0.5400  & RV\\
8740.58216118  &  104.28  &  5.27  &  0.452  &  0.002  &  0.3674  &  0.0005  &  0.0384  & RV\\
8740.80051021  &  -13.06  &  6.64  &  0.467  &  0.002  &  0.3738  &  0.0006  &  0.1219  & RV\\
8741.54042684  &  -87.90  &  6.58  &  0.438  &  0.002  &  0.3581  &  0.0007  &  -0.2952  & RV\\
8742.55027571  &  56.94  &  3.48  &  0.404  &  0.001  &  0.3414  &  0.0004  &  -0.2908  & RV\\
8744.75281186  &  -164.45  &  7.17  &  0.446  &  0.002  &  0.3633  &  0.0007  &  0.6132  & RV\\
8745.54903230  &  173.83  &  6.35  &  0.411  &  0.001  &  0.3465  &  0.0005  &  0.4092  & RV\\
8745.67414336  &  252.39  &  9.34  &  0.413  &  0.002  &  0.3471  &  0.0007  &  -0.0699  & RV\\
8746.52694681  &  10.21  &  5.21  &  0.437  &  0.002  &  0.3554  &  0.0006  &  0.0250  & RV\\
8746.64595849  &  27.44  &  6.14  &  0.437  &  0.002  &  0.3603  &  0.0006  &  -0.3990  & RV\\
8747.51584409  &  -96.80  &  6.70  &  0.430  &  0.002  &  0.3485  &  0.0006  &  -0.5570  & RV\\
8747.66231990  &  -135.06  &  7.89  &  0.429  &  0.003  &  0.3448  &  0.0008  &  -0.7235  & RV\\
8748.53623256  &  240.75  &  7.45  &  0.415  &  0.001  &  0.3463  &  0.0005  &  -0.0515  & RV\\
8748.70152699  &  273.57  &  6.97  &  0.429  &  0.001  &  0.3537  &  0.0004  &  -0.2823  & RV\\
8750.55367520  &  -134.64  &  6.32  &  0.410  &  0.001  &  0.3443  &  0.0004  &  0.1480  & RV\\
8754.61642240  &  110.85  &  5.44  &  0.427  &  0.001  &  0.3505  &  0.0004  &  0.0665  & RV\\
8755.74860732  &  -35.44  &  3.80  &  0.438  &  0.001  &  0.3621  &  0.0003  &  -0.1427 & RV \\
8756.61287509  &  -6.89  &  6.10  &  0.421  &  0.002  &  0.3451  &  0.0006  &  -0.0187 & RV \\
\hline\hline
\end{longtable}
}

\subsection{HARPS - Transit}\label{sec:data_harps_tr}
On the night 2019-08-10, within the same observing program described above, we observed the transit of DS\,Tuc\,A\,b in front of its star collecting 30 consecutive data points. Due to the large airmass of the target ($>1.8$) and the presence of veils just before the event, we could not monitor the pre-transit phase. We collected 15 spectra during the transit and 15 spectra during the post-transit phase. To improve the temporal resolution, the exposure time was decreased to 600 s, with a mean S/N of 100, thanks to the good quality of the observing conditions during the event. 

HARPS data were reduced as described in Sect. \ref{sec:data_harps_rv} and included in Table \ref{tab:harps} with the flag ``T'' (Transit). The RM effect detected with HARPS is depicted in Fig. \ref{fig:rml} with red and green dots showing the comparison between the RV extraction obtained from Terra and DRS, respectively. Further comments on the figure are provided in the next section.

The analysis of this transit is also reported in \cite{2020AJ....159..112M} (observed with the PFS spectrograph) and \cite{2020ApJ...892L..21Z} (observed with the CHIRON spectrograph).

\subsection{ESPRESSO}\label{sec:data_espresso}
%Transit duration = 189 min da benatti et al
We observed a second transit of DS\,Tuc\,A\,b with ESPRESSO at VLT during the night 2019-10-06 (programme 0104.C-0798, PI Borsa). 
We obtained 97 spectra with 300 s of exposure time that produced spectra with mean S/N of 111 at 5500 \AA. The mean uncertainty of the radial velocity is 2.5 \ms.
We collected 22 spectra before the transit event, 32 during the transit and 45 spectra were finally collected after the transit, for a total coverage of about ten hours. The object airmass at the beginning of the observation was 1.74, reaching the minimum value of 1.40 during the transit and 2.52 at the end of the night.
This transit is also reported in \cite{2020ApJ...892L..21Z} (observed with the PFS spectrograph).

\begin{figure}
    \centering
    \includegraphics[trim={1cm 3cm 2cm 2cm},clip,width=0.35\textwidth, angle=270]{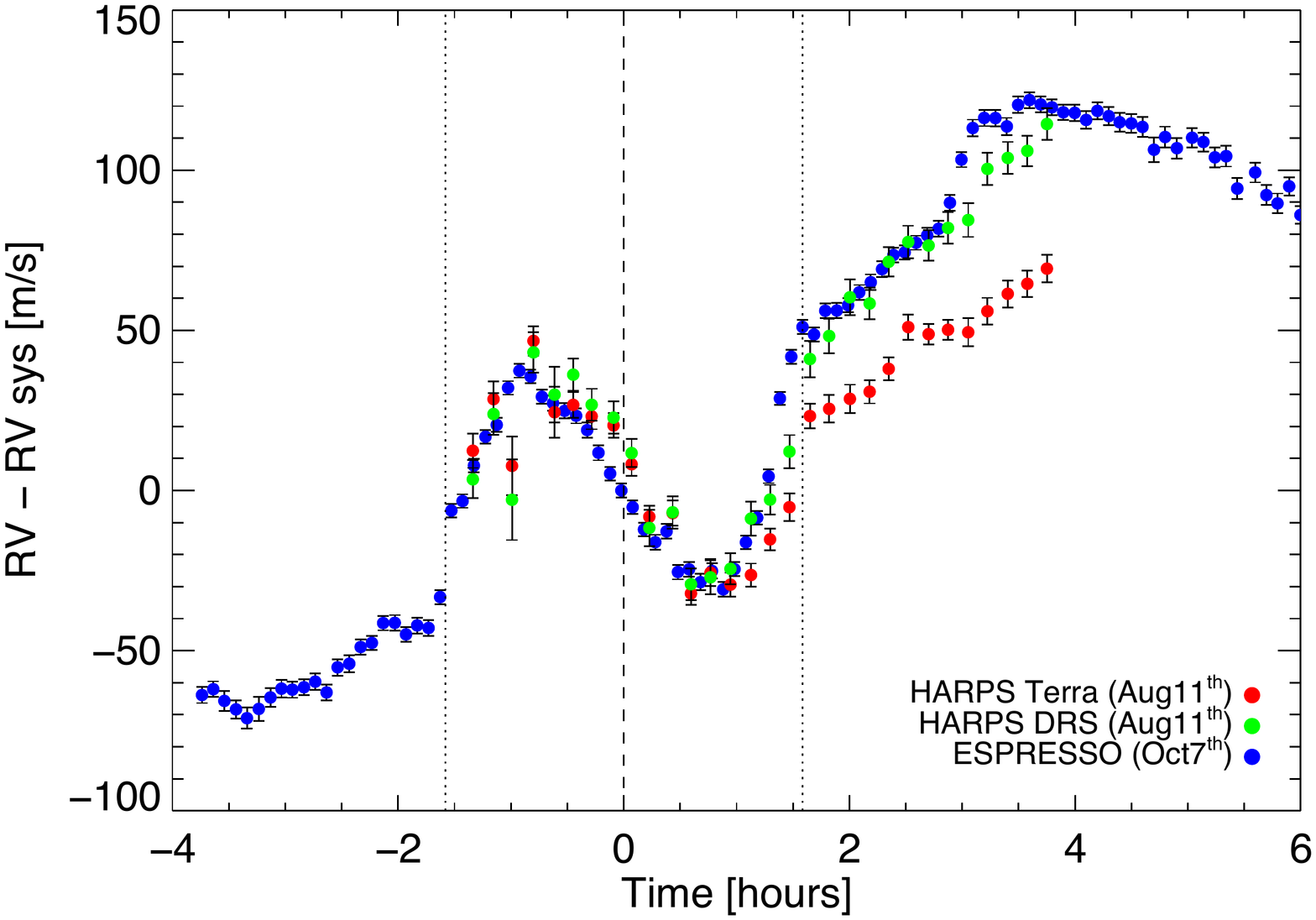}
    \caption{RM effect for the transits observed with HARPS (red dots, TERRA; green dots, DRS, systemic RV removed) and ESPRESSO (blue dots, DRS, systemic RV removed). Vertical lines indicate the first and fourth contact (dotted) and the centre of the transit (dashed), according to the ephemeris in \cite{2019ApJ...880L..17N}}
    \label{fig:rml}
\end{figure}

The ESPRESSO RVs were obtained from the ESPRESSO DRS (1.3.2), also based on the CCF method, by using the available G8 mask as a spectral template with the offline version\footnote{available at the ESO pipeline website \url{http://eso.org/sci/software/pipelines/}} and a larger width of the window for the evaluation of the CCF.
The ESPRESSO RVs are reported in Table \ref{tab:espresso} and are shown in Figure \ref{fig:rml} with blue dots as a function of time, compared with the ones obtained with HARPS on Aug 11. To obtain  Figure \ref{fig:rml}, we applied an offset of $\sim -25$ and $\sim -105$ \ms~for the HARPS RVs obtained with TERRA and DRS, respectively, and $\sim -33$ \ms~for the ESPRESSO RVs. 
The RM amplitude is is slightly lower than 50 \ms~and it is compatible for the two series of data (and consistent between the two different RV extractions for HARPS data). The out of transit series show a clear increasing trend which is in contrast with the expectation from Keplerian motion, and is dominated by the stellar activity.
Interestingly, for different transit events the slope of the HARPS out-of-transit series, as extracted from the official DRS (green dots), is the same as the one shown by ESPRESSO (blue dots). The two series deviate at about 1.5 hr after the egress, when the the ESPRESSO time series shows an RV "bump" (investigated in Sect. \ref{sec:rml}). Instead, the out-of-transit series obtained with TERRA (red dots) shows a lower steepness after the fourth contact with respect to the DRS series. This demonstrates that the usage of TERRA provides RV estimations that are less affected by stellar activity, since the disagreement  between TERRA and DRS occurs when the RV are dominated by the activity (out of the transit) and not by the  planet contribution (during the transit). 

Finally, the comparison between the RM effect for the transit of Aug 11 detected by HARPS and PFS (reported by \citealt{2020AJ....159..112M}) shows that the out-of-transit modulation is actually due to the stellar activity, since the data from two different instruments with different RV extraction (TERRA vs Iodine cell technique) are in full agreement. %(Fig. \ref{fig:harps-pfs}). 
No comparison can be performed with the transits reported in \cite{2020ApJ...892L..21Z}, namely Aug 11 with CHIRON and Aug 18 and Oct 7 with PFS, since the the data shown in their Fig. 3 are already corrected for the activity and have a shorter out-of-transit baseline.

\longtab{
\begin{longtable}{lllllll}
\caption{\label{tab:espresso} Time series of the ESPRESSO spectra for DS\,Tuc\,A: BJD$_{\rm TDB}$, radial velocity (RV) and radial velocity error (RV$_{\rm err}$), and CaII and H$\alpha$ with corresponding uncertainties as obtained with ACTIN.}\\
\hline\hline
BJD$_{\rm TDB}$--2\,450\,000 & RV & RV$_{\rm Err}$ & CaII & CaII$_{\rm Err}$ & H$\alpha$ & H$\alpha _{\rm Err}$  \\
 & [km\,s$^{-1}$] & [km\,s$^{-1}$]  & & & &  \\
\hline\hline
8763.48468831  &  7.7820  &  0.0025  &  0.578  &  0.001  &  0.3376  &  0.0002  \\
8763.48885959  &  7.7838  &  0.0024  &  0.578  &  0.001  &  0.3396  &  0.0002  \\
8763.49294730  &  7.7802  &  0.0031  &  0.582  &  0.001  &  0.3415  &  0.0002  \\
8763.49724229  &  7.7775  &  0.0029  &  0.576  &  0.001  &  0.3382  &  0.0003  \\
8763.50129564  &  7.7748  &  0.0033  &  0.630  &  0.001  &  0.3634  &  0.0002  \\
8763.50565712  &  7.7777  &  0.0038  &  0.631  &  0.001  &  0.3648  &  0.0003  \\
8763.50997977  &  7.7813  &  0.0029  &  0.580  &  0.001  &  0.3388  &  0.0002  \\
8763.51404318  &  7.7840  &  0.0027  &  0.615  &  0.001  &  0.3608  &  0.0002  \\
8763.51806832  &  7.7837  &  0.0025  &  0.572  &  0.001  &  0.3360  &  0.0002  \\
8763.52234925  &  7.7844  &  0.0025  &  0.586  &  0.001  &  0.3404  &  0.0002  \\
8763.52651548  &  7.7862  &  0.0025  &  0.618  &  0.001  &  0.3510  &  0.0002  \\
8763.53070295  &  7.7828  &  0.0025  &  0.638  &  0.001  &  0.3638  &  0.0002  \\
8763.53477641  &  7.7907  &  0.0026  &  0.632  &  0.001  &  0.3641  &  0.0003  \\
8763.53907901  &  7.7918  &  0.0026  &  0.590  &  0.001  &  0.3414  &  0.0002  \\
8763.54328764  &  7.7970  &  0.0024  &  0.622  &  0.001  &  0.3636  &  0.0003  \\
8763.54743106  &  7.7983  &  0.0022  &  0.614  &  0.001  &  0.3487  &  0.0002  \\
8763.55159798  &  7.8045  &  0.0023  &  0.591  &  0.001  &  0.3424  &  0.0002  \\
8763.55588670  &  7.8045  &  0.0024  &  0.638  &  0.001  &  0.3601  &  0.0002  \\
8763.56002593  &  7.8009  &  0.0024  &  0.582  &  0.001  &  0.3398  &  0.0002  \\
8763.56415816  &  7.8037  &  0.0025  &  0.573  &  0.001  &  0.3382  &  0.0002  \\
8763.56837395  &  7.8029  &  0.0025  &  0.577  &  0.001  &  0.3399  &  0.0002  \\
8763.57263062  &  7.8126  &  0.0022  &  0.584  &  0.001  &  0.3419  &  0.0002  \\
8763.57678706  &  7.8396  &  0.0021  &  0.624  &  0.001  &  0.3626  &  0.0002  \\
8763.58092606  &  7.8426  &  0.0021  &  0.618  &  0.001  &  0.3619  &  0.0002  \\
8763.58510390  &  7.8536  &  0.0021  &  0.587  &  0.001  &  0.3425  &  0.0002  \\
8763.58938240  &  7.8626  &  0.0021  &  0.589  &  0.001  &  0.3414  &  0.0002  \\
8763.59353079  &  7.8663  &  0.0021  &  0.579  &  0.001  &  0.3392  &  0.0002  \\
8763.59771878  &  7.8779  &  0.0021  &  0.612  &  0.001  &  0.3482  &  0.0002  \\
8763.60192375  &  7.8833  &  0.0021  &  0.638  &  0.001  &  0.3613  &  0.0002  \\
8763.60605201  &  7.8815  &  0.0021  &  0.583  &  0.001  &  0.3393  &  0.0002  \\
8763.61019871  &  7.8751  &  0.0022  &  0.614  &  0.001  &  0.3486  &  0.0002  \\
8763.61433442  &  7.8731  &  0.0024  &  0.631  &  0.001  &  0.3568  &  0.0002  \\
8763.61862714  &  7.8707  &  0.0024  &  0.579  &  0.001  &  0.3393  &  0.0002  \\
8763.62286022  &  7.8692  &  0.0024  &  0.615  &  0.001  &  0.3485  &  0.0002  \\
8763.62693358  &  7.8647  &  0.0023  &  0.570  &  0.001  &  0.3368  &  0.0002  \\
8763.63110747  &  7.8576  &  0.0024  &  0.584  &  0.001  &  0.3410  &  0.0002  \\
8763.63540993  &  7.8511  &  0.0021  &  0.585  &  0.001  &  0.3408  &  0.0002  \\
8763.63956188  &  7.8459  &  0.0022  &  0.577  &  0.001  &  0.3378  &  0.0003  \\
8763.64375009  &  7.8407  &  0.0021  &  0.575  &  0.001  &  0.3358  &  0.0002  \\
8763.64790664  &  7.8337  &  0.0022  &  0.578  &  0.001  &  0.3391  &  0.0002  \\
8763.65208815  &  7.8297  &  0.0023  &  0.576  &  0.001  &  0.3398  &  0.0002  \\
8763.65625831  &  7.8331  &  0.0023  &  0.640  &  0.001  &  0.3631  &  0.0002  \\
8763.66045677  &  7.8204  &  0.0023  &  0.586  &  0.001  &  0.3421  &  0.0002  \\
8763.66456446  &  7.8213  &  0.0023  &  0.640  &  0.001  &  0.3650  &  0.0002  \\
8763.66875571  &  7.8172  &  0.0026  &  0.584  &  0.001  &  0.3412  &  0.0002  \\
8763.67304310  &  7.8208  &  0.0024  &  0.624  &  0.001  &  0.3534  &  0.0002  \\
8763.67717311  &  7.8150  &  0.0023  &  0.584  &  0.001  &  0.3404  &  0.0002  \\
8763.68135733  &  7.8213  &  0.0022  &  0.580  &  0.001  &  0.3379  &  0.0002  \\
8763.68555534  &  7.8297  &  0.0022  &  0.577  &  0.001  &  0.3381  &  0.0002  \\
8763.68972567  &  7.8374  &  0.0021  &  0.587  &  0.001  &  0.3412  &  0.0002  \\
8763.69395469  &  7.8503  &  0.0021  &  0.585  &  0.001  &  0.3427  &  0.0002  \\
8763.69811797  &  7.8746  &  0.0021  &  0.585  &  0.001  &  0.3411  &  0.0002  \\
8763.70231511  &  7.8876  &  0.0022  &  0.582  &  0.001  &  0.3401  &  0.0002  \\
8763.70646524  &  7.8969  &  0.0022  &  0.591  &  0.001  &  0.3412  &  0.0002  \\
8763.71071530  &  7.8946  &  0.0023  &  0.579  &  0.001  &  0.3390  &  0.0002  \\
8763.71486075  &  7.9020  &  0.0023  &  0.601  &  0.001  &  0.3444  &  0.0002  \\
8763.71911475  &  7.9021  &  0.0024  &  0.583  &  0.001  &  0.3407  &  0.0002  \\
8763.72330570  &  7.9037  &  0.0022  &  0.568  &  0.001  &  0.3363  &  0.0002  \\
8763.72740725  &  7.9077  &  0.0023  &  0.579  &  0.001  &  0.3389  &  0.0002  \\
8763.73167455  &  7.9109  &  0.0024  &  0.620  &  0.001  &  0.3620  &  0.0002  \\
8763.73582112  &  7.9150  &  0.0024  &  0.585  &  0.001  &  0.3404  &  0.0002  \\
8763.74004020  &  7.9194  &  0.0023  &  0.591  &  0.001  &  0.3413  &  0.0002  \\
8763.74422129  &  7.9202  &  0.0022  &  0.632  &  0.001  &  0.3632  &  0.0003  \\
8763.74840879  &  7.9232  &  0.0022  &  0.590  &  0.001  &  0.3420  &  0.0002  \\
8763.75250591  &  7.9255  &  0.0024  &  0.586  &  0.001  &  0.3424  &  0.0002  \\
8763.75672833  &  7.9276  &  0.0025  &  0.620  &  0.001  &  0.3620  &  0.0002  \\
8763.76091618  &  7.9357  &  0.0025  &  0.581  &  0.001  &  0.3394  &  0.0002  \\
8763.76514891  &  7.9492  &  0.0023  &  0.586  &  0.001  &  0.3422  &  0.0002  \\
8763.76933939  &  7.9590  &  0.0026  &  0.616  &  0.001  &  0.3484  &  0.0002  \\
8763.77354141  &  7.9622  &  0.0025  &  0.586  &  0.001  &  0.3407  &  0.0002  \\
8763.77768807  &  7.9621  &  0.0026  &  0.592  &  0.001  &  0.3426  &  0.0002  \\
8763.78192522  &  7.9595  &  0.0027  &  0.588  &  0.001  &  0.3401  &  0.0002  \\
8763.78608207  &  7.9663  &  0.0026  &  0.613  &  0.001  &  0.3478  &  0.0002  \\
8763.79031076  &  7.9678  &  0.0025  &  0.574  &  0.001  &  0.3357  &  0.0002  \\
8763.79442721  &  7.9664  &  0.0024  &  0.584  &  0.001  &  0.3422  &  0.0002  \\
8763.79858642  &  7.9655  &  0.0025  &  0.582  &  0.001  &  0.3399  &  0.0002  \\
8763.80276425  &  7.9639  &  0.0024  &  0.569  &  0.001  &  0.3355  &  0.0002  \\
8763.80699358  &  7.9638  &  0.0025  &  0.577  &  0.001  &  0.3395  &  0.0002  \\
8763.81120316  &  7.9615  &  0.0027  &  0.616  &  0.001  &  0.3471  &  0.0002  \\
8763.81534918  &  7.9644  &  0.0027  &  0.639  &  0.001  &  0.3649  &  0.0003  \\
8763.81958090  &  7.9627  &  0.0029  &  0.590  &  0.001  &  0.3411  &  0.0002  \\
8763.82359114  &  7.9608  &  0.0029  &  0.589  &  0.001  &  0.3415  &  0.0002  \\
8763.82787800  &  7.9605  &  0.0029  &  0.591  &  0.001  &  0.3414  &  0.0002  \\
8763.83202066  &  7.9594  &  0.0031  &  0.582  &  0.001  &  0.3413  &  0.0002  \\
8763.83626790  &  7.9523  &  0.0038  &  0.584  &  0.001  &  0.3401  &  0.0002  \\
8763.84035912  &  7.9562  &  0.0032  &  0.589  &  0.001  &  0.3415  &  0.0002  \\
8763.84471415  &  7.9527  &  0.0032  &  0.582  &  0.001  &  0.3405  &  0.0002  \\
8763.85034655  &  7.9560  &  0.0030  &  0.587  &  0.001  &  0.3430  &  0.0002  \\
8763.85452356  &  7.9547  &  0.0029  &  0.588  &  0.001  &  0.3426  &  0.0002  \\
8763.85875066  &  7.9499  &  0.0031  &  0.568  &  0.001  &  0.3358  &  0.0002  \\
8763.86291629  &  7.9502  &  0.0032  &  0.575  &  0.001  &  0.3359  &  0.0002  \\
8763.86702102  &  7.9401  &  0.0033  &  0.585  &  0.001  &  0.3416  &  0.0002  \\
8763.87363471  &  7.9452  &  0.0031  &  0.614  &  0.001  &  0.3489  &  0.0002  \\
8763.87777435  &  7.9381  &  0.0031  &  0.594  &  0.001  &  0.3422  &  0.0002  \\
8763.88187546  &  7.9355  &  0.0031  &  0.630  &  0.001  &  0.3640  &  0.0002  \\
8763.88618500  &  7.9408  &  0.0029  &  0.633  &  0.001  &  0.3639  &  0.0002  \\
8763.89034833  &  7.9319  &  0.0027  &  0.581  &  0.001  &  0.3398  &  0.0002  \\
\hline\hline
\end{longtable}
}
%--------------------------------------------------------------------
\section{RV modelling and mass upper limit of DS\,Tuc\,A\,b} \label{sec:rvall}

\subsection{Frequency analysis} \label{sec:gls}
\begin{figure}
    \centering
    \includegraphics[trim={1 2.5cm 1 1},clip,width=0.43\textwidth, angle=270]{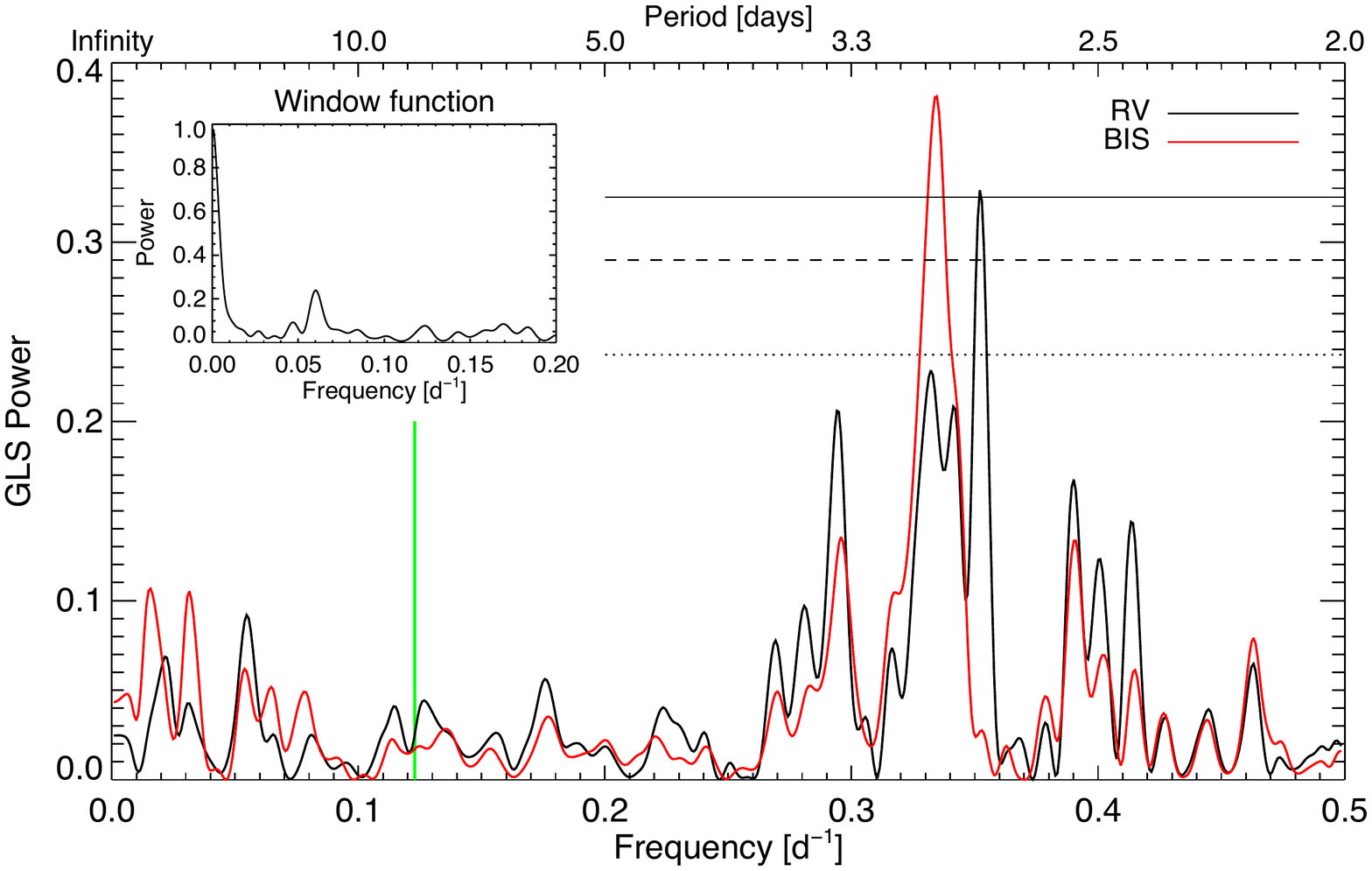}
    \caption{GLS of the HARPS RV (black line) and BIS (red line) time series of DS\,Tuc\,A. The green vertical line indicates the location of the planet orbital period, while the window function of the observations is reported in the inset plot. Dotted, dashed and solid black horizontal lines indicate the FAP thresholds at 1\%, 0.1\% and 0.01\%, respectively, for the RV periodogram.}
    \label{fig:gls}
\end{figure}

Fig. \ref{fig:gls} compares the Generalised Lomb-Scargle periodogram (GLS, \citealt{2009A&A...505..859Z}) of the HARPS RVs (black line) and BIS (red line) time series. The stellar activity dominates the behaviour of the RVs, since the two GLS periodograms show a similar shape with a significant periodicity around 3 days (False Alarm Probability lower than 0.01\%, evaluated with 10000 bootstrap random permutations), which is compatible with the rotation period of the star (2.85 days, see \citealt{2019A&A...630A..81B}). The presence of additional peaks between 0.29 and 0.4 d$^{-1}$ is mainly due to the spectral window of the observations (reported in the inset plot). The expected location of the orbital period of DS\,Tuc\,A\,b is marked with a green vertical line. No evidence of power is present around that frequency value.

The GLS of the other activity indicators extracted from TERRA show the similar periodic content, but only the BIS shows a strong correlation with the RVs (Spearman coefficient close to 0.9). 
The decorrelation of the RVs with the values of the BIS produces a time series with dispersion of about 90 \ms. The GLS of the residuals does not show any periodicity at the planet period, only some additional power around 3 days, suggesting that the RV contribution of the stellar activity must be modelled in a more sophisticated manner.

\subsection{Gaussian Process regression} \label{sec:rv}
\subsubsection{Quasi-periodic kernel}
We used a Gaussian process (GP) regression to model the activity correlated signal in the RVs. The fitting analysis was carried out using the publicly available Monte Carlo nested sampler and Bayesian inference tool \texttt{MultiNestv3.10} (e.g. \citealt{Feroz2019}), through the \texttt{pyMultiNest} wrapper \citep{Buchner2014}. The algorithm was setup to run with 1000 live points and a sampling efficiency of 0.3. We fitted the combined HARPS+ESPRESSO dataset, excluding the data taken during the transits. The zero-point corrections between the two datasets ($\gamma_{\rm HARPS}$ and $\gamma_{\rm ESPRESSO}$ in Table \ref{tab:rvgp}) are also fitted in the process. In what follows, we modelled the planetary signal with a sinusoid, assuming a circular orbit as in \cite{2019A&A...630A..81B}. To confirm this assumption, we performed a joint RV+\textit{TESS} photometry Monte-Carlo analysis, using the same transit light curves analysed in \cite{2019A&A...630A..81B}. We found that the use of the RVs does not help in constraining the eccentricity, which results to be compatible with zero within $\sim$\,1.5$\sigma$ ($e_{\rm b}=0.14^{+0.29}_{-0.10}$). Hereafter we present results from the analysis of only RVs, adopting the planet radius given by \cite{2019A&A...630A..81B}, and assuming as priors for the transit ephemeris the more precise values of $P_{\rm b}$ and $T_{\rm b,\, conj.}$, obtained by \cite{2019ApJ...880L..17N}.            
Initially, the activity term was modelled by adopting the quasi-periodic covariance matrix to model the correlated signal due to stellar activity, defined as follows:

\begin{eqnarray} \label{eq:eqgpkernel}
k(t, t^{\prime}) = h^2\cdot\exp\large[-\frac{(t-t^{\prime})^2}{2\lambda^2} - \frac{sin^{2}(\pi(t-t^{\prime})/\theta)}{2w^2}\large] + \nonumber \\
+\, (\sigma^{2}_{\rm RV}(t)+\sigma^{2}_{\rm jit})\cdot\delta_{t, t^{\prime}}
\end{eqnarray}
Here, $t$ and $t^{\prime}$ represent two different epochs of observations, $\sigma_{\rm RV}$ is the radial velocity uncertainty, and $\delta_{t, t^{\prime}}$ is the Kronecker delta. Our analysis takes into account other sources of uncorrelated noise -- instrumental and/or astrophysical -- by including a constant jitter term $\sigma_{\rm jit}$ which is added in quadrature to the formal uncertainties $\sigma_{\rm RV}$. The GP hyper-parameters are $h$, which denotes the scale amplitude of the correlated signal; $\theta$, which represents the periodic time-scale of the modelled signal, and corresponds to the stellar rotation period; $w$, which describes the "weight" of the rotation period harmonic content within a complete stellar rotation (i.e. a low value of $w$ indicates that the periodic variations contain a significant contribution from the harmonics of the rotation periods); and $\lambda$, which represents the decay timescale of the correlations, and is related to the temporal evolution of the magnetically active regions responsible for the correlated signal observed in the RVs.

The GP regression and kernel implementation was performed using the publicly available \textsc{python} module \textsc{GEORGE} v0.2.1 \citep{2015ITPAM..38..252A}. which was integrated within the nested sampling algorithm. A summary of priors and results is given in Table \ref{tab:rvgp}, and the posterior distributions for all the free parameters are shown in Fig. \ref{fig:corner}. The fitted planetary Doppler semi-amplitude $K_{\rm b}$ is consistent with zero, therefore we can only provide an upper limit for the planet mass  M$_{\rm b}]<14.4 M_{\oplus}$ at 1-$\sigma$ level. Fig. \ref{fig:spectorbitGPqp} shows the RV residuals (best-fit stellar signal and instrumental offsets subtracted from the original RVs) phase folded to the orbital period of DS Tuc A\,b. 
According to our results, the evolutionary time scale $\lambda$ of the quasi-periodic activity component is of the order of twice the rotation period. The quick evolution of the correlated activity signal is confirmed by analysing the H$\alpha$ and the S-index activity diagnostics, calculated with TERRA from the HARPS spectra, with the GP quasi-periodic kernel. By adopting for the hyper-parameters the same priors used for the RVs (except for the amplitude $h$), we got $\theta_{\rm H\alpha}=1.499^{+1.488}_{-0.008}$ d and $\lambda_{\rm H\alpha}=6.8\pm1.2$ d, $\theta_{\rm S-index}=2.9^{+0.1}_{-1.4}$ d and $\lambda_{\rm S-index}=3.7^{+1.6}_{-1.0}$ d. The posterior distribution of the rotation period for both activity indexes shows an over-density at the location of the first harmonic, as indicated by the asymmetric error bars, which is not found for the RVs.
This evidence suggested a test with a different kernel for modelling the activity signal in the RVs, presented in the following subsection.

\subsubsection{Simple Harmonic Oscillator kernel}
For this test, we considered a mixture of two stochastically driven, damped simple harmonic oscillators (SHOs) with undamped periods of $P_{\rm rot}$ and $P_{\rm rot}$/2. This can be done by combining two \texttt{SHOTerm} kernels included in the package \texttt{celerite} \citep{celerite} as done, for instance, by \cite{2019ApJ...885L..12D} to model the \textit{Kepler/K2} photometric variability of another very young star, V1298 Tau. 
The power spectral density corresponding to this kernel (also known as the `rotation kernel') is 
\begin{equation}
S(\omega) = \sqrt{\frac{2}{\pi}} \frac{S_1\omega_1^4}{(\omega^2-\omega_1^2)^2 +  2\omega_1^2\omega^2} + \sqrt{\frac{2}{\pi}} \frac{S_2\omega_2^4}{(\omega^2-\omega_2^2)^2 + 2\omega_2^2\omega^2/Q^2},
\end{equation}
where
\begin{gather} \label{eqn:sho1}
    S_{\rm 1}=\frac{A}{\omega_{\rm 1}Q_{\rm 1}}, \\ 
    S_{\rm 2}=\frac{A}{\omega_{\rm 2}Q_{\rm 2}}\times f, \\
    \omega_{\rm 1}=\frac{4 \pi Q_{\rm 1}}{P_{\rm rot}\sqrt{4Q_{\rm 1}^{2}-1}}, \\
    \omega_{\rm 2}=\frac{8 \pi Q_{\rm 2}}{P_{\rm rot}\sqrt{4Q_{\rm 2}^{2}-1}}, \\
    Q_{\rm 1}= \frac{1}{2}+Q_{\rm 0}+\Delta Q,\\
    \label{eqn:sho2}
    Q_{\rm 2}= \frac{1}{2}+Q_{\rm 0}. 
\end{gather}

The parameters in Eqs. [\ref{eqn:sho1}-\ref{eqn:sho2}], where the subscripts 1 and 2 refer to the primary ($P_{\rm rot}$) and secondary ($P_{\rm rot}$/2) modes, are the inputs to the \texttt{SHOTerm} kernels. However, instead of using them directly, we adopt a different parametrization using the following variables as free hyper-parameters in the MC analysis, from which $S_{\rm i}$, $Q_{\rm i}$, and $\omega_{\rm i}$ are derived through Eqs. [2-7]: the variability amplitude $A$ (\msq), the stellar rotation period $P_{rot}$ (the periodicity of the primary term), the quality factor $Q_{0}$, the difference $\Delta$Q between the quality factors of the first and second terms, and the fractional amplitude $f$ of the secondary mode relative to the primary\footnote{We notice that the kernel used in this work employs a slightly different parameterization than the version of the same ``rotation'' kernel defined in the most recent release of \texttt{celerite} (\url{https://github.com/exoplanet-dev/celerite2}).}.
We adopted priors in the natural logarithm space for all the hyper-parameters, except for $P_{rot}$. Priors and results of the fit are shown in Tab. \ref{tab:rvgp} and in the second panel of Fig. \ref{fig:corner}. Even in this case, we are able to provide only an upper limit for the planet mass (M$_{\rm b}]<40.3 M_{\oplus}$ at 1-$\sigma$ level), higher than that found using a quasi-periodic kernel. The instrumental jitters are lower for both HARPS and ESPRESSO, denoting a better fit to the data than the quasi-periodic activity model. However, the model with the mixture of SHOs is statistically mildly favored over the quasi-periodic ($\Delta \ln \mathcal{Z}$=3.4), therefore we do not have a strong argument to prefer this model over the other.   
Fig. \ref{fig:gpqpshoharps} shows a comparison between the best-fit activity model obtained with the quasi-periodic and rotation kernel from the HARPS+ESPRESSO joint analysis, for a subset of the HARPS RVs.

\begin{table*}[htbp]
    \centering
     \caption{Priors and percentiles (16$^{\rm th}$, 50$^{\rm th}$, and 84$^{\rm th}$) of the posterior distributions concerning the analysis of the RV datasets of DS\,Tuc\,A (HARPS + ESPRESSO).}
    \begin{tabular}{lccc}
    \hline
    \noalign{\smallskip}
        Parameter & Prior & \multicolumn{2}{c}{Best-fit values}  \\
                  &        & Quasi-periodic & SHO mixture \\
        \noalign{\smallskip}
        \hline \hline
        \noalign{\smallskip}
        \textbf{Stellar activity term}\\
        \noalign{\smallskip}
        $\theta$ [d]& $\mathcal{U}(0,4)$ & $2.95^{+0.01}_{-0.02}$ & --\\
        \noalign{\smallskip}
        $\lambda$ [d] & $\mathcal{U}(0,1000)$ & $8.2^{+1.6}_{-1.1}$ & --\\
        \noalign{\smallskip}
        $h$ [\ms] & $\mathcal{U}(0,300)$ & $140^{+16}_{-13}$ & -- \\
        \noalign{\smallskip}
        $w$ & $\mathcal{U}(0,1)$  & $0.14^{+0.02}_{-0.01}$ & -- \\
        \noalign{\smallskip}
        $ln(A)$ & $\mathcal{U}(0.05,20)$ & -- & $9.3^{+0.5}_{-0.4}$ \\
        \noalign{\smallskip}
        $P_{\rm rot}$ [d] & $\mathcal{U}(2,4)$ & -- & $2.90^{+0.04}_{-0.03}$ \\
        \noalign{\smallskip}
        $ln(Q_{0})$  & $\mathcal{U}(-10,10)$ & -- & $-2.9\pm0.4$  \\
        \noalign{\smallskip}
        $ln(\Delta Q)$ & $\mathcal{U}(0,1)$  & --& $3.4^{+0.7}_{-0.6}$  \\
        \noalign{\smallskip}
        $f$ & $\mathcal{U}(0,2)$  & -- & $0.8^{+0.5}_{-0.4}$  \\
        \noalign{\smallskip}
        \hline
        \noalign{\smallskip}
        \textbf{Planetary parameters}\\
        \noalign{\smallskip}
        $K_{\rm b}$ [\ms] & $\mathcal{U}(0,50)$ & $3.2^{+3.4}_{-2.2}$ ($<$4.7, 68$\%$)  & $8.6^{+10.7}_{-6.2}$ ($<$13.1, 68$\%$) \\
        \noalign{\smallskip}
        $P_{\rm b}$ [d] & $\mathcal{N}(8.138268,0.000011)$\tablefootmark{a} & $8.138268\pm0.000011$ & $8.138268\pm0.000011$ \\
        \noalign{\smallskip}
        $T_{\rm b,\:conj}$ [BJD-2\,450\,000] & $\mathcal{N}(8332.30997,0.00026)$\tablefootmark{a} & $8332.30997\pm0.00025$ & $8332.30998\pm0.00026$ \\
        \noalign{\smallskip}
        Mass, $m_{\rm b}$ [M$_{\oplus}$] & derived & <14.4 ($68\%$) & <40.3 ($68\%$) \\
        \noalign{\smallskip}
        Density, $\rho_{\rm b}$ [$g$ $cm^{-3}$] & derived & <0.44 ($68\%$) & <1.25 ($68\%$) \\
        \noalign{\smallskip}
        \hline
        \noalign{\smallskip}
        \textbf{Instrument-related parameters}\\
        $\sigma_{\rm jit, HARPS}$ [\ms] & $\mathcal{U}(0,100)$  & $6.2^{+3.1}_{-2.8}$ & $2.5^{+2.4}_{-1.7}$ \\
        \noalign{\smallskip}
        $\sigma_{\rm jit, ESPRESSO}$ [\ms] & $\mathcal{U}(0,100)$ & $4.0^{+0.7}_{-0.6}$  & $0.6^{+0.6}_{-0.4}$ \\
        \noalign{\smallskip}
        $\gamma_{\rm HARPS}$ [\ms] & $\mathcal{U}(-500,500)$  & $40\pm26$ & $46^{+13}_{-12}$ \\
        \noalign{\smallskip}
        $\gamma_{\rm ESPRESSO}$ [\ms] &  $\mathcal{U}(7500,8200)$ & $7917^{+81}_{-79}$ & $7815^{+100}_{-101}$ \\
        \noalign{\smallskip}
        \hline
        Bayesian evidence, $\ln\mathcal{Z}$ &  & -744.3 & -740.9 \\
        \hline \hline
    \end{tabular}
    \tablefoot{
    \tablefoottext{a}{After \cite{2019ApJ...880L..17N}}
    }
    \label{tab:rvgp}
\end{table*}

%\clearpage

\begin{figure*}
    \centering
    \begin{subfigure}[t]{\hsize}
    \includegraphics[width=0.9\textwidth, angle=0]{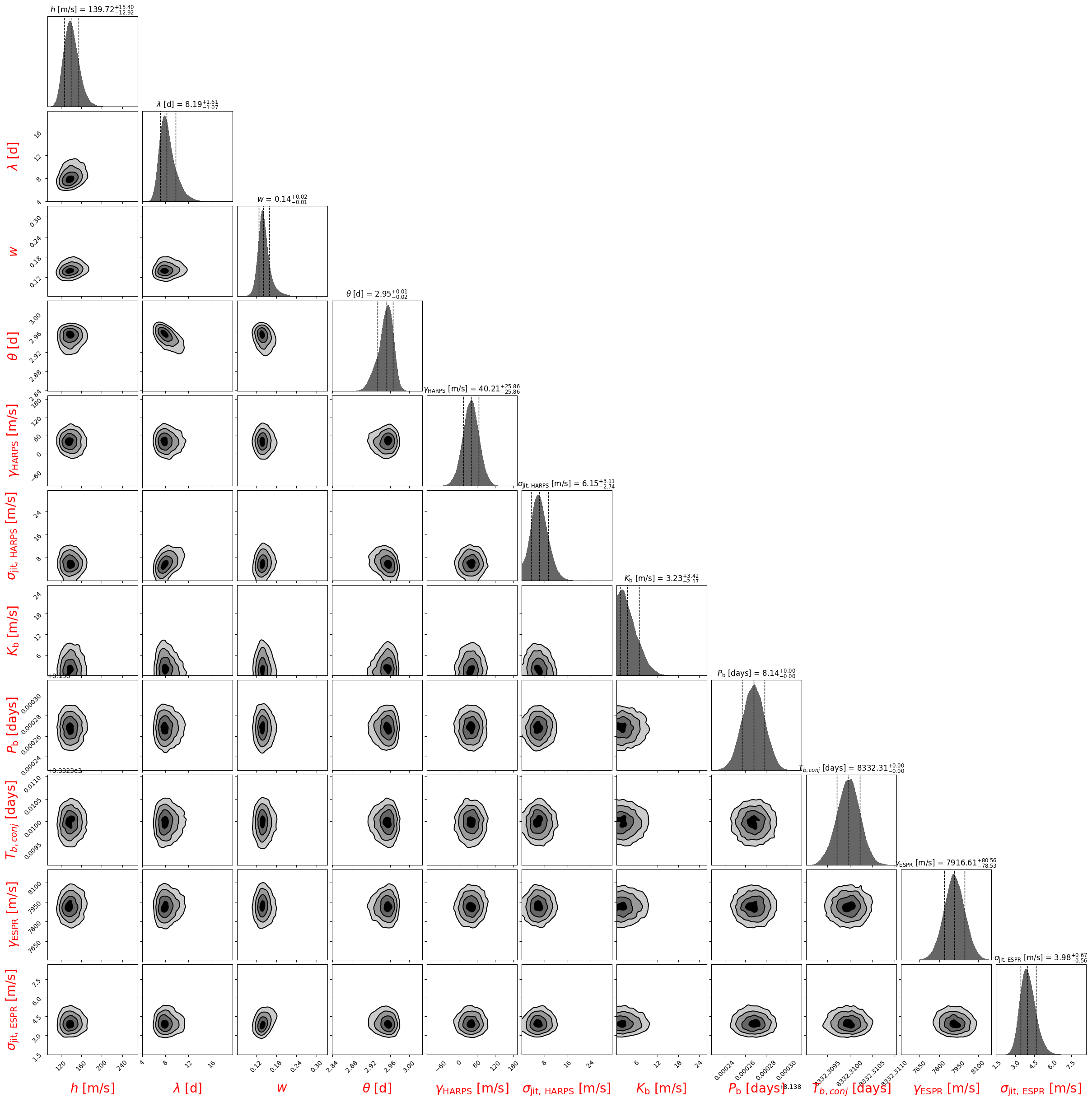}\\
    \end{subfigure}
     \caption{Corner plots for GP quasi-periodic (upper panel) and SHO mixture (lower panel) models used to fit the HARPS+ESPRESSO radial velocities.}
     \end{figure*}
      \clearpage   
    \begin{figure*}[tb]\ContinuedFloat
    \begin{subfigure}[t]{\hsize}
    \includegraphics[width=0.9\textwidth, angle=0]{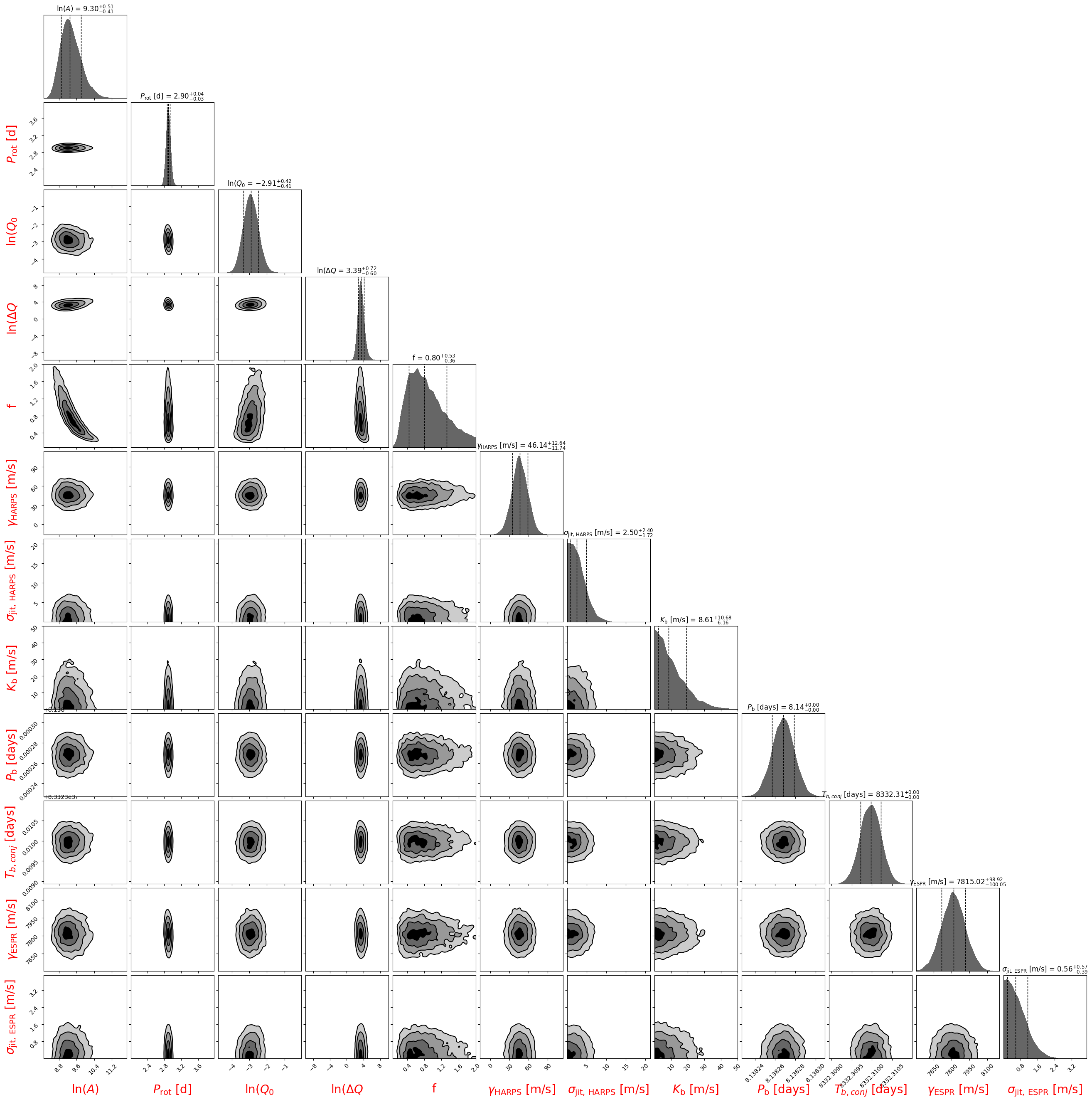}\\
    \end{subfigure}
   \caption{Continued}   
    \label{fig:corner}
\end{figure*}

%\clearpage 

\begin{figure}
    \centering
    \includegraphics[width=0.5\textwidth, angle=0]{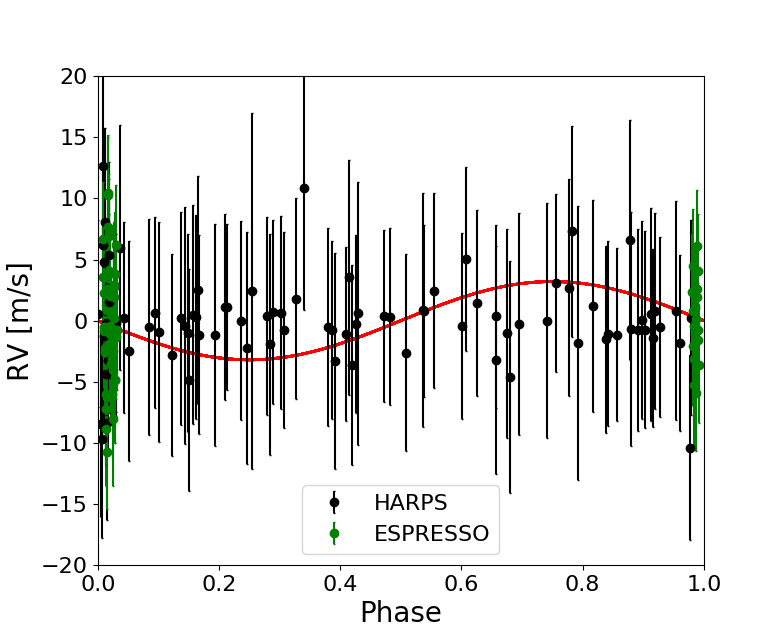}
    \caption{Spectroscopic orbit of planet DS Tuc A\,b. The instrumental offsets and the best-fit stellar activity signal (GP quasi-periodic model) have been subtracted from the HARPS and ESPRESSO original RVs. The error bars include the uncorrelated jitter terms added in quadrature to the formal uncertainties. The curve in red represents the best-fit Doppler signal (circular obit; $K_b$=3.2 \ms). }
    \label{fig:spectorbitGPqp}
\end{figure}

%\clearpage 

\begin{figure}
    \centering
    \includegraphics[trim={0.5cm 1 1cm 1},clip,width=0.52\textwidth, angle=0]{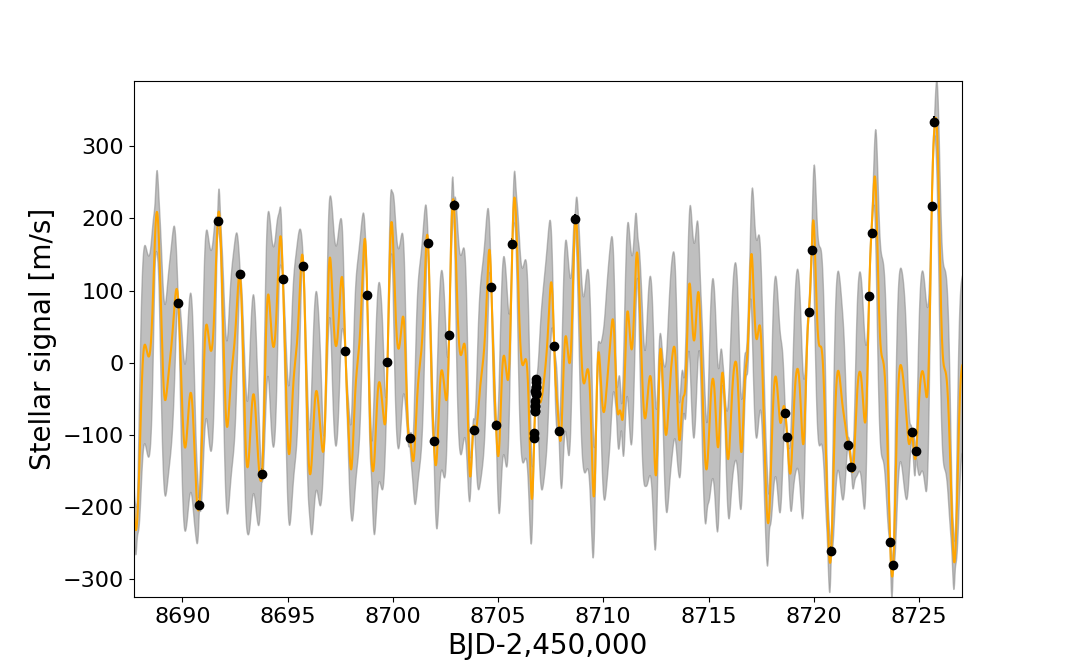}
    \includegraphics[trim={0.5cm 1 1cm 1},clip,width=0.52\textwidth, angle=0]{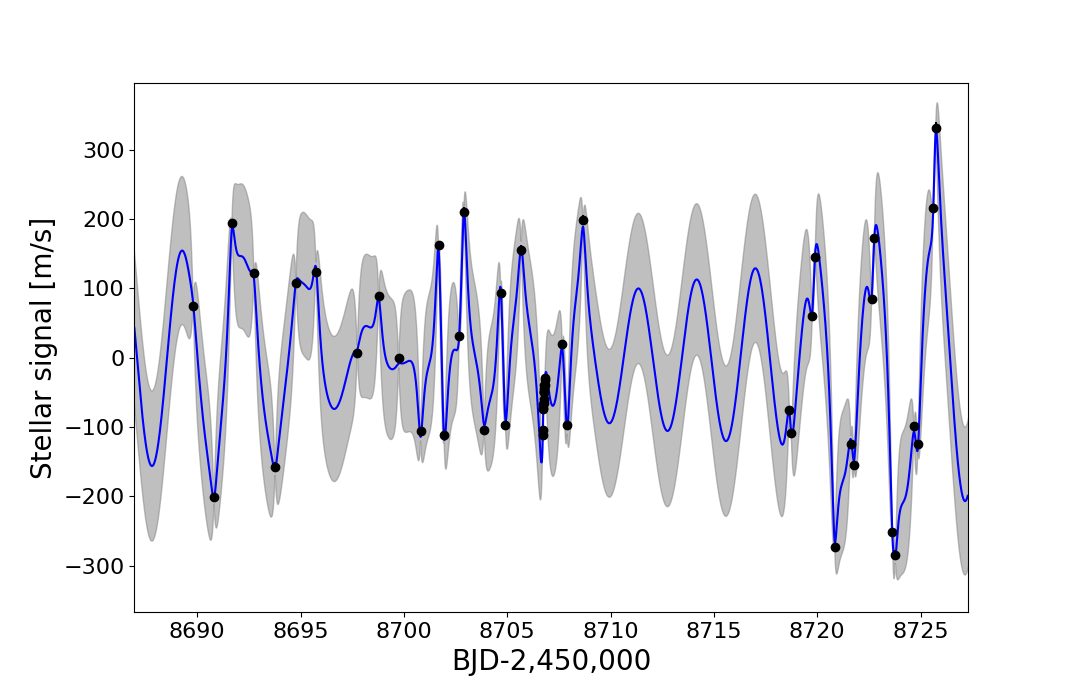}
    \caption{\textit{Upper panel.} GP quasi-periodic best-fit model (orange line), as obtained from the joint analysis of the HARPS+ESPRESSO RVs. Here we show a portion of the HARPS time series, after removing the best-fit signal induced by planet b. The shaded area in gray represents the $\pm1 \sigma$ confidence interval. \textit{Lower panel.} Same as above, but for the GP rotation kernel (blue line).}
    \label{fig:gpqpshoharps}
\end{figure}

%\clearpage

\subsection{Testing the mass upper limit of DS\,Tuc\,A\,b through injection/recovery simulations}
\label{sect:rvsimu}
The main conclusions from the RV analysis are that we are able to provide only an upper limit for the mass of DS\,Tuc\,A\,b, and this limit changes depending on how the correlated signal due to stellar activity is modelled. It may be that the GP quasi-periodic kernel is absorbing part of the planetary signal, causing the mass upper limit to be nearly 3 times lower than that found using the rotation kernel. This result has important consequences on characterising the planetary structure and studying the evolution of the planet atmosphere within the first Myrs from the formation, based on models of photo-evaporation that require the current mass as input. Statistical considerations suggest that slight preference should be given to the results obtained with the adoption of the rotation kernel, therefore to the higher mass limit that we found for DS\,Tuc\,A\,b, even if there is not a strong evidence in favor of it. Based on a similar statistical argument, the rotation kernel describes the time series of the H$\alpha$ and CaII H\&K activity diagnostics better than the quasi-periodic kernel. On the other hand, both kernels represent a priori a proper choice to treat the activity-related RV signals for a very active star such as DS\,Tuc\,A, and still very little is known in literature about their relative performances for this kind of stars. Since it is difficult to drawn general conclusions about this issue, tests should be done on a star-by-star basis. Therefore, we devised two different sets of injection-retrieval simulations, provided in the next subsections, to get further information on which kernel provides more reliable results for the specific case of DS\,Tuc\,A. 

\subsubsection{Injection/recovery test performed directly on the real data}
\begin{figure*}[ht]
    \centering
    \begin{minipage}[b]{0.45\textwidth}
   \includegraphics[width=1.07\textwidth]{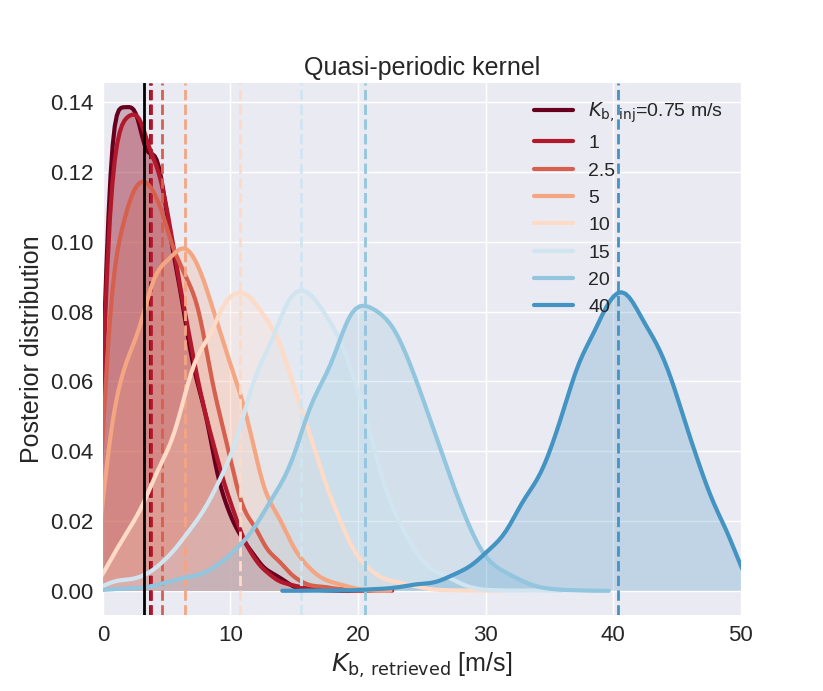}
    \end{minipage}
    \begin{minipage}[b]{0.45\textwidth}
  \includegraphics[width=1.05\textwidth]{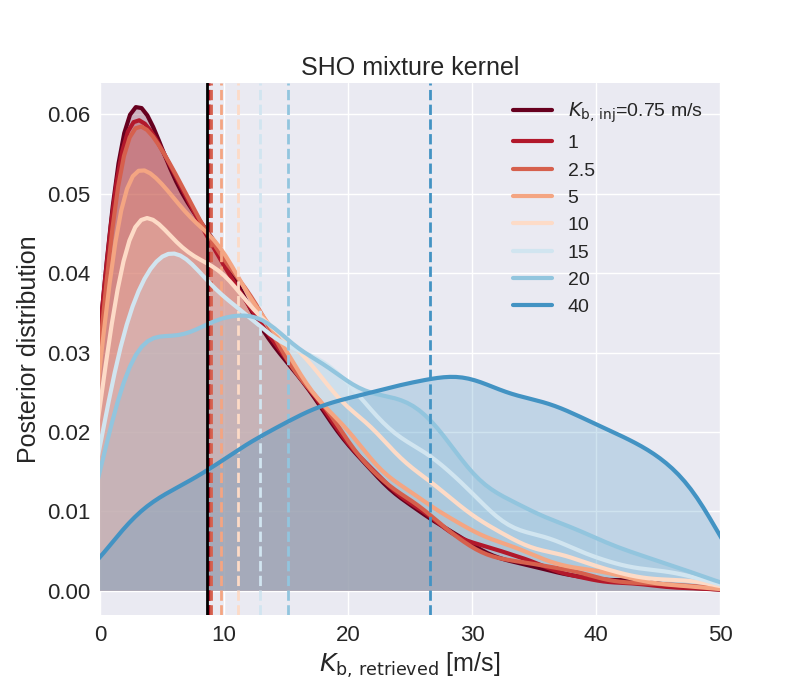}
    \end{minipage}
    \caption{Posteriors of the $K_{\rm b}$ parameter retrieved from the simulations involving the original RV datasets, with the injected $K_{\rm b,\, inj}$ values indicated in the legend. The vertical dashed lines represent the median of the posteriors of the corresponding colour. The vertical black lines indicates the median of the posteriors for the real data (3.2 and 8.6 \ms), as a reference.}
    \label{fig:kb_posteriors_simu_1}
\end{figure*}

First, we injected a planetary signal directly in the real data, namely using the same time stamps of the original RVs. This signal is represented by a sinusoid with period and phase fixed to the best-fit values found for DS\,Tuc\,A\,b from the analysis of the photometric light curve, and adopting a grid of values for the semi-amplitude $K_{\rm b}$ in the range 0.75-40 \ms. Then, each simulated dataset was fitted exactly within the same framework as for the real data, testing both the quasi-periodic and rotation kernels, and we compared the injected $K_{\rm b,\,inj}$ with the retrieved $K_{\rm b,\,retrieved}$, which is expressed as the 50th percentile of the posterior, as usual. This test can help to understand which model recovers better the semi-amplitude of the planetary signal. Since the transit ephemeris are very precise, the injection of the Doppler planetary signal should only amplify that already hidden in the data, which we could not constrain by fitting the original RVs. The results of this set of simulations are shown in Tab. \ref{tab:rv_simu_1}, and the posteriors of $K_{\rm b,\,retrieved}$ are shown in Fig. \ref{fig:kb_posteriors_simu_1}. There are some interesting features to point out. Concerning the quasi-periodic model, $K_{\rm b,\,retrieved}$ is only slightly overestimated and accurate within less than 1 \ms~for $K_{\rm b,\,inj}\geq$10 \ms, with a significance ranging from 2.4$\sigma$ to nearly 8.2$\sigma$, denoting a successful recovery of the injected signal. The overestimate does not exceeds 3 \ms~for the lower injected signal $K_{\rm b,\,inj}$=0.75 \ms. For the case of the rotation/SHO mixture kernel, the picture is different. $K_{\rm b,\,retrieved}$ is surprisingly underestimated for $K_{\rm b,\,inj}>$15 \ms, with the highest departure from the injected value occurring for $K_{\rm b,\,inj}$=40 \ms, which is supposed to be the easiest case for recovering the actual signal. While the overestimates observed for the quasi-periodic model %(despite they are one order of magnitude lower than the uncertainties on $K_{\rm b,\,retrieved}$) 
is an expected result, since the injected sinusoid adds up with the real signal due to planet b already present in the data, we interpret the underestimate observed for the case of the rotation kernel as a general failure of the model for values of $K_{\rm b,\,inj}$ sufficiently high. Further, in this case the error bars associated to $K_{\rm b,\,retrieved}$ are larger than for the quasi-periodic model, due to the density profile of the posteriors which follow an asymmetric distribution. For $K_{\rm b,\,inj}\leq$10 \ms, the recovered semi-amplitude is overestimated by an amount larger than for the quasi-periodic model. We remark that the best-fit values of the GP hyper-parameters derived from each mock dataset perfectly agree with those determined for the original RVs, indicating that the recovered stellar activity signals were not altered by the injection of a sinusoid into the data.  

The overall picture emerging from the first set of simulations lead us to conclude that, contrary to what suggested by the statistical evidence, the quasi-periodic model for the activity signal provides more reliable results and guarantees the best performance for DS\,Tuc\,A. This in turn makes us more confident that the mass of DS\,Tuc\,A\,b should be not larger than 14.4 M$_{\oplus}$ with 68$\%$ of confidence. 

\subsubsection{Injection/recovery test using mock RVs stemming from the best-fit solutions found for the real data}
In order to test further this conclusion, we devised a second type of simulations. We generated 10 mock RV time series using the best-fit solutions of the quasi-periodic and SHO mixture activity models, by randomly drawing the hyper-parameters from distributions based on the results in Tab. \ref{tab:rvgp}, and randomly sampling a list of predictions for the RV correlated signal at the real observing epochs using the \texttt{sample} module in \texttt{george} and \texttt{celerite}. We then injected a planetary sinusoid with $K_{\rm b}$=40 \ms~, $P_b$=8.138268 d, and $T_{b,\,conj}$=2458332.30997 BJD into the mock activity signals, including the best-fit RV instrumental zero-points $\gamma_{HARPS}$ and $\gamma_{ESPRESSO}$ in Tab. \ref{tab:rvgp} (the same constant values are adopted for all the mock dataset). Finally, we added to the RVs calculated so far a term (positive or negative) to take the instrumental uncorrelated jitters into account. For each time stamp, this term is randomly drawn from a normal distribution $\mathcal{N}(0,\sigma_{jitt}^{2})$ (for both HARPS and ESPRESSO), which corresponds to the operation $\sigma_{jitt}\times$\texttt{numpy.random.randn}($N$) using the syntax of \texttt{python}, where $N$ is the number of observing epochs. For each mock dataset, the jitter terms $\sigma_{jitt}$ are randomly drawn from the posterior distributions derived for the real data (Tab. \ref{tab:rvgp}). For the mock RVs we used the same error bars of the real data, and the instrumental zero-points and uncorrelated jitters were treated as free parameters in the Monte Carlo analysis, adopting in general the same framework for the analysis of the real RVs.
We fitted each mock RV dataset using both the quasi-periodic and rotation GP models. This gives a total of four tests: \textit{a)} generated: quasi-periodic; fitted: quasi-periodic; \textit{b)} generated: quasi-periodic; fitted: rotation; \textit{c)} generated: rotation; fitted: quasi-periodic; \textit{d)} generated: rotation; fitted: rotation. The idea behind this test is to evaluate the general performance of the two models in retrieving the Doppler planetary signal with a cross check, taking advantage of the fact that we know the underlying (mock) stellar activity signal in each dataset. 

\begin{figure*}
    \centering
    \begin{minipage}[b]{0.45\textwidth}
    \includegraphics[width=\textwidth]{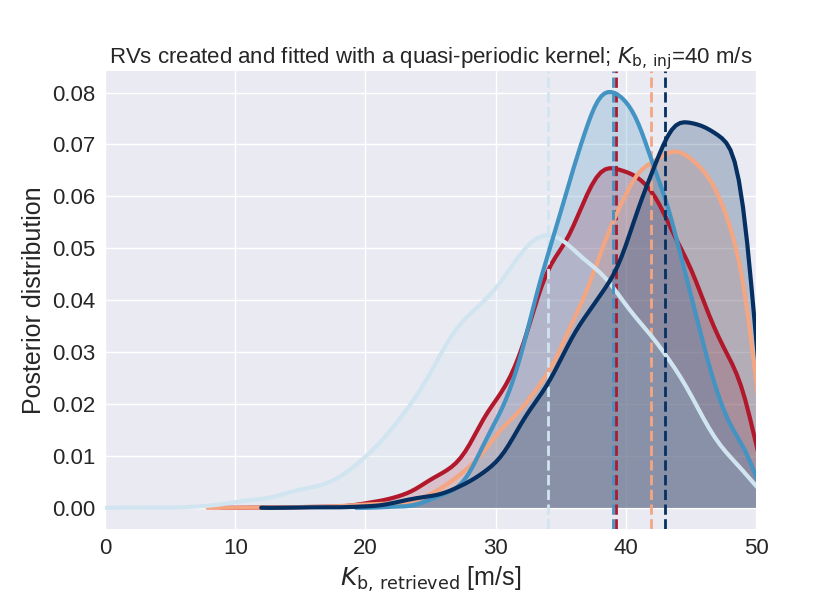}
    \end{minipage}
    \begin{minipage}[b]{0.45\textwidth}
    \includegraphics[width=\textwidth]{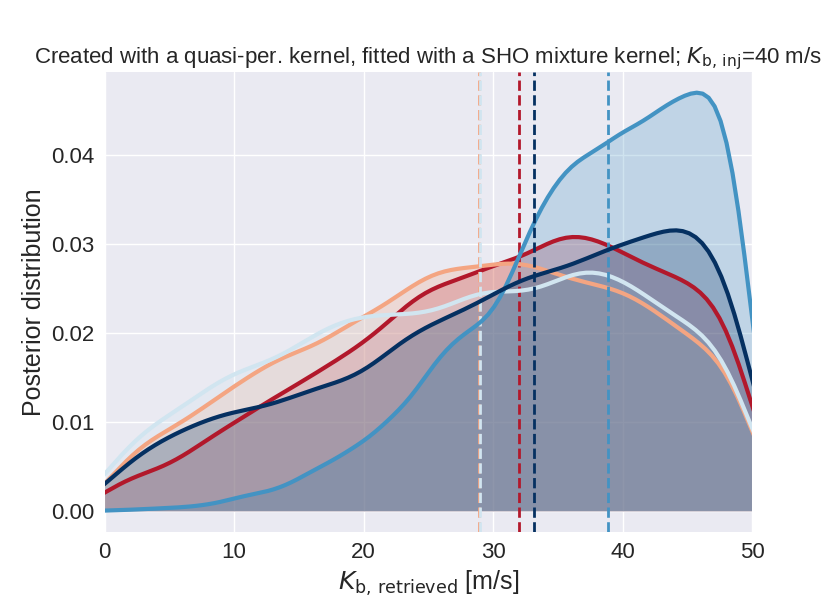}
    \end{minipage}
    \begin{minipage}[b]{0.45\textwidth}
    \includegraphics[width=\textwidth]{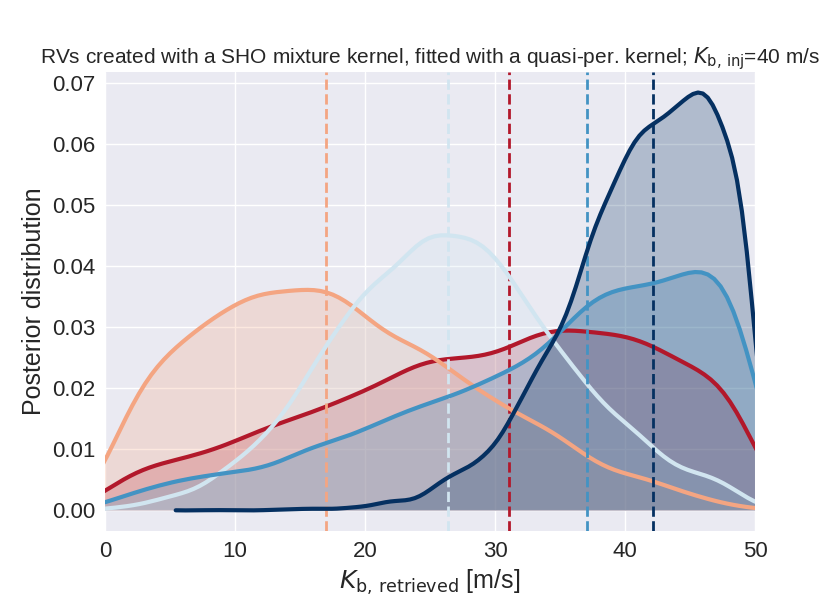}
    \end{minipage}
    \begin{minipage}[b]{0.45\textwidth}
    \includegraphics[width=\textwidth]{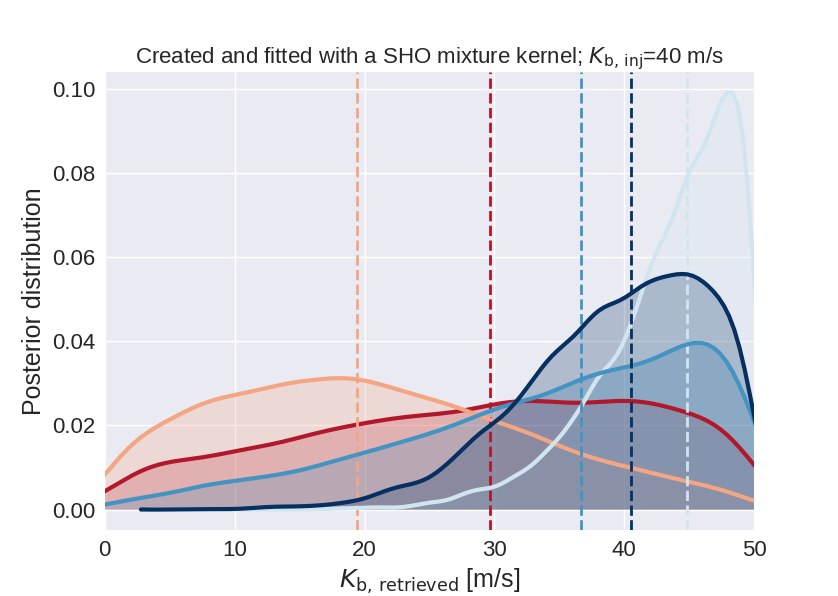}
    \end{minipage}
    \caption{Posteriors of the $K_{\rm b}$ parameter retrieved from the simulations involving mock RV datasets build on the best-fit stellar activity signals based on a quasi-periodic and SHO mixture models (see the text for details). The simulated RV time series include an injected planetary signal of semi-amplitude $K_{b,\,inj}$=40 \ms. Only five out of a total of ten posteriors are shown for the sake of a better the readability \textit{First row.} Results from the fit of data generated using a quasi-periodic kernel. \textit{Second row.} Results from the fit of data generated using a SHO mixture kernel. In both rows, each colour identifies one mock dataset, therefore the posterior distributions with the same colour can be compared. The vertical lines indicate the median of the posteriors.}
    \label{fig:kb_posteriors_simu_2}
\end{figure*}

For illustrative purposes, part of the 10 posteriors of $K_{\rm b}$ obtained for each test is shown in Fig. \ref{fig:kb_posteriors_simu_2}. It can be seen that the injected planet is well recovered only for the case \textit{a} ($K_{\rm b,\,retrieved}$=38.3$\pm$2.9 \ms) with a median detection significance of 5.1$\sigma$ over the 10 mock dataset. For the case \textit{b}, the semi-amplitude is systematically underestimated ($K_{\rm b,\,retrieved}$=32.2$\pm$2.9 \ms) and the posteriors are wide, with a median detection significance of 2.1$\sigma$. These results are consistent with those in Tab. \ref{tab:rv_simu_1} for $K_{\rm b,\, inj}$=40 \ms~obtained using the real data. The fitting procedure equally fails in recovering the planetary signal when the mock dataset are generated using the SHO mixture kernel ($K_{\rm b,\,retrieved}$=29.9$\pm$9.4 \ms~and $K_{\rm b,\,retrieved}$=33.1$\pm$9.1 \ms, for case \textit{c} and \textit{d}, respectively). 

The result of this cross-check further support the conclusion that using a quasi-periodic kernel to model the stellar activity better describes the temporal variations observed in the real RV data.  

\begin{table}
    \centering
    \onehalfspacing
    \tiny
    \caption{Results of the first set of injection-retrieval simulations described in Sect.\ref{sect:rvsimu}. Here, a sinusoid with semi-amplitude $K_{\rm b,\, inj}$ chosen from a grid of values is injected in the original HARPS and ESPRESSO RVs.}
    \begin{tabular}{c|cc|cc}
    \hline
    $K_{\rm b,\, inj}$ & \multicolumn{2}{|c|}{$K_{\rm b,\, retrieved}$} & \multicolumn{2}{|c}{$K_{\rm b,\, retrieved}-K_{\rm b,\, inj}$}\\
    \hline
    & Quasi-periodic & SHO mixture & Quasi-periodic & SHO mixture \\
    %& & (rotation kernel) & & (rotation kernel) \\
    \hline
    40 & 40.4$^{+4.4}_{-4.9}$ & 26.5$^{+13.8}_{-14.2}$ & +0.4 & -13.5 \\
    20 & 20.5$^{+4.6}_{-5.0}$ & 15.1$^{+13.5}_{-10.2}$ & +0.5 & -4.9  \\
    15 & 15.5$^{+4.5}_{-4.8}$ & 12.9$^{+13.3}_{-8.7}$ & +0.5 & -2.1  \\
    10 & 10.9$^{+4.5}_{-4.6}$ & 11.2$^{+12.4}_{-7.8}$ & +0.9 & +1.2 \\
    5 & 6.4$^{+4.2}_{-3.7}$ & 9.7$^{+11.7}_{-7.0}$ & +1.4 & +4.7 \\
    2.5 & 4.6$^{+3.8}_{-3.0}$ & 9.0$^{+11.0}_{-6.4}$ & +2.1 & +6.5 \\
    1 & 3.8$^{+3.5}_{-2.5}$ & 8.8$^{+11.0}_{-6.4}$ & +2.8 & +7.8 \\
    0.75 & 3.73$^{+3.44}_{-2.54}$ & 8.79$^{+10.86}_{-6.35}$ & +2.98 & +8.04 \\
    \hline\hline
    \end{tabular}
    \label{tab:rv_simu_1}
\end{table}
    
\section{RM modelling}
\label{sec:rml}
\begin{figure}
    \centering
    \includegraphics[trim={1cm 3cm 3 6cm},clip,width=0.55\textwidth, angle=0]{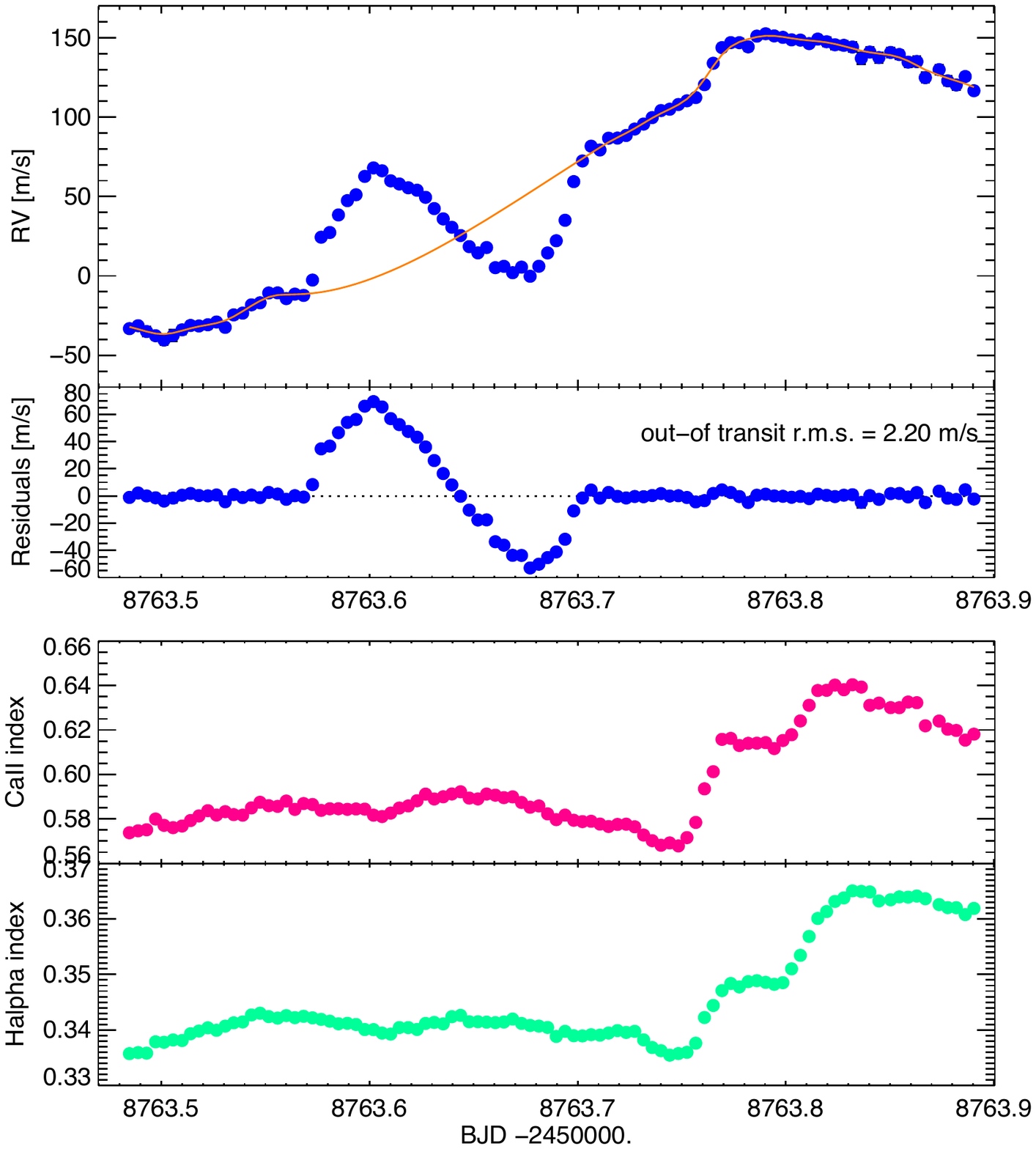}
    \caption{Upper panel: GP model of the stellar activity (orange line) during the transit of Oct 7 of DS\,Tuc\,A\,b observed with ESPRESSO (blue dots), and the corresponding residuals. Lower panel: CaII H\&K and H$\alpha$ time series extracted from the same spectra.}
    \label{fig:rmlcorr}
\end{figure}

The employment of the GP regression allowed us not only to model the stellar activity related to the time series, but also to make  predictions on the activity behaviour during the transit of the night of Oct 6, observed with ESPRESSO, for which we have a dense monitoring before and after the event. The activity model obtained with the SHO kernel was more suitable than the quasi-periodic kernel to describe the out-of-transit modulation. In particular, the quasi-periodic kernel failed in the modelling of the RV ``bump" occurring at about JD2458763.76 (see Fig. \ref{fig:rmlcorr}, upper panel), while this feature was better treated by the SHO. It is possible that the quasi-periodic kernel is more adapt to describe the global behaviour of the activity, while the SHO better manages events on a short-term basis. The SHO model is shown in the upper panel of Fig. \ref{fig:rmlcorr} (orange line), where the residuals are also reported. 
The RV bump seems to be connected with a strong activity event like a stellar flare, as confirmed by the time series of the CaII H\&K and H$\alpha$ indices, obtained with the ACTIN code (and reported in Table \ref{tab:espresso}). At the time corresponding to the RV bump, a sudden increase of the value of both indices can be observed (lower panel of Fig. \ref{fig:rmlcorr}), followed by a second peak and a slow decay, possibly down to the value before this event. 

Since the HARPS monitoring missed the pre-transit/ingress phase, it was not possible to reconstruct a similar activity model for the transit of Aug 11. For this reason we only considered the ESPRESSO time series to fit the RM effect, showing higher temporal resolution and lower RV uncertainties. However, we stress that HARPS data allowed us to confirm the amplitude of the RM effect detected with ESPRESSO.
%We fitted a RM model \citep{2005ApJ...622.1118O} to the activity-corrected RVs using the IDL routine \textsc{MPFIT} (see Fig. \ref{fig:rml1fit}). 

We fitted a RM model \citep{2005ApJ...622.1118O} to the activity-corrected RVs in a Bayesian framework by employing a differential evolution Markov chain Monte Carlo (DE-MCMC) technique \citep{TerBraak2006, Eastmanetal2013}, running ten DE-MCMC chains of 100,000 steps and discarding the burn-in. The medians and the 15.86\% and 84.14\% quantiles of the posterior distributions were taken as the best values and $1\sigma$ uncertainties. 
The first six in-transit RV measurements were masked from the fit procedure, as they were clearly deviating from the other RVs and could have biased the retrieved value of the limb darkening parameter $\mu$. We took into account the error given by the stellar activity correction performed by adding it in quadrature to the one of the RVs extracted by the DRS pipeline.
% In the fit, we left as free parameters the linear limb darkening coefficient $\mu$, the spin-orbit angle $\lambda$, \vsini.
Gaussian priors for $i$, a/Rs, and Rp/Rs were set to the values in \citet{2019A&A...630A..81B}, leaving as free parameters the linear limb darkening coefficient $\mu_1$, \vsini\ and the projected spin-orbit angle $\lambda_{\rm PSO}$ (we use this notation to distinguish this parameter from the GP hyper-parameter $\lambda$ used in Sect. \ref{sec:rv}).
The \vsini\ was left free because different RM models are known to produce \vsini\ measurements that are often in disagreement with each other and with estimates obtained from spectral line broadening \citep{2017MNRAS.464..810B}, while for the linear limb darkening parameter $\mu$ we left it free because it is sensitive to the stellar activity, and often the theoretical values are in disagreement with the observed ones for active and spotted stars \citep[e.g.][]{csizmadia2013}.

\begin{figure}
    \centering
    \includegraphics[trim={0.5cm 0 0 0},clip,width=0.5\textwidth, angle=0]{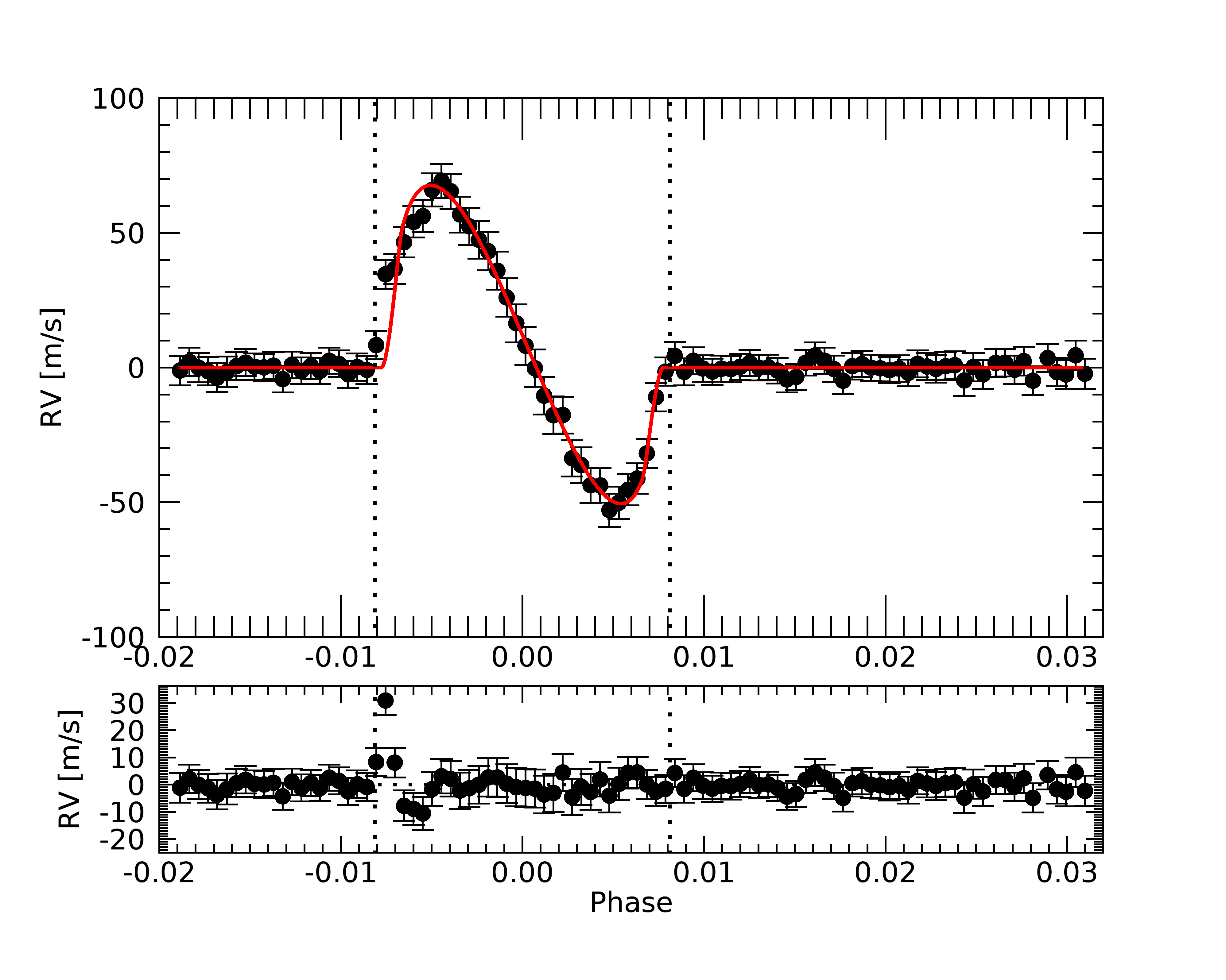}
    \caption{Upper panel: RM effect for DS\,Tuc\,A\,b from ESPRESSO data modelled by our fit (red line). Vertical dotted lines indicate the transit ingress and egress. Lower panel: Corresponding residuals.}
    \label{fig:rml1fit}
\end{figure}

RVs and the best fit are shown in Fig. \ref{fig:rml1fit}, with the best-fit values presented in Table \ref{tab:rml}.
We note that $\mu$ takes a value which is quite distant from the theoretical one calculated for the star on the ESPRESSO wavelength range using \texttt{LDTK} \citep[$\mu$=0.704$\pm$0.002,][]{ldtk}, but still compatible within 3$\sigma$.
When setting a prior to the \texttt{LDTK} estimated value, the returned fit is clearly not matching the observations even by eye, but we note that the resulting value of $\lambda_{\rm PSO}$ is perfectly compatible the one here reported.

The measurement of the projected spin-orbit obliquity, $\lambda_{\rm PSO}$, is in well agreement with the one of \citet[][12$\pm$13 degrees]{2020AJ....159..112M}, thus confirming an almost (but not perfectly) aligned system.
\cite{2020ApJ...892L..21Z} reported a value of $\lambda_{\rm PSO}$ of 2.5$\pm$1 deg, which is not compatible with our value.
Indeed, the determination of $\lambda_{\rm PSO}$ through the RM effect is sensitive to stellar activity \citep{2018A&A...619A.150O}, so it is possible (and likely) that the differences are due to the activity correction performed.
While in the RM fit procedure \citet{2020AJ....159..112M} and \citet{2020ApJ...892L..21Z} use different spot-modeling techniques, we let all the residual activity left from the Gaussian process model (which does not take into account the in-transit RV values) to be included in the limb-darkening estimation. 
%We note that our RV error-bars are much lower than the ones presented in \citep{2020ApJ...892L..21Z}. 
A common analysis of all available datasets, which are taken with different instruments and have different quality, will help in check the stellar activity impact on the retrieved $\lambda_{\rm PSO}$.

\begin{table}[]
    \centering
        \onehalfspacing
    \tiny
        \caption{Results of the RM fit for the transit of DS\,Tuc\,A\,b observed with ESPRESSO. }
    \begin{tabular}{l|c}
    \hline
        Parameter & Value \\
        \hline
        % \noalign{\smallskip}
        Inclination, $i$ [deg] & 88.73$^{+0.18}_{-0.17}$  \\
        %\smallskip
        a/R$_{\star}$ & 19.79$^{+0.49}_{-0.50}$ \\
        %\smallskip
        Limb darkening, $u_{\rm 1}$ & 0.93$^{+0.04}_{-0.08}$\\
        %\smallskip
        R$_{\rm p}$/R$_{\star}$ & 0.05972$^{+0.00065}_{-0.00064}$ \\
        %\smallskip
        \vsini\ [km s $^{-1}$] & 27.75$^{+1.73}_{-1.54}$ \\
        %\smallskip
        Obliquity, $\lambda_{\rm PSO}$ [deg] & 12.1$^{+2.6}_{-2.1}$\\
        %\noalign{\smallskip}
        \hline
    \end{tabular}
   % \tablefoot{*The errorbar on $\lambda_{\rm PSO}$ does not take into account the systematic error budget given stellar activity}
    \label{tab:rml}
\end{table}

\begin{figure}
    \centering
    \includegraphics[trim={0 0 0 0},clip,width=0.5\textwidth, angle=0]{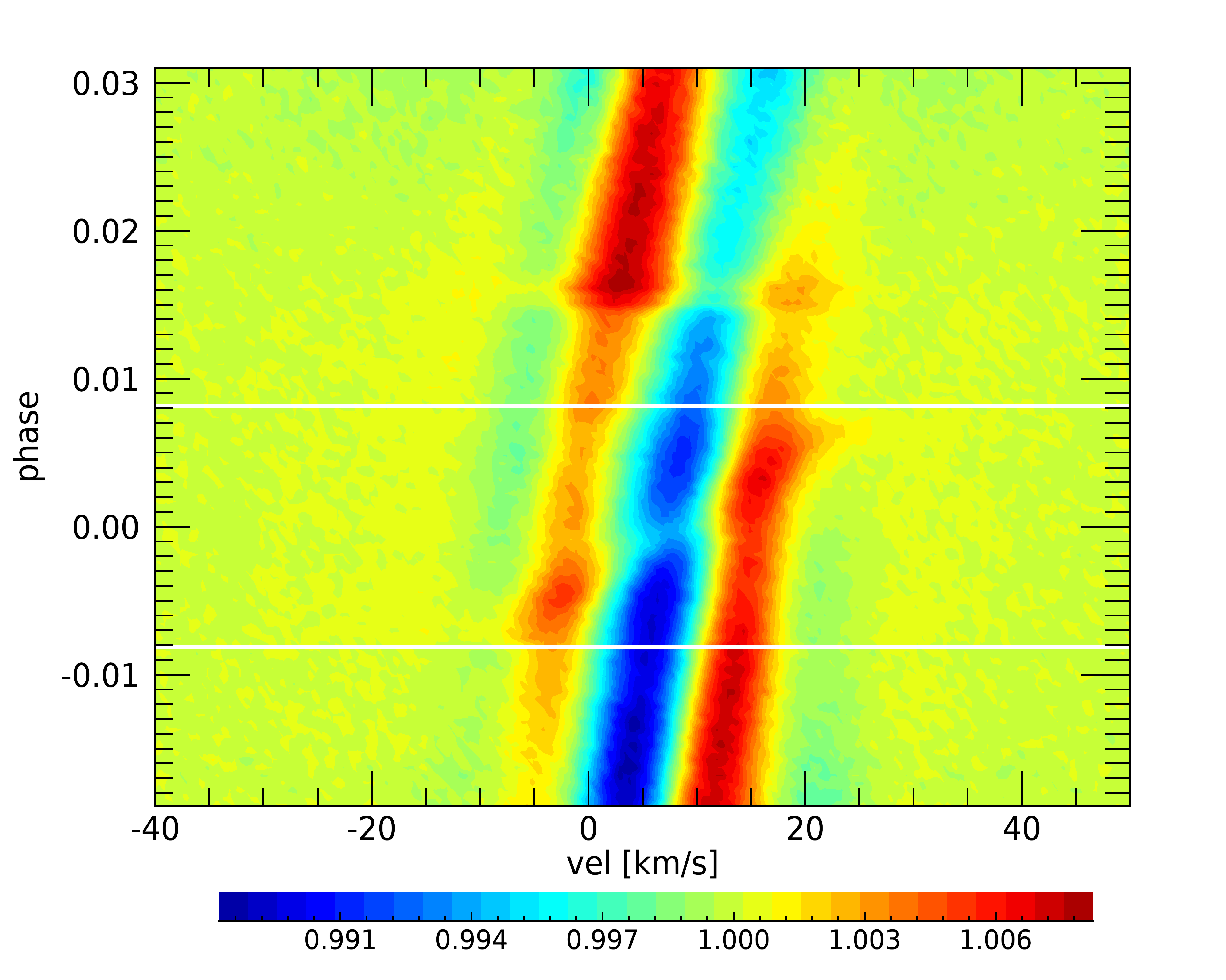}
    \caption{CCF residuals during the transit of DS\,Tuc\,A\,b  in the stellar restframe. Horizontal white lines indicate the transit ingress and egress. }
    \label{fig:ccf_tom}
\end{figure}

We also performed a line profile tomography aiming to reveal changes in the CCF morphology before, during, and after the transit of DS\,Tuc\,A\,b. As in the case of AU Mic (Fig. 4 in \citealt{PalleAUMic}), our Doppler tomography  shows the very high activity level of the host (Fig. \ref{fig:ccf_tom}), that prevented us from obtaining a reference CCF for a proper subtraction, and in turn to fit the planetary Doppler shadow. However, this feature can be noticed as a slight ripple occurring during the transit phase, compatible with an aligned planet. 
The flare event previously identified can be observed also in Fig. \ref{fig:ccf_tom} as a clear increase of the distortion of the line profile (depicted in dark red) at phase $\sim0.015$.
We note that methods such as the reloaded RM \citep{2016A&A...588A.127C}, which in principle are less affected by stellar activity in determining the projected spin-orbit angle, would also likely fail with such a level of activity.
%Masuda \& Winn (2020)

\section{Atmospheric characterization} \label{sec:atmo}
In the following sections, we provide our analysis of the ESPRESSO spectra taken during the transit of DS\,Tuc\,A\,b occurred on Oct 7, 2019, aiming to detect traces of its atmospheric compounds.

\subsection{Transmission spectrum}
\begin{figure}
    \centering
    \includegraphics[trim={1 1 1 1},clip,width=0.43\textwidth, angle=0]{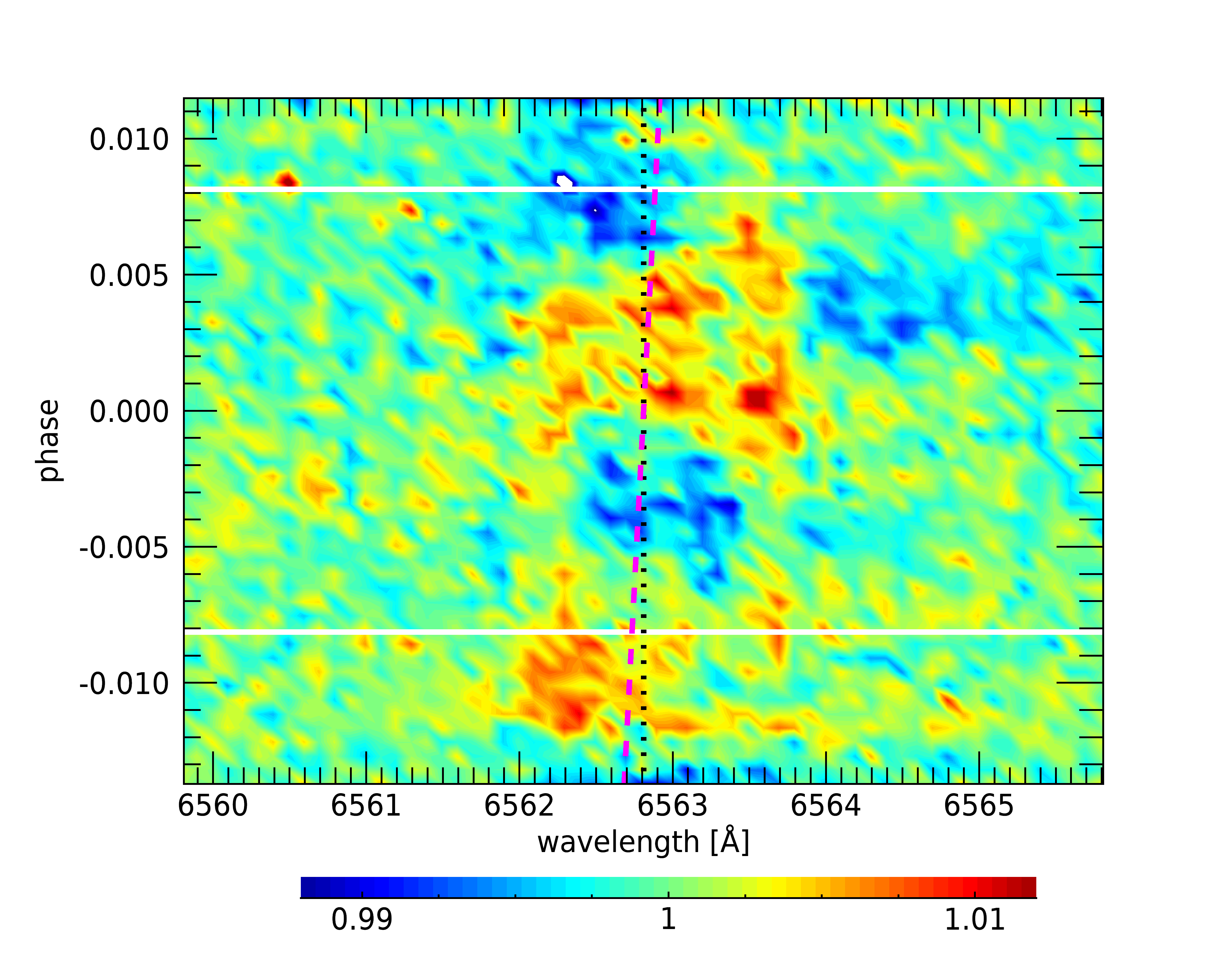}
    \caption{2D tomography of H$\alpha$ in the stellar restframe. The vertical dotted line shows the stellar restframe. The diagonal magenta dashed line shows the expected planetary restframe. The horizontal white lines mark the beginning and end of the transit.}
    \label{fig:tomo_halpha}
\end{figure}

\begin{figure*}
    \centering
    \includegraphics[trim={1 1 1 1},clip,width=\textwidth, angle=0]{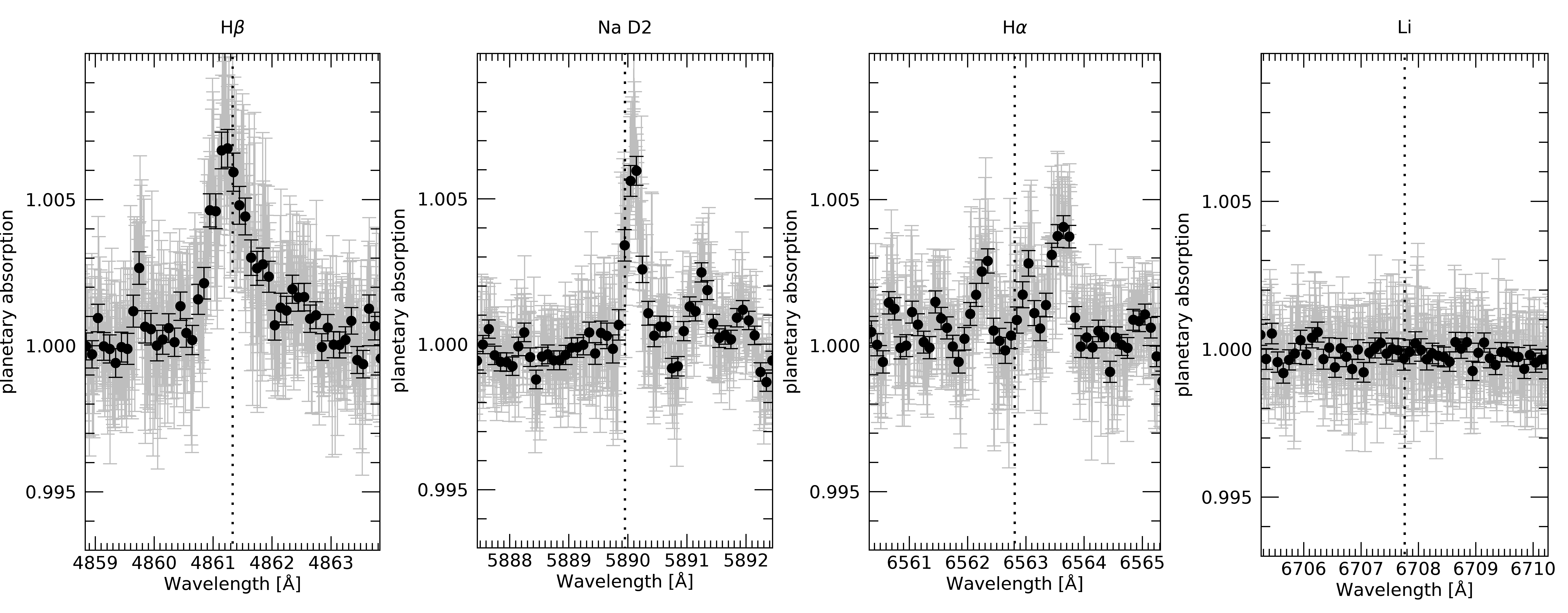}
    \caption{Transmission spectrum extracted in the zone of the H$\beta$ \textit{(left panel)}, Na D2 \textit{(central left panel)}, H$\alpha$ \textit{(central right panel)} and Li \textit{(right panel)} lines. The vertical dashed lines show the planetary restframe. Black points refer to 0.1 Angstrom binning. Different lines we searched for are affected by the contamination given by the stellar activity in different ways.}
    \label{fig:ts_halpha}
\end{figure*}

We extracted the planetary transmission spectrum from the ESPRESSO data following the classic procedure used in high-resolution transmission spectroscopy \citep[e.g.][]{wyttenbach,borsa2020}.
We shifted the spectra to the stellar restframe by using the Keplerian model of the system with the upper limits found in Table \ref{tab:rvgp}, then we normalised each spectrum.
Telluric correction was performed by exploiting the scaling relation between airmass and telluric line strength \citep[e.g.][]{2008A&A...487..357S,2013A&A...557A..56A}, rescaling all the spectra as if they were observed at the airmass of the transit center. 
Then we divided all the spectra by a master stellar spectrum created by averaging the out-of-transit spectra.
To avoid being too much contaminated by stellar activity, we used only the out-of-transit data close to the transit, this means that we discarded the first 10 and the last 33 spectra of the ESPRESSO timeseries. 
The \textit{wiggles} pattern \citep[e.g.][]{tabernero_esp} is corrected in the same way presented in \citet{borsa2020}, namely by fitting a sinusoid with varying period, phase, and amplitude for each spectrum, with the fit performed independently for each wavelength region where we looked for spectral features.
At the end we moved to the planetary restframe by shifting all the residual spectra for the theoretical planetary radial velocity. 
The planetary signal during the transit is always within the stellar line profile (Fig. \ref{fig:tomo_halpha}).
The transmission spectrum was then created by averaging all the full-in-transit residual spectra (see e.g. Fig. \ref{fig:ts_halpha} for H$\alpha$, H$\beta$, Na D2 and Li lines).

Since this star has a very high level of activity, its line profile changes rapidly, thus the consequence is that the stellar activity deeply contaminates the transmission spectrum and prevents us from probing the planetary atmosphere as shown by Fig. \ref{fig:ts_halpha}. This was also noted for the case of AU Mic b \citep{PalleAUMic}, another planet transiting a young star. 
The movement of active regions on the star is sufficiently fast to give well-detectable changes in the line profiles during the time span of our observations. This is well visible in Fig. \ref{fig:ccf_tom}, where these active regions are the blue and red stripes (flux decrease and increase with respect to the average out-of-transit CCF, respectively) moving along the residuals of the CCF, which can be considered as an average stellar line profile. 
The slope of these stripes in the velocity-phase 2D plot ($\sim$4-5 \kms~in $\sim$4 hours) is compatible with regions moving on the stellar surface, when assuming a \vsini$\sim$20 \kms~and stellar rotation of 2.9 days (Table \ref{tab:rvgp}).
In order to extract the transmission spectrum, we are forced to set some baseline out-of-transit, but we could not find any combination of spectra for which the out-of- vs. in-transit active regions configuration is the same.
Interestingly, the effect of the contamination in the transmission spectrum looks wavelength dependent (Fig. \ref{fig:ts_halpha}), being more prominent toward blue wavelengths and going to almost disappear toward the red edge of the spectrograph. We expect that features like starspots are the main responsible for this effect, since the temperature difference with the photosphere (500-2000 K) is much larger than the one predicted with the faculae ($\sim 100$ K in the solar case).  

We searched for possible absorption given by H$\alpha$ ($R_{eff}<1.22 R_p$), H$\beta$ ($R_{eff}<1.3 R_p$), Na D doublet ($R_{eff}<1.28 R_p$), K ($R_{eff}<1.16 R_p$), Mg ($R_{eff}<1.21 R_p$), Li ($R_{eff}<1.12 R_p$), without finding any trace of significant absorption  beyond the variation caused by stellar activity.
We estimated indicative upper limits to the effective planetary radius $R_{eff}$ by using the standard deviation of the transmission spectrum in a range of $\pm$1 \AA, centered at the rest frame wavelength of the element. We translated this value to $R_{eff}$ by assuming $R^2_{\rm eff}/ R^2_{\rm p}=(\delta+h)/\delta$, where $\delta$ is the transit depth and $h$ the calculated standard deviation.

\subsection{Cross-correlation with theoretical templates}
In order to try to overcome the stellar variability issue, we tried to look for elements/molecules that we expect to find possibly more in the planetary atmosphere than in the star.
We generated model templates for V, Cr, AlO, VO, TiO (Plez linelists), Y by using \texttt{petitRADTRANS} \citep{pRT}, assuming an isothermal atmospheric profile with T=1000 K, continuum pressure level of 10 mbar and solar abundances.  The atmospheric models were then translated into flux (R$_{\rm p}$/R$_{\rm s}$)$^2$, convolved at the ESPRESSO resolving power and continuum normalised. 
Cross-correlation between model templates and the single transmission spectra was then performed in the same way as in \citet{borsa2020}. Unfortunately we find no evidence in the planetary atmosphere for the presence of any of the species we looked for.

\section{X-ray emission} \label{sec:x}
\subsection{XMM-Newton data} \label{x_data}
We observed DS Tuc with \xmm~(obsid:0863400901, PI S. Wolk) aiming 
to assess the spectrum and time variability of the host star in X-rays, and hence to characterise the coronal activity level and the dose of high-energy radiation received by its young planet. A measure of the strong irradiation by the host star can give us useful constraints to model the evolution of the planetary atmosphere due to photo-evaporation processes.

\begin{figure}[ht]
\begin{center}
% \hbox{
\includegraphics[width=\hsize]{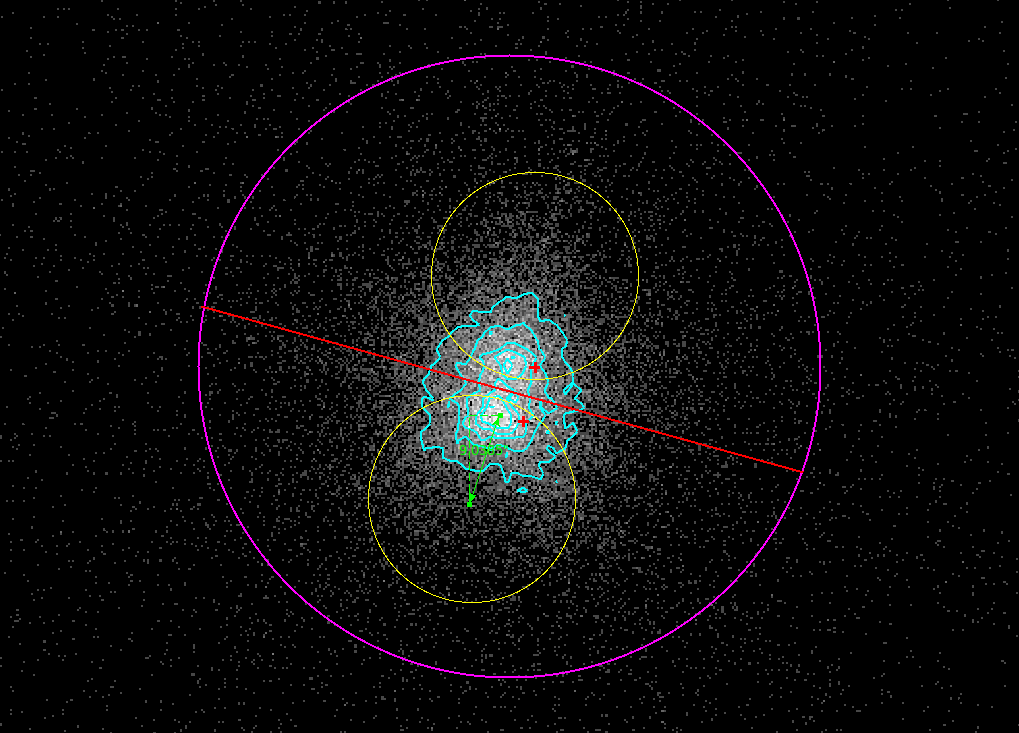} 
% \vspace{-1.0cm}
% }
\end{center}
\caption{\footnotesize{
XMM/Newton MOS1 image of \dstuc. The two red points indicate the ICRS positions of the two stars from SIMBAD, the centroid of
X-ray emission is offset because of the proper motions of the system. 
The contours show the emission of the two sources, marginally resolved. 
The large circle (in red) has 30\arcsec~radius, while the smaller circles (in yellow) are 10\arcsec~wide. They were used to extract
the spectra and light curves of the A and B system components (see text).}}
\label{mos1}
\end{figure}

The exposure time of the observation was about 40 ks and the prime instrument was EPIC with full frame window imaging mode and the {\it Medium} filter. 
The Observation Data Files were retrieved from the \xmm\ archive and reduced with the Science Analysis System (SAS, ver.18.0.0\footnote{\url{https://www.cosmos.esa.int/web/xmm-newton/sas}}). We obtained FITS lists of photons detected with the MOS1, MOS2, and pn CCD cameras, calibrated in energy, arrival time and astrometry with the {\it evselect} SAS
task.
Since the two components of the \dstuc\ system are separated by about 4.8\arcsec, they are partially resolved in the MOS images (2\arcsec~pixel
resolution), but not with the pn CCD (about 4\arcsec~pixel resolution).
Figure \ref{mos1}  shows the  MOS1 image of \dstuc, with the system primary (A) being the source at the bottom, and the secondary (B) at the top. The two sources appear with similar intensities, but
the tight separation implies a cross-contamination of order of 16\% in estimating their fluxes.

In order to estimate the contamination and extract the correct flux for each component we compared 
the events extracted from two different sets of regions.  The first set is made of two semicircular regions with radius 30\arcsec~centered half way between the two sources. These regions collect most of the counts of the two sources, 
but they suffer from a high cross-talk contamination. The second set is made of two smaller circles of 10\arcsec~radius, displaced by $\sim$9\arcsec~from each emission peak along the direction
connecting the pair. In this way we minimise the cross-talk contamination still retaining enough counts for
light curves and spectra. 

\begin{figure}[ht]
\begin{center}
% \hbox{
\includegraphics[width=\hsize]{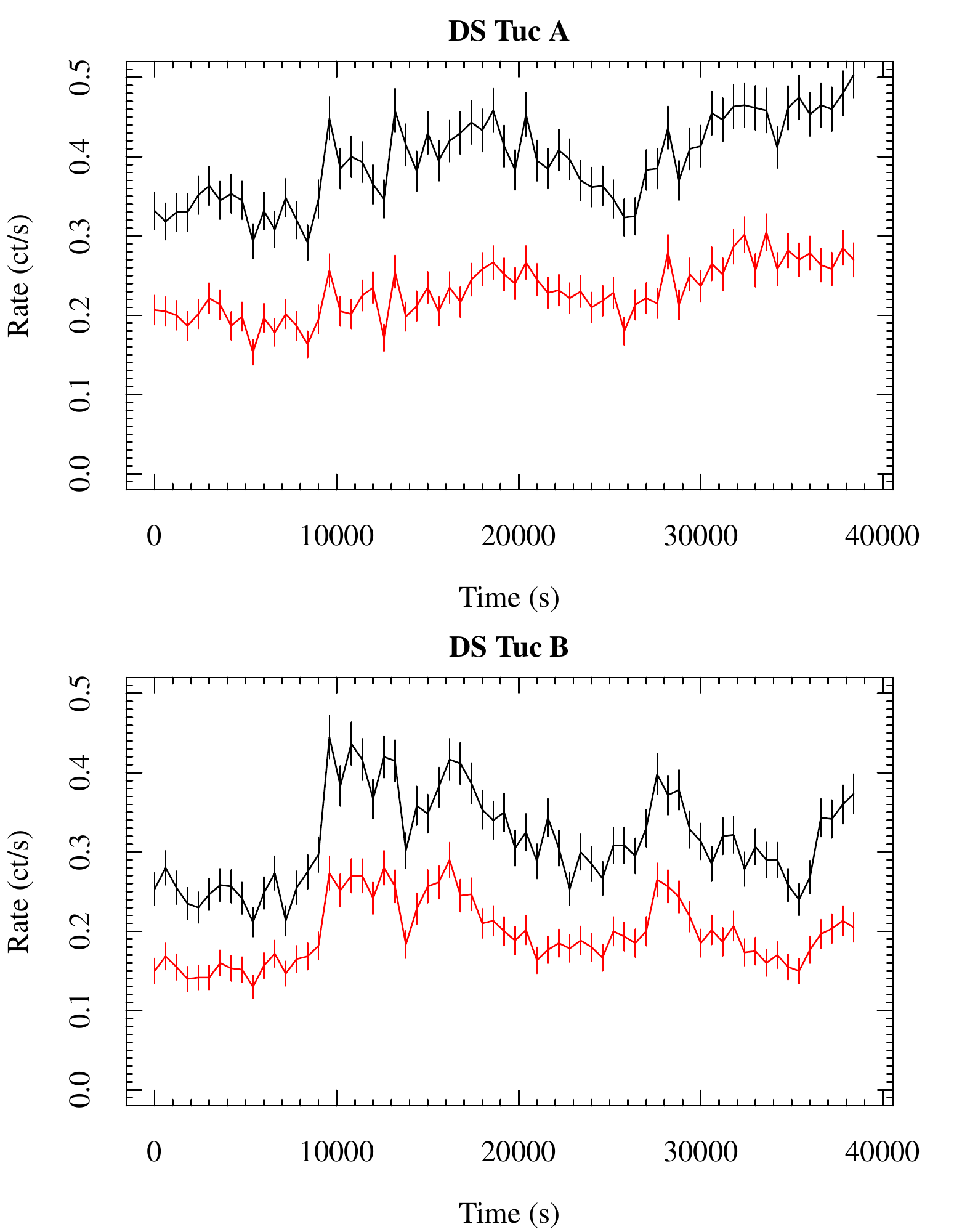} 
% \vspace{-1.0cm}
% }
\end{center}
\caption{\footnotesize{
\label{xraylc} Top: Mos1 light curves of \dstuc\,A. 
Black line is the light curve from the semi-circular region of radius 30\arcsec, red line is the
light curve from the circular region of radius 10\arcsec. Error bars on rates are 
at 68\% confidence level. Bottom: same for \dstuc\,B. 
}}
\end{figure}

The extracted light curves are shown in Fig. \ref{xraylc}. These plots show that the star A is less variable and slightly brighter than the star B, with the latter characterised by   
at least a couple of flares of duration 7--10 ks, that
contaminate the light curve of A accumulated in the large semicircular region.

\begin{figure}[ht]
\begin{center}
% \hbox{
\includegraphics[width=\hsize]{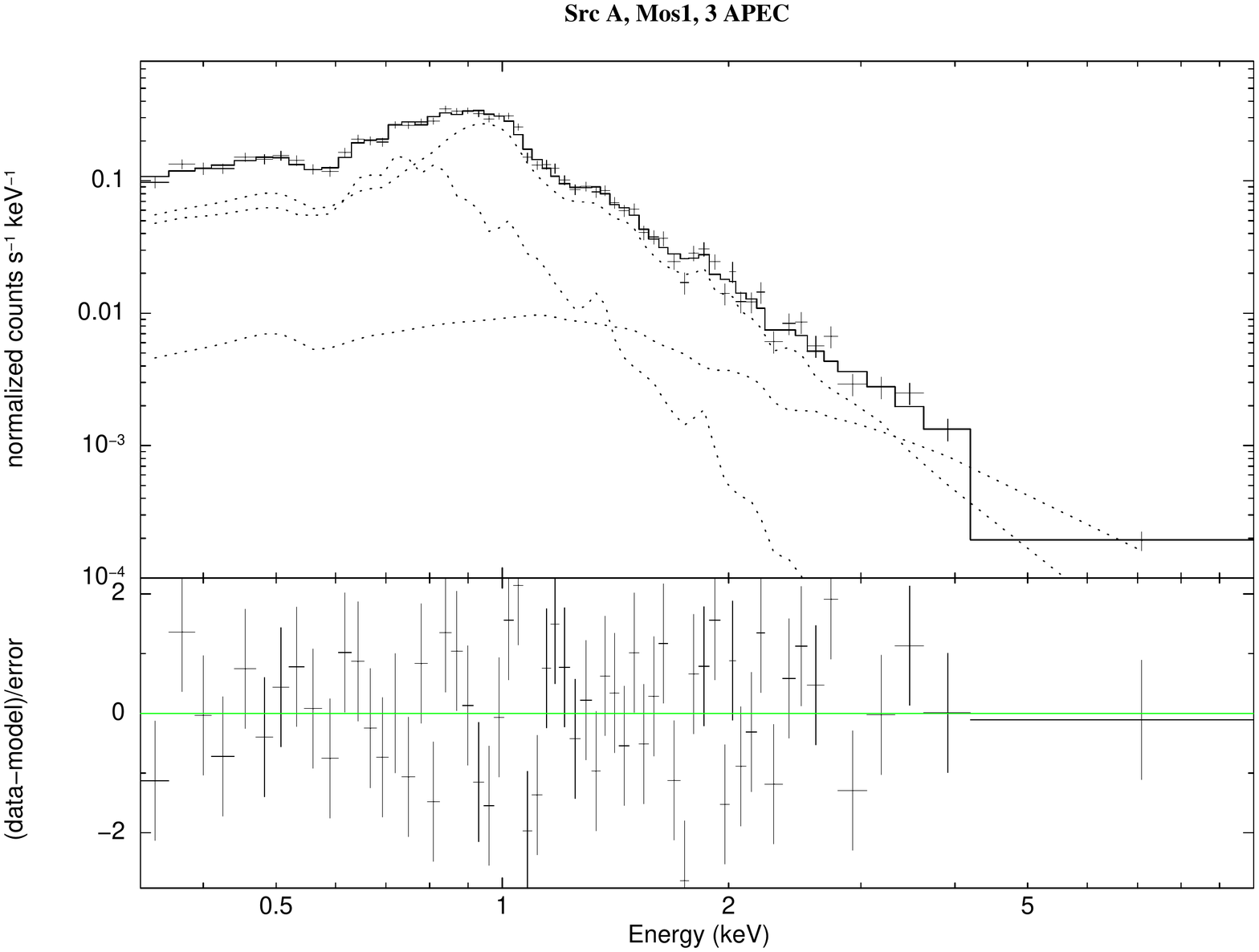}
\includegraphics[width=\hsize]{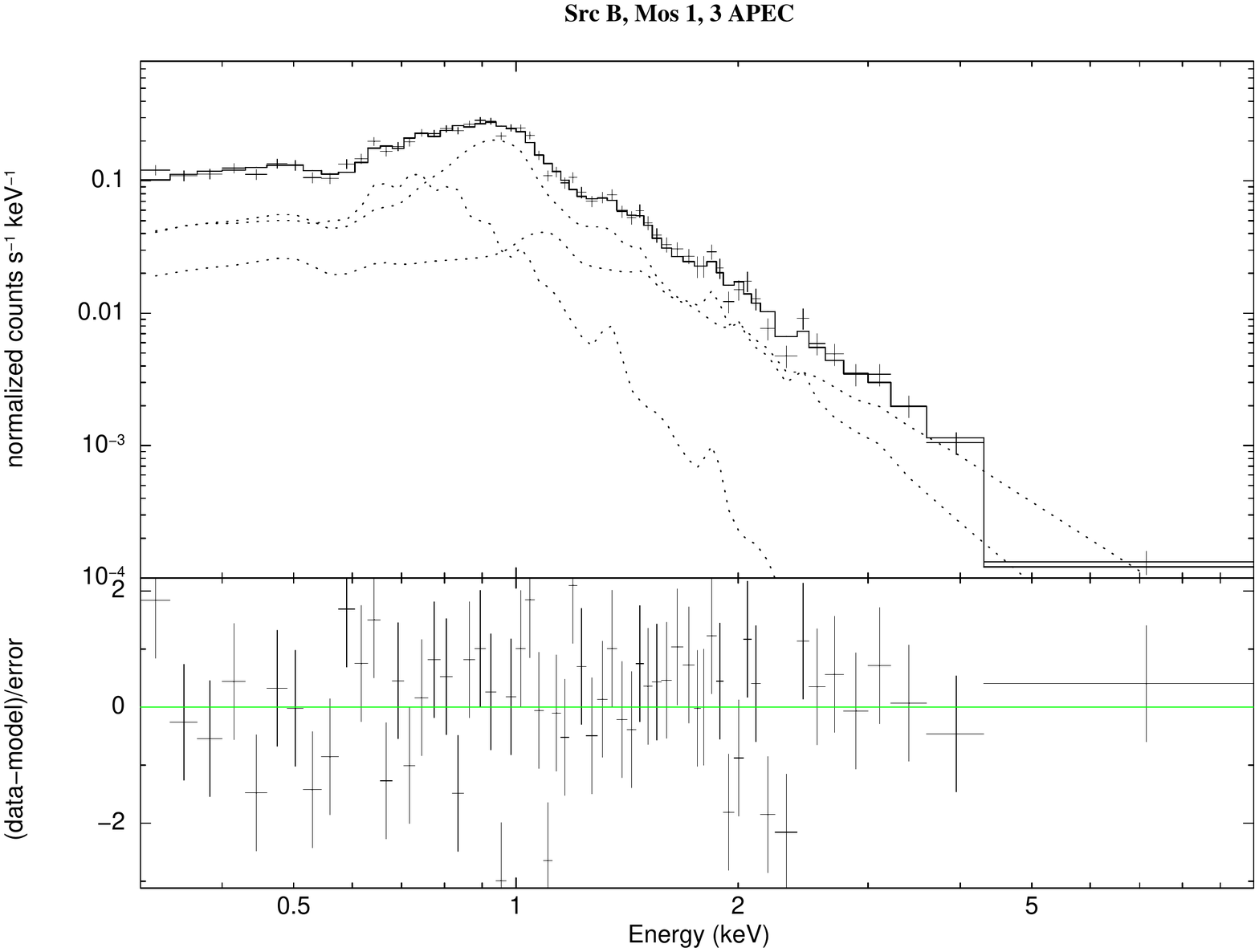}
% \vspace{-1.0cm}
% }
\end{center}
\caption{\footnotesize{
\label{xraysp} X-ray spectra of DS\,Tuc\,A (top panel) and DS Tuc B (bottom panel). Each panel is subdivided in two plots: the spectrum with the best fit model (a sum of 3 APEC components, each one indicated by the dotted lines) and the terms of $\chi^2$ for each bin in units of $\chi^2$ standard deviation. 
}}
\end{figure}

From the events in the 10\arcsec\ regions we extracted spectra and relative instrument response files with
SAS. The spectra for the A and B sources are shown in Fig.\ref{xraysp}. 
We fitted the spectra with {\sc xspec}, adopting a multi-temperature optically-thin coronal emission model ({\sc apec}) multiplied by a global interstellar absorption component ({\sc tbabs}).  
The results are reported in Table \ref{tabxspec}. In order to obtain a statistically acceptable fit, we need to describe the corona of each source with at least three isothermal components: a cool temperature component at $\sim0.3$ keV, 
another component at $\sim0.95$  keV and a hot component at $\sim 2-5.5$ keV. 
From the estimates of the plasma emission measure, the 0.95 keV component is the most dominant,
followed by the cool component at 0.3 keV, and then the hot component, which is less constrained.

\subsection{Results}\label{x_results}
We evaluated the luminosity for both sources also in the \textit{ROSAT} band (0.1-2.4 keV), for a direct comparison with the previous measurements. 
In this band, the luminosity for the primary (A) is  $9.57\times10^{29}$ \lxu, and  $7.27\times10^{29}$ \lxu~ for the secondary (B). 
Since the X-ray luminosity evolution is usually calibrated in the \textit{ROSAT} band (e.g. \citealt{Penz08a}), we considered this value for DS\,Tuc\,A to test photo-evaporation models of the planetary atmosphere (see the next section).

The gas absorption is better constrained from the spectrum of \dstuc\,A, with a best-fit interstellar hydrogen column density
$N_\mathrm H \sim 2.4\times10^{20}$ cm$^{-2}$, which is compatible with the low value
of optical $A_\mathrm V = 0.15$ from the \textit{Gaia} DR1 data analysis by \cite{2018MNRAS.475.1121G}, typical for nearby stars within $\sim 50$\,pc.

We have verified that the total observed X-ray flux of the combined A+B source
is about 70\% the one derived from the detection reported in the ROSAT All-Sky Survey Point Source Catalogue (2RXS), assuming the same 3-component thermal model best-fitting the XMM-Newton data. 
This result implies very little variability on a time scale of about 30 years, being DS\,Tuc\,A an active star, and comparable to the short-term variability that we observed in the time frame of the XMM-Newton observation ($\Delta t \sim 0.5$\,d). Since we have now resolved for the first time the two stellar components, we can recompute more precisely the ratios of X-ray to bolometric luminosities: assuming $L_{\rm bol,A} = 2.80 \times 10^{33}$\,erg s$^{-1}$ and $L_{\rm bol,B} = 1.26 \times 10^{33}$\,erg s$^{-1}$ \citep{2019ApJ...880L..17N}, we derive $\log L_{\rm x}/L_{\rm bol} = -3.55$ and $-3.31$ for the G6V primary and the K3V secondary, respectively. This result implies that \dstuc\,A shows X-ray emission just below the threshold  activity level for saturated coronal sources \citep{Pizz03}, that solar-type stars maintain until their rotation period remains below $\sim 2$ days. In fact, $P_{\rm rot,A} = 2.85 \pm 0.02$\,d \citep{2019A&A...630A..81B}. Instead \dstuc\,B is likely at the saturation level, possibly due to its lower mass, although we lack a measure of $P_{\rm rot,B}$.
Finally, we have checked that the X-ray luminosities of both components are consistent with the expectation for stars in open clusters of similar age, such as the 50\,Myr old cluster of $\alpha$~Per \citep{2013AJ....145..143P}.
These results allow us to model the evolution of the average high-energy irradiation received by the planet, and to estimate its atmospheric evaporation in the last few tens million years, namely since the dissipation of the parent circumstellar disc.

%-----------------------------Figure Start------------------------------
%-----------------------------Figure End--------------------------------

%-----------------------------Figure Start------------------------------
%-----------------------------Figure End--------------------------------

\begin{table*}
\caption{\label{tabxspec} Best fit parameters from modelling of the Mos1 spectra of \dstuc~A (top) and B (bottom). Errors are quoted at the 90\% confidence level. }
\resizebox{\textwidth}{!}{
\begin{tabular}{cccccccccccc}\\\hline\hline
NH & 	kT1 & 	EM1 & 	kT2 & 	EM2 & 	kT3 & 	EM3 & 	Z/Zsun & 	chi2 & 	d.o.f. & 	Prob & $f_{\rm{X}}$ \\
10$^{20}$ cm$^{-2}$ & 	keV	& $10^{52}$ cm$^{-3}$	& keV &	$10^{52}$ cm$^{-3}$ &	keV	& $10^{52}$ cm$^{-3}$ & & & & & erg$~$s$^{-1}~$cm$^{-2}$ \\

\multicolumn{10}{c}{--- Src A ---} \\
2.4 (0.7-4.4)	& 0.35 (0.31-0.39) & 	3.73 (2.80-5.36) & 	0.96 (0.93-0.99) & 	4.89 (3.73 6.06) & 	5.4 (2.7-64) & 	0.54 (0.30-1.00) & 	0.18 (0.14-0.23) & 	68.0 & 	53 &	0.08 & 3.41 (3.19-3.67) 10$^-12$\\ 
% \hline
% ---	---	---	---	---	---	---	---	---	--- \\
\multicolumn{10}{c}{--- Src B ---}  \\
$\le 2.0$	 & 0.31 (0.29-0.35)	& 2.3 (1.9-3.3) &	0.95(0.91-1.00) & 	2.8 (2.1-4.0) & 	2.0 (1.7-3.1) & 	1.40 (0.75-1.90) & 	0.23 (0.17-0.30) &	70.2 &	53 &	0.056 & 2.63 (2.58-2.79) 10$^-12$ \\
\hline
\end{tabular}
}
\end{table*}

\section{Atmospheric evolution induced by photo-evaporation}\label{sec:models}
In Sect. \ref{sec:rv} we obtained for DS\,Tuc\,A\,b a mass upper limit of 14.4 M$_{\oplus}$.
We investigated the predicted atmospheric evolution induced by the stellar high energy radiation derived in Sect. \ref{x_results} (hereafter XUV) for different values of the planetary mass, in order to evaluate the different scenarios.

Proceeding similarly as in \cite{2021A&A...645A..71C} and Georgieva et al. 2021 (submitted), we evaluated the mass loss rate of the planetary atmosphere using the hydrodynamic-based approximation developed by~\cite{Kub18b}. 
For the X-ray evolution we use two different laws both based on cluster and field star data, namely, those proposed by \cite{Penz08a}, hereafter P08a, and \cite{2012MNRAS.422.2024J}, hereafter J12. 
Both describe the evolution of the X-ray luminosity, L$_{\rm x}$, with a saturation and a decay phase. The main difference is that J12 assume a constant X-ray emission up to an age of $\sim 70$\,Myr, while P08a predict high-energy irradiation decreasing with a shallow power-law up to 600 Myr; for older ages (decay phase), the steeper decrease of L$_{\rm x}$ is also described with slightly different slopes.

In order to evaluate the stellar irradiation in extreme ultraviolet band we adopted the scaling law derived by~\cite{SF11}, providing a relation between EUV and X-ray luminosity, calibrated on the stars observed in both bands.

We took into account the evolution of the planetary radius through the relation given in~\cite{LopFor14}, as already adopted, for instance, by~\cite{Pop20}. This relation was developed for H-He dominated atmospheres, and provides the envelope radius $R_{\rm env}$, as a function of the planetary mass, the bolometric flux received, the atmospheric mass fraction $f_{\rm atm}$, and the age of the system, allowing to account also for the gravitational shrinking.
Considering an upper limit for the mass of DS\,Tuc\,A\,b of 15 M$_\oplus$, we performed several simulations for different values of the mass in the range 5-15 $M_\oplus$. For each value, we estimated the core radius, $R_{\rm core}$, through the following analytic relation, derived by fitting the data in Figure 1 of~\cite{Mocq14} for planets with Earth-like composition of the core:
\begin{equation}
R_{\rm core}= R_\oplus \biggl( \frac{M_{\rm P}}{M_\oplus} \biggr)^\frac{1}{3.7} e^{ \bigl[w \bigl(\frac{M_{\rm P}-M_\oplus}{M_{\rm P}} \bigr)\bigr]},
\label{eq:rcore}
\end{equation}
where  $M_{\rm P}$ is the mass of the planet, and $w=-0.014$ is the fitting parameter. Given  $R_{\rm core}$ and the measured planetary radius $R_{\rm P} = 5.6$ R$_{\oplus}$, we estimated the initial $R_{\rm env}$ as the difference between $R_{\rm P}$ and  $R_{\rm c}$. Inverting  the relation of~\cite{LopFor14}, we calculated the initial atmospheric mass fraction for each assumed planetary mass. For each time step of the simulation, we updated $f_{\rm atm}$  and the planetary mass in response to the mass loss, obtaining a new value of $R_{\rm env}$. The latter quantity added to the core radius provided the updated planetary radius.  For each mass of our grid we let the system to evolve from the current age, supposed to be 40 Myr, until 5 Gyr (Fig.\ref{fig:massrad_evolution}, upper panel).
We stopped the simulations when the cumulative percentage of the mass loss reached the corresponding atmospheric mass fraction, that is when the planet has lost completely its atmosphere.

\subsection{Results}
In all cases the planet loses its atmosphere within 1 Gyr.
The two X-ray evolution laws give results fully compatible when lower masses are considered.
The most discrepant case is for a planet of 15 $M_\oplus$. Following the P08a evolution, such a planet is subject to a complete atmospheric loss in 600 Myr, while J12 foresees that at such an age the planet retains still an atmosphere of $\sim 1$\% of the planetary mass, which is completely dissipated at $\sim 3$ Gyr.

In the particular case in which the mass of the planet is $ \sim 8 \ \ M_\oplus$, or lower, the atmosphere is completely lost in less than 1\,My, which is the time step of our simulations. In this case, if a planet atmosphere would be detected, we were in the fortunate position to witness the final stages of a planetary atmosphere's life. In other words, if DS\,Tuc\,A\,b still has an atmosphere that we can detect, we can conclude that its mass is very likely greater than 8 $M_\oplus$, given the known age of the system. However, the scenario in which the planet has a mass up to 8 $M_\oplus$ and no atmosphere still remains plausible. These results can be compared with those presented by \cite{2020MNRAS.498.5030O}, which excluded a planetary mass lower than $\sim$4.5 M$_{\oplus}$. Instead, a mass up to 7 M$_{\oplus}$ requires a boil-off phase during planet formation, while masses between 7 and 10 M$_{\oplus}$ would be consistent with a standard picture of core accretion.

\begin{figure} 
\centering
\includegraphics[trim={1cm 3cm 2cm 2cm},clip,width=.35\textwidth,angle=270]{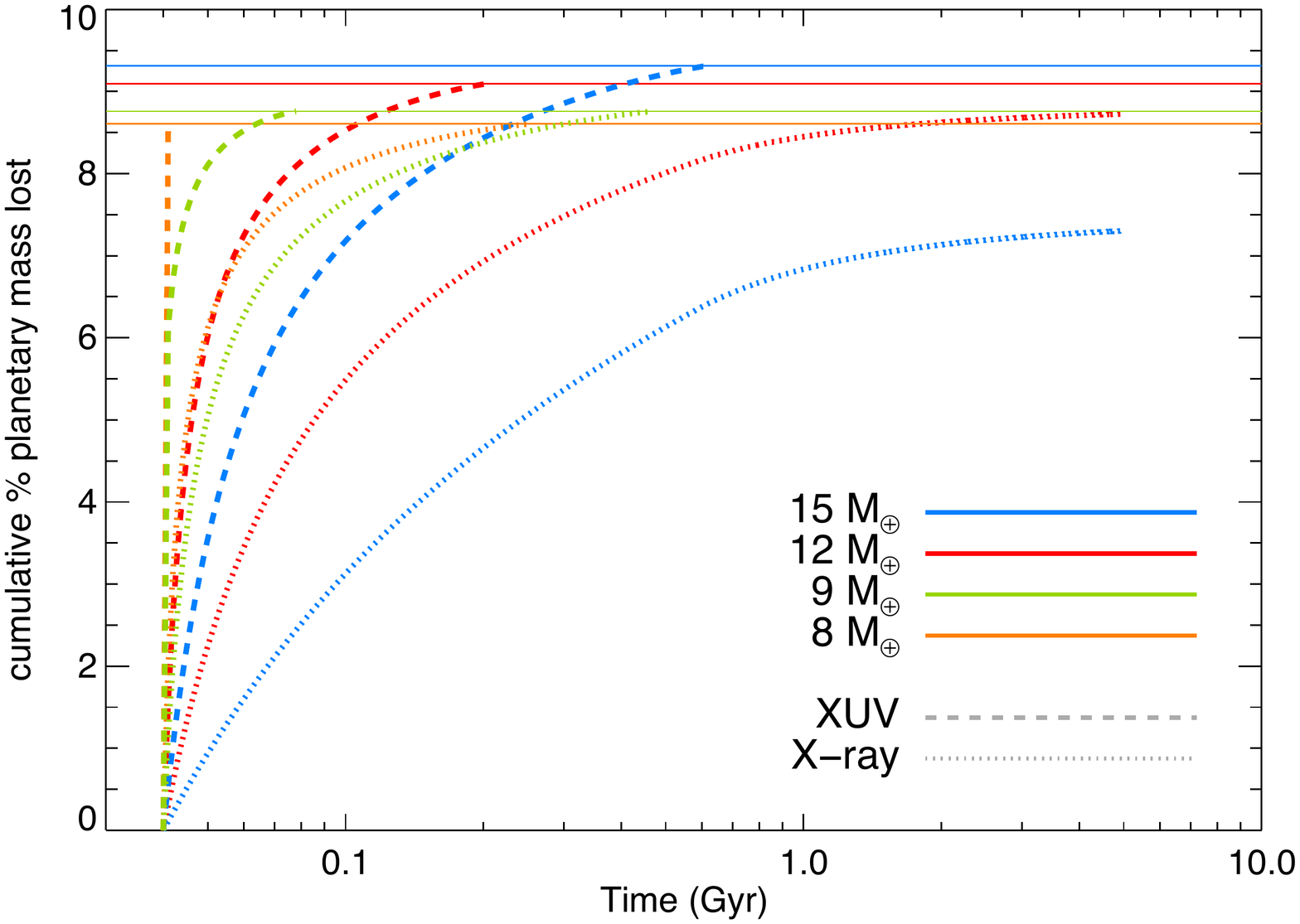}
\includegraphics[trim={1cm 3cm 2cm 2cm},clip,width=.35\textwidth,angle=270]{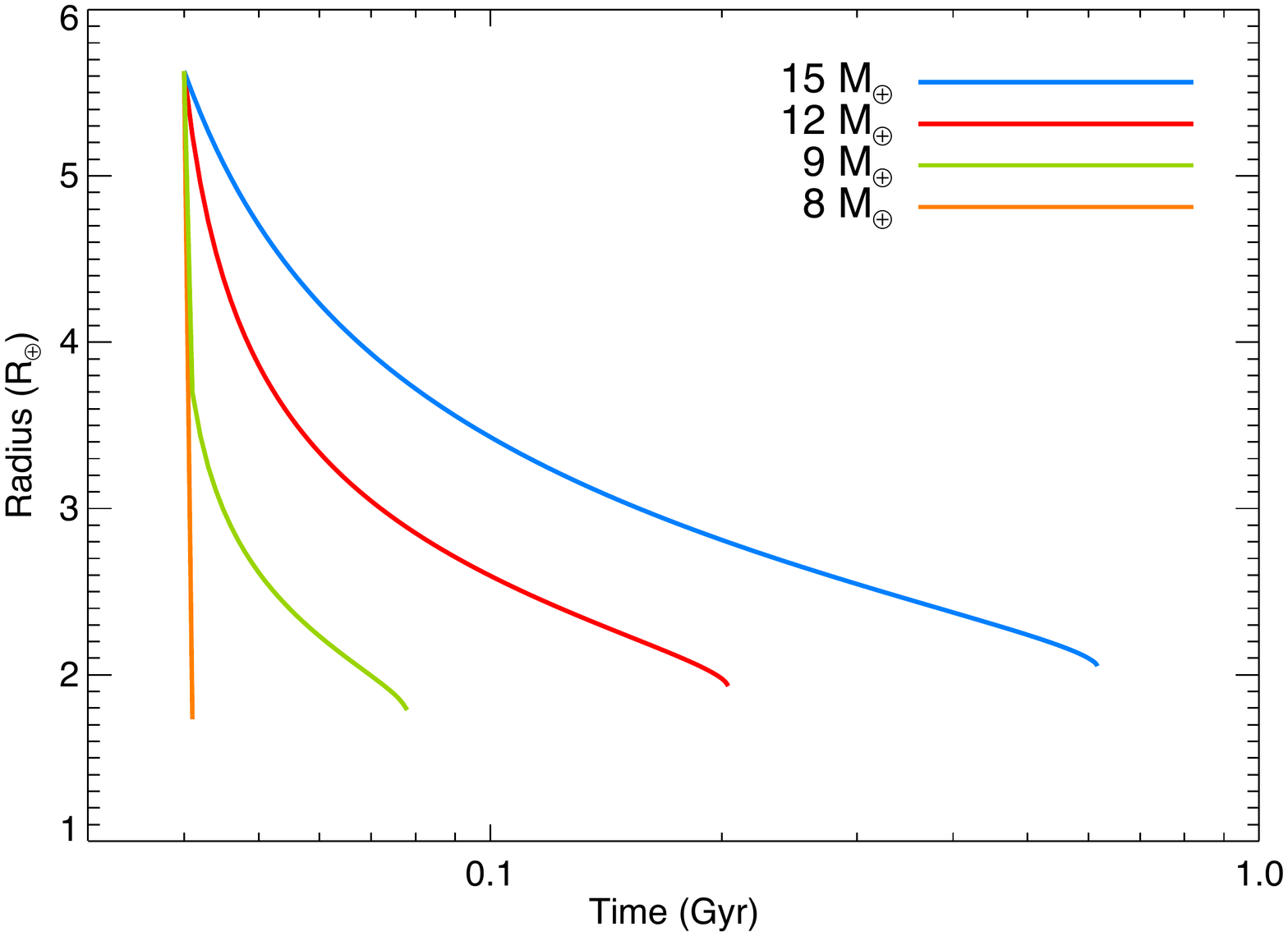}
\caption{Upper panel: cumulative percentage of the planetary mass loss evolution considering the X+EUV irradiation (dashed) or X-ray irradiation alone (dotted). We used the results obtained with the evolution by \cite{Penz08a}. Horizontal solid lines represent the limits of the atmospheric mass concerning four representative cases with different starting masses, colour-coded as in the legend. 
Lower panel: Radius evolution corresponding to the  representative cases according to our models from the present age of the system to the end of the photo-evaporation phase. }
\label{fig:massrad_evolution}
\end{figure}

We also investigated the separate contribution of the X-ray irradiation with respect to the X+EUV irradiation.
Fig.\ref{fig:massrad_evolution} (upper panel) shows the previous simulations, in the case of the P08a evolution (dashed lines), compared to the same model evaluated considering the X-ray flux only (dotted lines). Clearly, the EUV contribution to the evaporation is dominant, since the atmosphere of planets with masses $> 10 M_{\oplus}$ would be retained up to 5 Gyr in absence of it. In the case of a 8 $M_{\oplus}$ complete evaporation would occur in 200-300 Myr.

According to our models, at the end of the evaporation phase the planetary radius will be about 2.0, 1.9, 1.8 and 1.7 R$_{\oplus}$, namely coincident with the core radius, in the simulations with initial mass of 15, 12, 9 and 8 M$_{\oplus}$, respectively. 
Figure \ref{fig:massrad_evolution} (lower panel) shows the radius evolution foreseen with the P08a evolution.  Similarly, the final planetary mass is evaluated as 13.6, 11.0, 8.2 and 7.3 M$_{\oplus}$, respectively.

In all of the cases explored, at the end of the evaporation phase the planet would lose approximately $\sim 10\%$ of its total mass, whereas its radius would be reduced by a factor of $\sim 3$. Our models foresee that DS\,Tuc\,A\,b would conclude in any case its evolution within the gap of the Fulton radius distribution \citep{2017AJ....154..109F}, where approximately half of all planets have lost completely their atmospheres \citep{2020ApJ...891..158M}.
The final density will be about 20 times higher than the initial one.

\section{Discussion}
\label{sec:discussion}
\subsection{Young planets' parameters in context: mass-radius}
\begin{figure*}
    \centering
    \includegraphics[trim={1cm 3.5cm 3cm 2cm},clip,width=0.345\textwidth,angle=270]{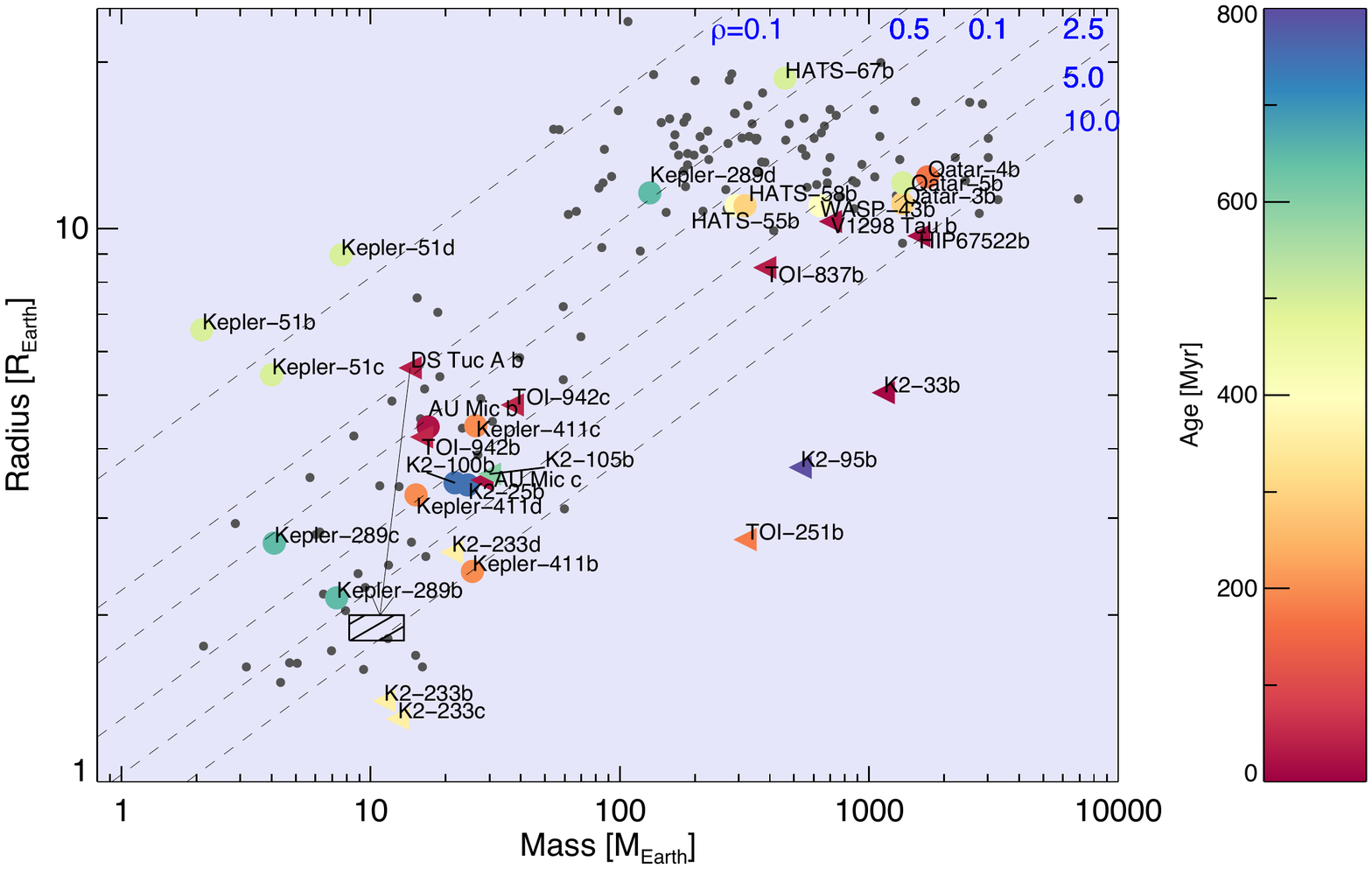}
    \includegraphics[trim={1cm 3.5cm 3cm 2cm},clip,width=0.345\textwidth,angle=270]{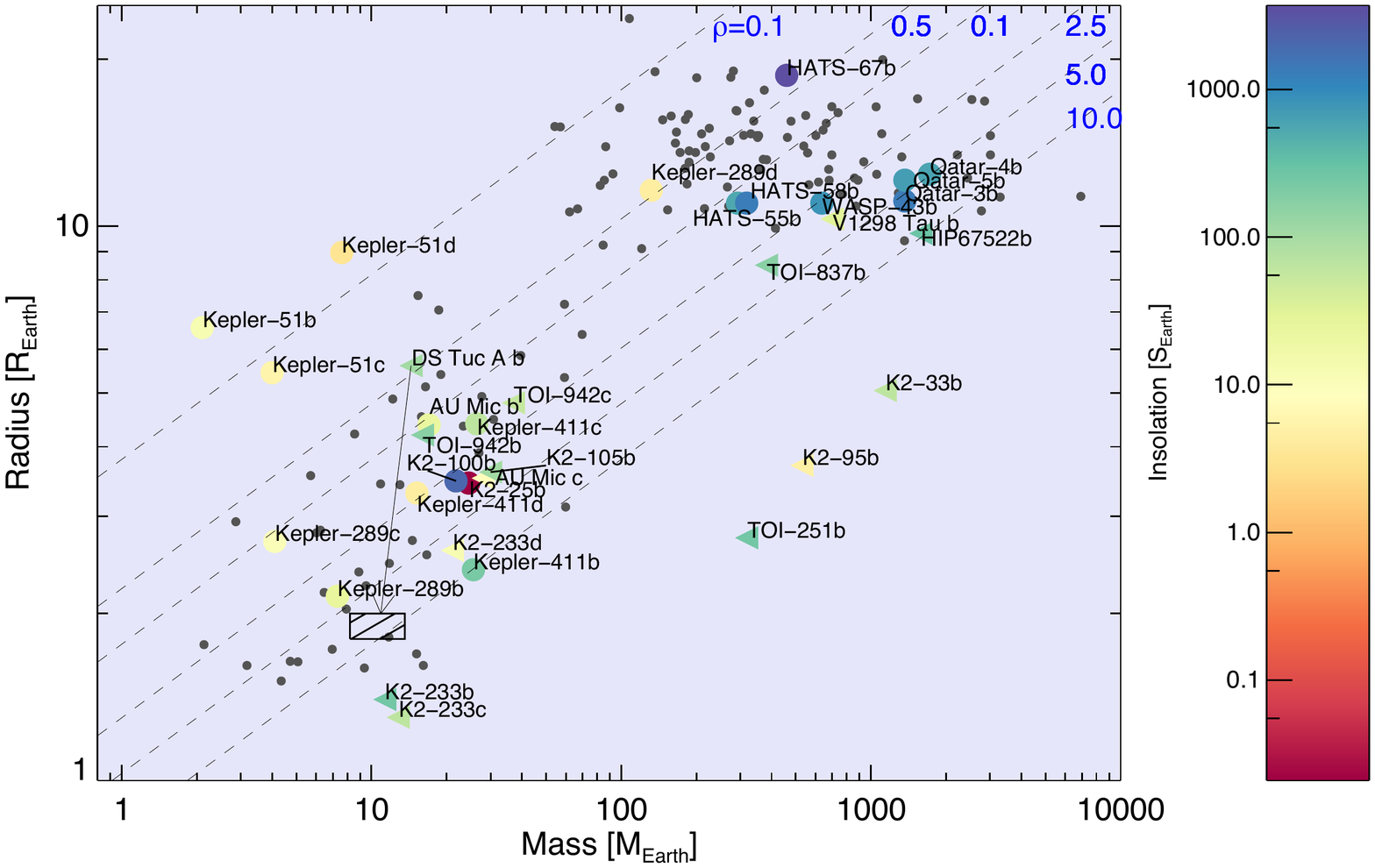}
    \caption{Planetary Mass-Radius diagram for a sample of characterised young transiting planets as a function of the stellar age (left) and the planet insolation (right). Coloured circles indicate planet with measured mass while triangles indicate planets with mass upper limit only. Grey dots represent a sample of well-characterised mature exoplanets for comparison. Dashed lines represent loci of equal density, as indicated at the top of the figure. Hatched rectangle indicates the locus of the final mass and radius of DS\,Tuc\,A\,b according to our planetary evolution models. }
    \label{fig:transiting}
\end{figure*}

The ongoing characterization of transiting exoplanets at young age is providing us the first clues of their population distribution, at least with respect to the mature systems. By using the available data we aimed to update the emerging picture with a sample of well-constrained adult systems. We selected those exoplanets having measured masses and radii with uncertainty better than 20\% and 10\%, respectively, by using the Exo-MerCat tool \citep{2020A&C....3100370A}\footnote{User interface available at \url{https://gitlab.com/eleonoraalei/exo-mercat-gui}}. We used these systems to plot a mass-radius diagram (grey dots) in Fig. \ref{fig:transiting}. % and a period-radius diagram (grey dots) in Fig. \ref{fig:transiting2}. 
We positioned the young transiting planets with measured mass known to date as a function of their age (left panel) and planet insolation in Earth units (right panel). We also placed dashed lines indicating the traces with equal planet density. For planets younger than $\sim$100 Myr we only have a mass upper limit, as in the case of DS\,Tuc\,A\,b, and for this reason they are indicated in the plot with triangles pointing toward lower values of the planetary mass. Their actual location should be then shifted more on the left. This condition is clearly related to the difficult RV modelling due to the contribution of the stellar activity. To built these plots, we considered those planets found by \textit{Kepler/K2} in young associations and the ones detected by TESS, besides DS\,Tuc\,A\,b. Among them, HIP67522 b (Sco-Cen, age $\sim 17$ Myr, \citealt{rizzuto2020}), TOI-251 b (age $\sim 18$ Myr, \citealt{2021AJ....161....2Z}), AU Mic b and c (age $\sim 22$ Myr, \citealt{2020Natur.582..497P,2020arXiv201213238M}), TOI-837 b (age $\sim 35$ Myr, \citealt{2020AJ....160..239B}) and TOI-942 b and c (age $\sim 50$ Myr, \citealt{2021A&A...645A..71C,2021AJ....161....2Z}).
From Fig. \ref{fig:transiting} (left panel) we notice that all the planets younger than 100 Myr show radii larger than $\sim 4 R_{\oplus}$, and they are located in the upper portion of the distribution for intermediate age/mature planets with similar masses. DS\,Tuc\,A\,b is the low-mass planet with the lowest density in the diagram, except for the system of Kepler-51\footnote{Kepler-51 (500 Myr) is a peculiar system with three transiting planets with periods between 45 and 103 days. They show a very low density and low insolation, with planetary masses obtained through the Transit-Timing Variation technique.  In Fig. \ref{fig:transiting} we considered the revised radii reported in \cite{2018ApJ...866...99B}.}.
Currently, its density (< 0.44 g cm$^{-3}$) is similar to the Jupiter-like planet Kepler-289 d and, according to our simulations, its mass-radius evolution is expected to follow the black arrow down to the hatched rectangle. The expected density at this location ranges from 5.6 to 12.8 g cm$^{-3}$ and according to the models by \cite{2013PASP..125..227Z} this should correspond to a composition between 100\% MgSiO$_3$ and 50\%Fe + 50\% MgSiO$_3$. The most curious result is that the final radius of DS\,Tuc\,A\,b is expected to be within the Fulton gap. 

The right panel of Fig. \ref{fig:transiting} shows that DS\,Tuc\,A\,b is the less dense planet in the insolation range between $\sim$10 and $\sim$500 S$_{\oplus}$. Again, we must recall that some of those planets without a robust measured mass could be less dense as well.
More in general, Fig. \ref{fig:transiting} shows that, except for a few outliers\footnote{The masses of K2-33, K2-95 b and TOI-251 b are constrained with a handful of RV data. Furthermore, K2-33 and K2-95 have faint V magnitude preventing for a proper monitoring. Kepler-51 system was previously mentioned.}, young planets seems to lie within or close to the distribution of relatively older planets, with DS\,Tuc\,A\,b located on the low-density edge of it. 

\subsection{Young planets' parameters in context: period-radius}
\begin{figure}
    \centering
    \includegraphics[trim={1.5cm 3cm 2cm 2cm},clip,width=0.35\textwidth,angle=270]{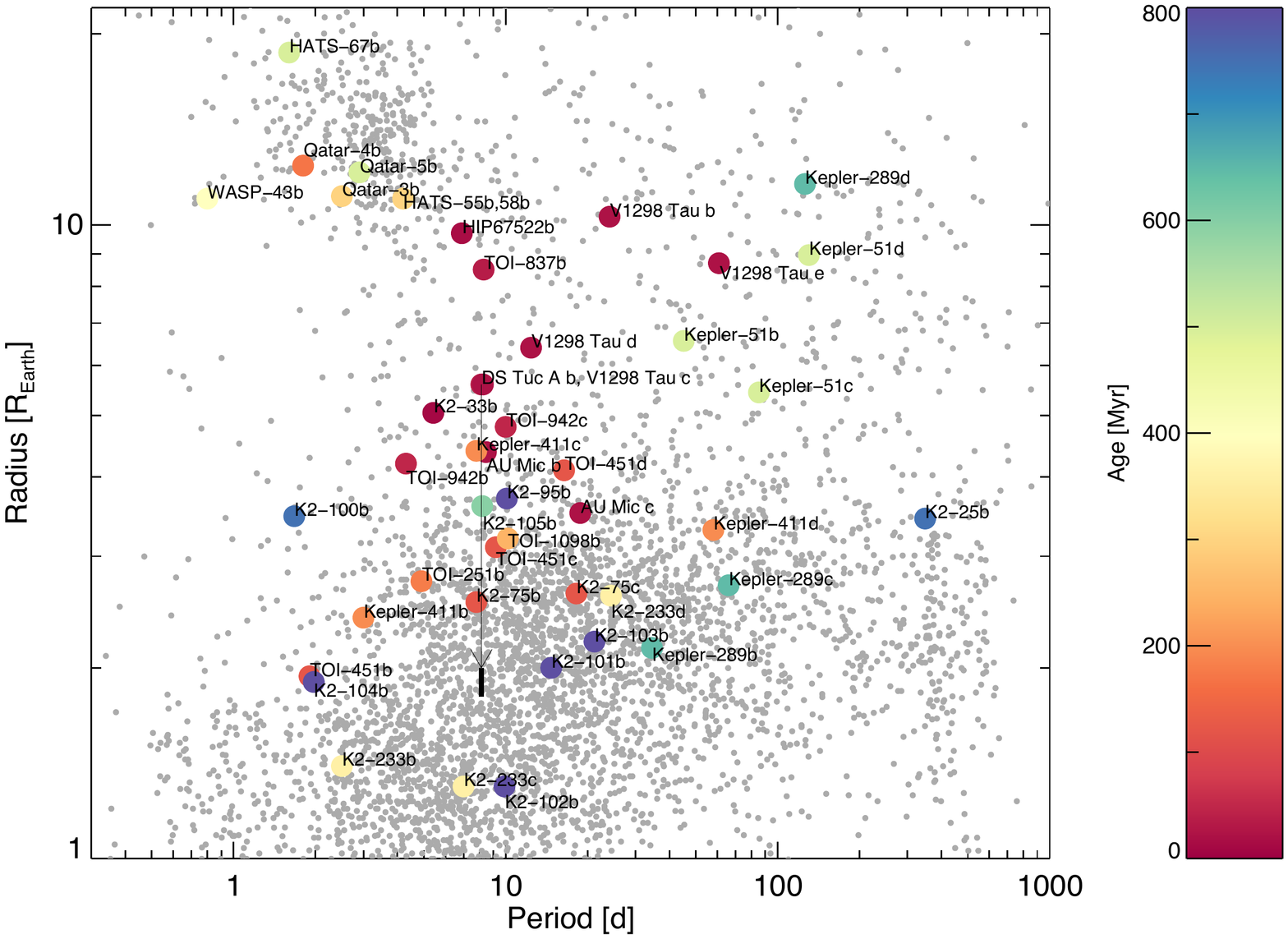}
    \caption{Planetary period-radius diagram for the sample of characterised young transiting planets as a function of the stellar age. Grey dots represent all the known planets with measured period and radius. Black bar line indicates the locus of the final radius of DS\,Tuc\,A\,b according to our planetary evolution models.}
    \label{fig:transiting2}
\end{figure}

Similarly to Fig. \ref{fig:transiting}, we produced a period-radius diagram in Fig. \ref{fig:transiting2}.
It shows a more clear separation between young/larger radii planets (red dots) with respect to intermediate-age/smaller radii planets (other colours), further noticing that almost all the young planets lay in the less populated regions of the plot. 
We must take into account that this distribution could be the result of observation biases. For instance, transiting planets with shorter radii can hardly be detectable when orbiting young stars affected by high photometric variability. This case includes a potential population of non-inflated young planets. As a consequence, these type of planets possibly have lower masses that are more difficult to measure with RVs, as clearly demonstrated in this work.
To investigate this issue, a frequency/population synthesis study should be done, like the one presented by \cite{2010ApJ...719.1454M} in the case of low-mass planets at wide separations. However, this topic is out of the purpose of the present study.
%mann + 2010 apj 719 1454 simula pianeti piccoli formati alla ice line che migrano all'interno ma al max fino a 1 au, quindi alcuni di loro si rivelano con difficolta'.

Another possible bias concerns the timescale of the low-mass planet formation if they follow the same formation path of the terrestrial planets in the Solar System. According to the Grand Tack theory, the latter are formed from the material left behind by the violent destruction of the first generation planets, due to the gravitational interaction produced by Jupiter \citep{2015PNAS..112.4214B}. If the timescale of the second generation planet formation ranges between 30 and 100 Myr \citep{2009GeCoA..73.5150K}, planets around younger stars are still not present in the observed distribution.  %This second hypothesis is true only for those systems for which a debris disc is still present after some tens of Myr, as in the case of AU Mic, while this is not possible for DS Tuc since no IR excess is detected.
Nevertheless, the diagram in Fig. \ref{fig:transiting2} clearly shows the expected radius evolution toward lower values as the system ages, but it can also be used to evaluate possible orbital evolution that could modify the planet period. In the case of DS\,Tuc\,A\,b we do not expect a further inward migration through the disc, since no IR excess is detected \citep{2011ApJ...732...61Z} and therefore the disc is already dissipated. On a larger timescale we must consider dynamic interactions with additional bodies in the system able to interact with DS\,Tuc\,A\,b. From \cite{2019A&A...630A..81B} we excluded planets with 2--3 Jupiter masses and period shorter than 10 days and 7 M$_{\rm J}$ within 100 days, from RV data, and 7-8 M$_{\rm J}$ at separation larger than 40\,AU from direct imaging. DS Tuc B resides at 240 au apart. From the available information, we can not confirm the presence of additional perturbing bodies in the system. Finally, star-planet tidal interactions can occur and produce modifications of the semi-major axis of the orbit. According to the models by \cite{2016CeMDA.126..275B}, tidal interactions between a solar-type star and a close-in planet with mass of 10 M$_{\oplus}$ could shrink its orbit if the planet semi-major axis is $\lesssim 0.04$ au after the disc dissipation.
Being the distance between DS\,Tuc\,A and its planetary companion about 0.09 au, we do not expect a significant modification of the planetary orbit through this specific channel.

The expected location of DS\,Tuc\,A\,b in the diagram at the end of the photo-evaporation phase is therefore within the known distribution (indicated with a black bar line).

\subsection{The challenging detection of the young planets' atmosphere} 
As reported in Sect. \ref{sec:atmo}, no evidence for atmospheric compounds in the atmosphere of DS\,Tuc\,A\,b is found in our high-resolution ESPRESSO data. The lack of H$\alpha$ detection could be explained by a large fraction of ionised planetary escaping material produced by the interaction with the strong XUV stellar radiation (see e.g. \cite{2020A&A...639A.109S} in the case of Ly$\alpha$ in $\pi$ Men c). 
Alternatively, a suppression of the planetary mass-loss rate can be observed in presence of strong magnetic field both of the star and the planet \citep{2014MNRAS.444.3761O}.

Recent studies seem to confirm that a flat spectrum is a characteristic of young planetary atmospheres, despite their large scale height.
\cite{2020AJ....159...57L} and \cite{2020AJ....159...32T} explained the featureless spectra of the intermediate-age planets Kepler-51, b and d, and K2-25 b, respectively, as due to the presence of photochemical hazes. This condition is predicted for Neptune-size companions by \cite{2017AJ....154..261C}, since the equilibrium temperature of those planets is less than 800 K.
In the case of AU Mic b, \cite{PalleAUMic} indicated the stellar activity as the main responsible for the non-detection of atomic species. We must note that the latter case is more similar to the one we have analysed in this work. Because of its young age (about 20 Myr), a high level of stellar activity is present in AU Mic, possibly comparable to the one of DS\,Tuc\,A. Their equilibrium temperatures are 850 K (DS\,Tuc\,A\,b, \citealt{2019ApJ...880L..17N}) and 600-800 K (AU Mic, \citealt{PalleAUMic}), which are close to the transition threshold set by \cite{2017AJ....154..261C} from obscuring hazes to clear atmosphere. This fact complicates possible considerations unlike the case of Kepler-51 and K2-25. 
However, the possible presence of clouds or hazes with Rayleigh scattering masking the H$\alpha$ detection could be further explored with low resolution transmission spectroscopy.

We can conclude that at very young age the contribution of the activity from the star dominates the behaviour of planetary spectral features with respect to all the other components of the planet atmosphere.

\section{Conclusions} \label{sec:concl}
In this work we performed a first attempt to characterise the young star-planet system of DS\,Tuc\,A\,by using different techniques. We monitored this target with HARPS and ESPRESSO aiming to search for a RV signal of the planetary companion and the detection of its atmospheric compounds. Despite the fact that the high level of the stellar activity in such a young host made our analysis very challenging, we were able to provide a robust mass upper limit of the planetary companion (14.4 M$_{\oplus}$) that allowed us to confirm its low density. We confirm a certain degree of alignment of the system through the modelling of the RM effect. Stellar activity also impacted on the atmospheric detection, reinforcing the evidence that the atmosphere of young planets are not easily accessible. We also observed this target with \xmm~and we were able to estimate for the first time the flux, luminosity and coronal plasma temperatures for each stellar component separately. The X-ray luminosity of DS\,Tuc\,A served as input for models of mass evolution of young planets in order to estimate mass evaporation as as a function of age.  
From the photometric point of view, a deeper characterisation of DS\,Tuc\,A\,b and its host will be provided by the new available observations of the TESS cycle-2 (Sectors 27 and 28), and from a dedicated monitoring with CHEOPS (AO-1:ID-0020, PI Desidera, Scandariato et al. in prep.). 
To overcome the effect of the stellar activity in the detection of atmospheric compounds, observation in the near-Infrared band is an option. Besides lowering the impact of the activity, this spectral range offers the possibility to recover Helium \citep[e.g.][]{2019A&A...623A..58A,2020A&A...639A..49G} and several molecules. Spectrographs like CRIRES+ \citep{2014SPIE.9147E..19F} at VLT and NIRPS \citep{2017Msngr.169...21B} at ESO 3.6m telescope could give an important contribution to this issue.

%-----------------------------------------------------------------
\begin{acknowledgements}
We thank the anonymous Referee for the useful comments and suggestions that improved the quality and the clarity of the paper. We thank F. Bouchy and X. Dumusque for managing the sharing of observing time between several programs approved for HARPS in P103 and the observers of these programs for carrying out observations of our target in their nights. We thank C. Lovis for the support in the reprocessing of the HARPS data. 
The authors acknowledge the support of the ARIEL ASI-INAF agreement n.2018-22-HH.0. We acknowledge financial support from the ASI-INAF agreement n.2018-16-HH.0.
We acknowledge the support by INAF/Frontiera through the "Progetti Premiali" funding scheme of the Italian Ministry of Education, University, and Research. 
\end{acknowledgements}

\bibliographystyle{aa} % style aa.bst

\bibliography{dstuc_RML} % your references Yourfile.bib

\begin{thebibliography}{92}
\expandafter\ifx\csname natexlab\endcsname\relax\def\natexlab#1{#1}\fi

\bibitem[{{Alei} {et~al.}(2020){Alei}, {Claudi}, {Bignamini}, \&
  {Molinaro}}]{2020A&C....3100370A}
{Alei}, E., {Claudi}, R., {Bignamini}, A., \& {Molinaro}, M. 2020, Astronomy
  and Computing, 31, 100370

\bibitem[{{Allart} {et~al.}(2019){Allart}, {Bourrier}, {Lovis}, {Ehrenreich},
  {Aceituno}, {Guijarro}, {Pepe}, {Sing}, {Spake}, \&
  {Wyttenbach}}]{2019A&A...623A..58A}
{Allart}, R., {Bourrier}, V., {Lovis}, C., {et~al.} 2019, \aap, 623, A58

\bibitem[{{Ambikasaran} {et~al.}(2015){Ambikasaran}, {Foreman-Mackey},
  {Greengard}, {Hogg}, \& {O'Neil}}]{2015ITPAM..38..252A}
{Ambikasaran}, S., {Foreman-Mackey}, D., {Greengard}, L., {Hogg}, D.~W., \&
  {O'Neil}, M. 2015, IEEE Transactions on Pattern Analysis and Machine
  Intelligence, 38

\bibitem[{{Anglada-Escud{\'e}} \& {Butler}(2012)}]{2012ApJS..200...15A}
{Anglada-Escud{\'e}}, G. \& {Butler}, R.~P. 2012, \apjs, 200, 15

\bibitem[{{Astudillo-Defru} \& {Rojo}(2013)}]{2013A&A...557A..56A}
{Astudillo-Defru}, N. \& {Rojo}, P. 2013, \aap, 557, A56

\bibitem[{{Bashi} {et~al.}(2017){Bashi}, {Helled}, {Zucker}, \&
  {Mordasini}}]{2017A&A...604A..83B}
{Bashi}, D., {Helled}, R., {Zucker}, S., \& {Mordasini}, C. 2017, \aap, 604,
  A83

\bibitem[{{Battley} {et~al.}(2020){Battley}, {Pollacco}, \&
  {Armstrong}}]{2020MNRAS.496.1197B}
{Battley}, M.~P., {Pollacco}, D., \& {Armstrong}, D.~J. 2020, \mnras, 496, 1197

\bibitem[{{Batygin} \& {Laughlin}(2015)}]{2015PNAS..112.4214B}
{Batygin}, K. \& {Laughlin}, G. 2015, Proceedings of the National Academy of
  Science, 112, 4214

\bibitem[{{Benatti} {et~al.}(2019){Benatti}, {Nardiello}, {Malavolta},
  {Desidera}, {Borsato}, {Nascimbeni}, {Damasso}, {D'Orazi}, {Mesa}, {Messina},
  {Esposito}, {Bignamini}, {Claudi}, {Covino}, {Lovis}, \&
  {Sabotta}}]{2019A&A...630A..81B}
{Benatti}, S., {Nardiello}, D., {Malavolta}, L., {et~al.} 2019, \aap, 630, A81

\bibitem[{{Berger} {et~al.}(2018{\natexlab{a}}){Berger}, {Howard}, \&
  {Boesgaard}}]{2018ApJ...855..115B}
{Berger}, T.~A., {Howard}, A.~W., \& {Boesgaard}, A.~M. 2018{\natexlab{a}},
  \apj, 855, 115

\bibitem[{{Berger} {et~al.}(2018{\natexlab{b}}){Berger}, {Huber}, {Gaidos}, \&
  {van Saders}}]{2018ApJ...866...99B}
{Berger}, T.~A., {Huber}, D., {Gaidos}, E., \& {van Saders}, J.~L.
  2018{\natexlab{b}}, \apj, 866, 99

\bibitem[{{Bolmont} \& {Mathis}(2016)}]{2016CeMDA.126..275B}
{Bolmont}, E. \& {Mathis}, S. 2016, Celestial Mechanics and Dynamical
  Astronomy, 126, 275

\bibitem[{{Borsa} {et~al.}(2021){Borsa}, {Allart}, {Casasayas-Barris},
  {Tabernero}, {Zapatero Osorio}, {Cristiani}, {Pepe}, {Rebolo}, {Santos},
  {Adibekyan}, {Bourrier}, {Demangeon}, {Ehrenreich}, {Pall{\'e}}, {Sousa},
  {Lillo-Box}, {Lovis}, {Micela}, {Oshagh}, {Poretti}, {Sozzetti}, {Allende
  Prieto}, {Alibert}, {Amate}, {Benz}, {Bouchy}, {Cabral}, {Dekker},
  {D'Odorico}, {Di Marcantonio}, {Figueira}, {Genova Santos}, {Gonz{\'a}lez
  Hern{\'a}ndez}, {Lo Curto}, {Manescau}, {Martins}, {M{\'e}gevand}, {Mehner},
  {Molaro}, {Nunes}, {Riva}, {Su{\'a}rez Mascare{\~n}o}, {Udry}, \&
  {Zerbi}}]{borsa2020}
{Borsa}, F., {Allart}, R., {Casasayas-Barris}, N., {et~al.} 2021, \aap, 645,
  A24

\bibitem[{{Borucki} {et~al.}(2003){Borucki}, {Koch}, {Lissauer}, {Basri},
  {Caldwell}, {Cochran}, {Dunham}, {Geary}, {Latham}, {Gilliland}, {Caldwell},
  {Jenkins}, \& {Kondo}}]{2003SPIE.4854..129B}
{Borucki}, W.~J., {Koch}, D.~G., {Lissauer}, J.~J., {et~al.} 2003, in Society
  of Photo-Optical Instrumentation Engineers (SPIE) Conference Series, Vol.
  4854, Future EUV/UV and Visible Space Astrophysics Missions and
  Instrumentation., ed. J.~C. {Blades} \& O.~H.~W. {Siegmund}, 129--140

\bibitem[{{Bouchy} {et~al.}(2017){Bouchy}, {Doyon}, {Artigau}, {Melo},
  {Hernandez}, {Wildi}, {Delfosse}, {Lovis}, {Figueira}, {Canto Martins},
  {Gonz{\'a}lez Hern{\'a}ndez}, {Thibault}, {Reshetov}, {Pepe}, {Santos}, {de
  Medeiros}, {Rebolo}, {Abreu}, {Adibekyan}, {Bandy}, {Benz}, {Blind},
  {Bohlender}, {Boisse}, {Bovay}, {Broeg}, {Brousseau}, {Cabral}, {Chazelas},
  {Cloutier}, {Coelho}, {Conod}, {Cumming}, {Delabre}, {Genolet}, {Hagelberg},
  {Jayawardhana}, {K{\"a}ufl}, {Lafreni{\`e}re}, {de Castro Le{\~a}o}, {Malo},
  {de Medeiros Martins}, {Matthews}, {Metchev}, {Oshagh}, {Ouellet}, {Parro},
  {Rasilla Pi{\~n}eiro}, {Santos}, {Sarajlic}, {Segovia}, {Sordet}, {Udry},
  {Valencia}, {Vall{\'e}e}, {Venn}, {Wade}, \&
  {Saddlemyer}}]{2017Msngr.169...21B}
{Bouchy}, F., {Doyon}, R., {Artigau}, {\'E}., {et~al.} 2017, The Messenger,
  169, 21

\bibitem[{{Bouma} {et~al.}(2020){Bouma}, {Hartman}, {Brahm}, {Evans},
  {Collins}, {Zhou}, {Sarkis}, {Quinn}, {de Leon}, {Livingston}, {Bergmann},
  {Stassun}, {Bhatti}, {Winn}, {Bakos}, {Abe}, {Crouzet}, {Dransfield},
  {Guillot}, {Marie-Sainte}, {M{\'e}karnia}, {Triaud}, {Tinney}, {Henning},
  {Espinoza}, {Jord{\'a}n}, {Barbieri}, {Nandakumar}, {Trifonov}, {Vines},
  {Vuckovic}, {Ziegler}, {Law}, {Mann}, {Ricker}, {Vanderspek}, {Seager},
  {Jenkins}, {Burke}, {Dragomir}, {Levine}, {Quintana}, {Rodriguez}, {Smith},
  \& {Wohler}}]{2020AJ....160..239B}
{Bouma}, L.~G., {Hartman}, J.~D., {Brahm}, R., {et~al.} 2020, \aj, 160, 239

\bibitem[{{Brown} {et~al.}(2017){Brown}, {Triaud}, {Doyle}, {Gillon}, {Lendl},
  {Anderson}, {Collier Cameron}, {H{\'e}brard}, {Hellier}, {Lovis}, {Maxted},
  {Pepe}, {Pollacco}, {Queloz}, \& {Smalley}}]{2017MNRAS.464..810B}
{Brown}, D.~J.~A., {Triaud}, A.~H.~M.~J., {Doyle}, A.~P., {et~al.} 2017,
  \mnras, 464, 810

\bibitem[{{Buchner} {et~al.}(2014){Buchner}, {Georgakakis}, {Nandra}, {Hsu},
  {Rangel}, {Brightman}, {Merloni}, {Salvato}, {Donley}, \&
  {Kocevski}}]{Buchner2014}
{Buchner}, J., {Georgakakis}, A., {Nandra}, K., {et~al.} 2014, \aap, 564, A125

\bibitem[{{Carleo} {et~al.}(2018){Carleo}, {Benatti}, {Lanza}, {Gratton},
  {Claudi}, {Desidera}, {Mace}, {Messina}, {Sanna}, {Sissa}, {Ghedina},
  {Ghinassi}, {Guerra}, {Harutyunyan}, {Micela}, {Molinari}, {Oliva}, {Tozzi},
  {Baffa}, {Baruffolo}, {Bignamini}, {Buchschacher}, {Cecconi}, {Cosentino},
  {Endl}, {Falcini}, {Fantinel}, {Fini}, {Fugazza}, {Galli}, {Giani},
  {Gonz{\'a}lez}, {Gonz{\'a}lez-{\'A}lvarez}, {Gonz{\'a}lez}, {Hernandez},
  {Hernandez Diaz}, {Iuzzolino}, {Kaplan}, {Kidder}, {Lodi}, {Malavolta},
  {Maldonado}, {Origlia}, {Perez Ventura}, {Puglisi}, {Rainer}, {Riverol},
  {Riverol}, {San Juan}, {Scuderi}, {Seemann}, {Sokal}, {Sozzetti}, \&
  {Sozzi}}]{2018A&A...613A..50C}
{Carleo}, I., {Benatti}, S., {Lanza}, A.~F., {et~al.} 2018, \aap, 613, A50

\bibitem[{{Carleo} {et~al.}(2021){Carleo}, {Desidera}, {Nardiello},
  {Malavolta}, {Lanza}, {Livingston}, {Locci}, {Marzari}, {Messina}, {Turrini},
  {Baratella}, {Borsa}, {D'Orazi}, {Nascimbeni}, {Pinamonti}, {Rainer}, {Alei},
  {Bignamini}, {Gratton}, {Micela}, {Montalto}, {Sozzetti}, {Squicciarini},
  {Affer}, {Benatti}, {Biazzo}, {Bonomo}, {Claudi}, {Cosentino}, {Covino},
  {Damasso}, {Esposito}, {Fiorenzano}, {Frustagli}, {Giacobbe}, {Harutyunyan},
  {Leto}, {Magazz{\`u}}, {Maggio}, {Mainella}, {Maldonado}, {Mallonn},
  {Mancini}, {Molinari}, {Molinaro}, {Pagano}, {Pedani}, {Piotto}, {Poretti},
  {Redfield}, \& {Scandariato}}]{2021A&A...645A..71C}
{Carleo}, I., {Desidera}, S., {Nardiello}, D., {et~al.} 2021, \aap, 645, A71

\bibitem[{{Carleo} {et~al.}(2020){Carleo}, {Malavolta}, {Lanza}, {Damasso},
  {Desidera}, {Borsa}, {Mallonn}, {Pinamonti}, {Gratton}, {Alei}, {Benatti},
  {Mancini}, {Maldonado}, {Biazzo}, {Esposito}, {Frustagli},
  {Gonz{\'a}lez-{\'A}lvarez}, {Micela}, {Scandariato}, {Sozzetti}, {Affer},
  {Bignamini}, {Bonomo}, {Claudi}, {Cosentino}, {Covino}, {Fiorenzano},
  {Giacobbe}, {Harutyunyan}, {Leto}, {Maggio}, {Molinari}, {Nascimbeni},
  {Pagano}, {Pedani}, {Piotto}, {Poretti}, {Rainer}, {Redfield}, {Baffa},
  {Baruffolo}, {Buchschacher}, {Billotti}, {Cecconi}, {Falcini}, {Fantinel},
  {Fini}, {Galli}, {Ghedina}, {Ghinassi}, {Giani}, {Gonzalez}, {Gonzalez},
  {Guerra}, {Hernandez Diaz}, {Hernandez}, {Iuzzolino}, {Lodi}, {Oliva},
  {Origlia}, {Perez Ventura}, {Puglisi}, {Riverol}, {Riverol}, {San Juan},
  {Sanna}, {Scuderi}, {Seemann}, {Sozzi}, \& {Tozzi}}]{2020A&A...638A...5C}
{Carleo}, I., {Malavolta}, L., {Lanza}, A.~F., {et~al.} 2020, \aap, 638, A5

\bibitem[{{Cegla} {et~al.}(2016){Cegla}, {Lovis}, {Bourrier}, {Beeck},
  {Watson}, \& {Pepe}}]{2016A&A...588A.127C}
{Cegla}, H.~M., {Lovis}, C., {Bourrier}, V., {et~al.} 2016, \aap, 588, A127

\bibitem[{{Chen} \& {Kipping}(2017)}]{2017ApJ...834...17C}
{Chen}, J. \& {Kipping}, D. 2017, \apj, 834, 17

\bibitem[{{Crossfield} \& {Kreidberg}(2017)}]{2017AJ....154..261C}
{Crossfield}, I. J.~M. \& {Kreidberg}, L. 2017, \aj, 154, 261

\bibitem[{{Csizmadia} {et~al.}(2013){Csizmadia}, {Pasternacki}, {Dreyer},
  {Cabrera}, {Erikson}, \& {Rauer}}]{csizmadia2013}
{Csizmadia}, S., {Pasternacki}, T., {Dreyer}, C., {et~al.} 2013, \aap, 549, A9

\bibitem[{{Damasso} {et~al.}(2020){Damasso}, {Lanza}, {Benatti}, {Rajpaul},
  {Mallonn}, {Desidera}, {Biazzo}, {D'Orazi}, {Malavolta}, {Nardiello},
  {Rainer}, {Borsa}, {Affer}, {Bignamini}, {Bonomo}, {Carleo}, {Claudi},
  {Cosentino}, {Covino}, {Giacobbe}, {Gratton}, {Harutyunyan}, {Knapic},
  {Leto}, {Maggio}, {Maldonado}, {Mancini}, {Micela}, {Molinari}, {Nascimbeni},
  {Pagano}, {Piotto}, {Poretti}, {Scandariato}, {Sozzetti}, {Capuzzo Dolcetta},
  {Di Mauro}, {Carosati}, {Fiorenzano}, {Frustagli}, {Pedani}, {Pinamonti},
  {Stoev}, \& {Turrini}}]{2020A&A...642A.133D}
{Damasso}, M., {Lanza}, A.~F., {Benatti}, S., {et~al.} 2020, \aap, 642, A133

\bibitem[{{David} {et~al.}(2019){David}, {Petigura}, {Luger}, {Foreman-Mackey},
  {Livingston}, {Mamajek}, \& {Hillenbrand}}]{2019ApJ...885L..12D}
{David}, T.~J., {Petigura}, E.~A., {Luger}, R., {et~al.} 2019, \apjl, 885, L12

\bibitem[{{Donati} {et~al.}(2020){Donati}, {Bouvier}, {Alencar}, {Moutou},
  {Malo}, {Takami}, {M{\'e}nard}, {Dougados}, {Hussain}, \& {The Matysse
  Collaboration}}]{2020MNRAS.491.5660D}
{Donati}, J.~F., {Bouvier}, J., {Alencar}, S.~H., {et~al.} 2020, \mnras, 491,
  5660

\bibitem[{{Donati} {et~al.}(2016){Donati}, {Moutou}, {Malo}, {Baruteau}, {Yu},
  {H{\'e}brard}, {Hussain}, {Alencar}, {M{\'e}nard}, {Bouvier}, {Petit},
  {Takami}, {Doyon}, \& {Cameron}}]{donati2016}
{Donati}, J.~F., {Moutou}, C., {Malo}, L., {et~al.} 2016, \nat, 534, 662

\bibitem[{{Donati} {et~al.}(2017){Donati}, {Yu}, {Moutou}, {Cameron}, {Malo},
  {Grankin}, {H{\'e}brard}, {Hussain}, {Vidotto}, {Alencar}, {Haywood},
  {Bouvier}, {Petit}, {Takami}, {Herczeg}, {Gregory}, {Jardine}, {Morin}, \&
  {MaTYSSE Collaboration}}]{2017MNRAS.465.3343D}
{Donati}, J.~F., {Yu}, L., {Moutou}, C., {et~al.} 2017, \mnras, 465, 3343

\bibitem[{{Eastman} {et~al.}(2013){Eastman}, {Gaudi}, \&
  {Agol}}]{Eastmanetal2013}
{Eastman}, J., {Gaudi}, B.~S., \& {Agol}, E. 2013, \pasp, 125, 83

\bibitem[{{Feroz} {et~al.}(2019){Feroz}, {Hobson}, {Cameron}, \&
  {Pettitt}}]{Feroz2019}
{Feroz}, F., {Hobson}, M.~P., {Cameron}, E., \& {Pettitt}, A.~N. 2019, The Open
  Journal of Astrophysics, 2, 10

\bibitem[{{Follert} {et~al.}(2014){Follert}, {Dorn}, {Oliva}, {Lizon},
  {Hatzes}, {Piskunov}, {Reiners}, {Seemann}, {Stempels}, {Heiter}, {Marquart},
  {Lockhart}, {Anglada-Escude}, {L{\"o}winger}, {Baade}, {Grunhut}, {Bristow},
  {Klein}, {Jung}, {Ives}, {Kerber}, {Pozna}, {Paufique}, {Kaeufl}, {Origlia},
  {Valenti}, {Gojak}, {Hilker}, {Pasquini}, {Smette}, \&
  {Smoker}}]{2014SPIE.9147E..19F}
{Follert}, R., {Dorn}, R.~J., {Oliva}, E., {et~al.} 2014, in Society of
  Photo-Optical Instrumentation Engineers (SPIE) Conference Series, Vol. 9147,
  Ground-based and Airborne Instrumentation for Astronomy V, ed. S.~K.
  {Ramsay}, I.~S. {McLean}, \& H.~{Takami}, 914719

\bibitem[{Foreman-Mackey {et~al.}(2017)Foreman-Mackey, Agol, Ambikasaran, \&
  Angus}]{celerite}
Foreman-Mackey, D., Agol, E., Ambikasaran, S., \& Angus, R. 2017, The
  Astronomical Journal, 154, 220

\bibitem[{{Fulton} {et~al.}(2017){Fulton}, {Petigura}, {Howard}, {Isaacson},
  {Marcy}, {Cargile}, {Hebb}, {Weiss}, {Johnson}, {Morton}, {Sinukoff},
  {Crossfield}, \& {Hirsch}}]{2017AJ....154..109F}
{Fulton}, B.~J., {Petigura}, E.~A., {Howard}, A.~W., {et~al.} 2017, \aj, 154,
  109

\bibitem[{{Gao} \& {Zhang}(2020)}]{2020ApJ...890...93G}
{Gao}, P. \& {Zhang}, X. 2020, \apj, 890, 93

\bibitem[{{Gomes da Silva} {et~al.}(2018){Gomes da Silva}, {Figueira},
  {Santos}, \& {Faria}}]{2018JOSS....3..667G}
{Gomes da Silva}, J., {Figueira}, P., {Santos}, N., \& {Faria}, J. 2018, The
  Journal of Open Source Software, 3, 667

\bibitem[{{Gontcharov} \& {Mosenkov}(2018)}]{2018MNRAS.475.1121G}
{Gontcharov}, G.~A. \& {Mosenkov}, A.~V. 2018, \mnras, 475, 1121

\bibitem[{{Guilluy} {et~al.}(2020){Guilluy}, {Andretta}, {Borsa}, {Giacobbe},
  {Sozzetti}, {Covino}, {Bourrier}, {Fossati}, {Bonomo}, {Esposito},
  {Giampapa}, {Harutyunyan}, {Rainer}, {Brogi}, {Bruno}, {Claudi}, {Frustagli},
  {Lanza}, {Mancini}, {Pino}, {Poretti}, {Scandariato}, {Affer}, {Baffa},
  {Baruffolo}, {Benatti}, {Biazzo}, {Bignamini}, {Boschin}, {Carleo},
  {Cecconi}, {Cosentino}, {Damasso}, {Desidera}, {Falcini}, {Martinez
  Fiorenzano}, {Ghedina}, {Gonz{\'a}lez-{\'A}lvarez}, {Guerra}, {Hernandez},
  {Leto}, {Maggio}, {Malavolta}, {Maldonado}, {Micela}, {Molinari},
  {Nascimbeni}, {Pagano}, {Pedani}, {Piotto}, \&
  {Reiners}}]{2020A&A...639A..49G}
{Guilluy}, G., {Andretta}, V., {Borsa}, F., {et~al.} 2020, \aap, 639, A49

\bibitem[{{Howell} {et~al.}(2014){Howell}, {Sobeck}, {Haas}, {Still},
  {Barclay}, {Mullally}, {Troeltzsch}, {Aigrain}, {Bryson}, {Caldwell},
  {Chaplin}, {Cochran}, {Huber}, {Marcy}, {Miglio}, {Najita}, {Smith},
  {Twicken}, \& {Fortney}}]{2014PASP..126..398H}
{Howell}, S.~B., {Sobeck}, C., {Haas}, M., {et~al.} 2014, \pasp, 126, 398

\bibitem[{{Jackson} {et~al.}(2012){Jackson}, {Davis}, \&
  {Wheatley}}]{2012MNRAS.422.2024J}
{Jackson}, A.~P., {Davis}, T.~A., \& {Wheatley}, P.~J. 2012, \mnras, 422, 2024

\bibitem[{{Kleine} {et~al.}(2009){Kleine}, {Touboul}, {Bourdon}, {Nimmo},
  {Mezger}, {Palme}, {Jacobsen}, {Yin}, \& {Halliday}}]{2009GeCoA..73.5150K}
{Kleine}, T., {Touboul}, M., {Bourdon}, B., {et~al.} 2009, \gca, 73, 5150

\bibitem[{{Kraus} {et~al.}(2015){Kraus}, {Cody}, {Covey}, {Rizzuto}, {Mann}, \&
  {Ireland}}]{2015ApJ...807....3K}
{Kraus}, A.~L., {Cody}, A.~M., {Covey}, K.~R., {et~al.} 2015, \apj, 807, 3

\bibitem[{{Kubyshkina} {et~al.}(2018){Kubyshkina}, {Fossati}, {Erkaev},
  {Cubillos}, {Johnstone}, {Kislyakova}, {Lammer}, {Lendl}, \&
  {Odert}}]{Kub18b}
{Kubyshkina}, D., {Fossati}, L., {Erkaev}, N.~V., {et~al.} 2018, \apjl, 866,
  L18

\bibitem[{{Libby-Roberts} {et~al.}(2020){Libby-Roberts}, {Berta-Thompson},
  {D{\'e}sert}, {Masuda}, {Morley}, {Lopez}, {Deck}, {Fabrycky}, {Fortney},
  {Line}, {Sanchis-Ojeda}, \& {Winn}}]{2020AJ....159...57L}
{Libby-Roberts}, J.~E., {Berta-Thompson}, Z.~K., {D{\'e}sert}, J.-M., {et~al.}
  2020, \aj, 159, 57

\bibitem[{{Linder} {et~al.}(2019){Linder}, {Mordasini}, {Molli{\`e}re},
  {Marleau}, {Malik}, {Quanz}, \& {Meyer}}]{2019A&A...623A..85L}
{Linder}, E.~F., {Mordasini}, C., {Molli{\`e}re}, P., {et~al.} 2019, \aap, 623,
  A85

\bibitem[{{Lopez} \& {Fortney}(2014)}]{LopFor14}
{Lopez}, E.~D. \& {Fortney}, J.~J. 2014, \apj, 792, 1

\bibitem[{{Lopez} {et~al.}(2012){Lopez}, {Fortney}, \&
  {Miller}}]{2012ApJ...761...59L}
{Lopez}, E.~D., {Fortney}, J.~J., \& {Miller}, N. 2012, \apj, 761, 59

\bibitem[{{Mann} {et~al.}(2010){Mann}, {Gaidos}, \&
  {Gaudi}}]{2010ApJ...719.1454M}
{Mann}, A.~W., {Gaidos}, E., \& {Gaudi}, B.~S. 2010, \apj, 719, 1454

\bibitem[{{Mann} {et~al.}(2016){Mann}, {Newton}, {Rizzuto}, {Irwin}, {Feiden},
  {Gaidos}, {Mace}, {Kraus}, {James}, {Ansdell}, {Charbonneau}, {Covey},
  {Ireland}, {Jaffe}, {Johnson}, {Kidder}, \&
  {Vanderburg}}]{2016AJ....152...61M}
{Mann}, A.~W., {Newton}, E.~R., {Rizzuto}, A.~C., {et~al.} 2016, \aj, 152, 61

\bibitem[{{Martioli} {et~al.}(2020){Martioli}, {H{\'e}brard}, {Correia},
  {Laskar}, \& {Lecavelier des Etangs}}]{2020arXiv201213238M}
{Martioli}, E., {H{\'e}brard}, G., {Correia}, A.~C.~M., {Laskar}, J., \&
  {Lecavelier des Etangs}, A. 2020, arXiv e-prints, arXiv:2012.13238

\bibitem[{{Mayor} {et~al.}(2003){Mayor}, {Pepe}, {Queloz}, {Bouchy},
  {Rupprecht}, {Lo Curto}, {Avila}, {Benz}, {Bertaux}, {Bonfils}, {Dall},
  {Dekker}, {Delabre}, {Eckert}, {Fleury}, {Gilliotte}, {Gojak}, {Guzman},
  {Kohler}, {Lizon}, {Longinotti}, {Lovis}, {Megevand}, {Pasquini}, {Reyes},
  {Sivan}, {Sosnowska}, {Soto}, {Udry}, {van Kesteren}, {Weber}, \&
  {Weilenmann}}]{2003Msngr.114...20M}
{Mayor}, M., {Pepe}, F., {Queloz}, D., {et~al.} 2003, The Messenger, 114, 20

\bibitem[{{McLaughlin}(1924)}]{1924ApJ....60...22M}
{McLaughlin}, D.~B. 1924, \apj, 60, 22

\bibitem[{{Mocquet} {et~al.}(2014){Mocquet}, {Grasset}, \& {Sotin}}]{Mocq14}
{Mocquet}, A., {Grasset}, O., \& {Sotin}, C. 2014, Philosophical Transactions
  of the Royal Society of London Series A, 372, 20130164

\bibitem[{{Modirrousta-Galian} {et~al.}(2020){Modirrousta-Galian}, {Locci}, \&
  {Micela}}]{2020ApJ...891..158M}
{Modirrousta-Galian}, D., {Locci}, D., \& {Micela}, G. 2020, \apj, 891, 158

\bibitem[{{Molli{\`e}re} {et~al.}(2019){Molli{\`e}re}, {Wardenier}, {van
  Boekel}, {Henning}, {Molaverdikhani}, \& {Snellen}}]{pRT}
{Molli{\`e}re}, P., {Wardenier}, J.~P., {van Boekel}, R., {et~al.} 2019, \aap,
  627, A67

\bibitem[{{Montet} {et~al.}(2020){Montet}, {Feinstein}, {Luger}, {Bedell},
  {Gully-Santiago}, {Teske}, {Wang}, {Butler}, {Flowers}, {Shectman}, {Crane},
  \& {Thompson}}]{2020AJ....159..112M}
{Montet}, B.~T., {Feinstein}, A.~D., {Luger}, R., {et~al.} 2020, \aj, 159, 112

\bibitem[{{Nardiello}(2020)}]{2020MNRAS.498.5972N}
{Nardiello}, D. 2020, \mnras, 498, 5972

\bibitem[{{Nardiello} {et~al.}(2020){Nardiello}, {Piotto}, {Deleuil},
  {Malavolta}, {Montalto}, {Bedin}, {Borsato}, {Granata}, {Libralato}, \&
  {Manthopoulou}}]{2020MNRAS.495.4924N}
{Nardiello}, D., {Piotto}, G., {Deleuil}, M., {et~al.} 2020, \mnras, 495, 4924

\bibitem[{{Newton} {et~al.}(2019){Newton}, {Mann}, {Tofflemire}, {Pearce},
  {Rizzuto}, {Vanderburg}, {Martinez}, {Wang}, {Ruffio}, {Kraus}, {Johnson},
  {Thao}, {Wood}, {Rampalli}, {Nielsen}, {Collins}, {Dragomir}, {Hellier},
  {Anderson}, {Barclay}, {Brown}, {Feiden}, {Hart}, {Isopi}, {Kielkopf},
  {Mallia}, {Nelson}, {Rodriguez}, {Stockdale}, {Waite}, {Wright}, {Lissauer},
  {Ricker}, {Vanderspek}, {Latham}, {Seager}, {Winn}, {Jenkins}, {Bouma},
  {Burke}, {Davies}, {Fausnaugh}, {Li}, {Morris}, {Mukai}, {Villase{\~n}or},
  {Villeneuva}, {De Rosa}, {Macintosh}, {Mengel}, {Okumura}, \&
  {Wittenmyer}}]{2019ApJ...880L..17N}
{Newton}, E.~R., {Mann}, A.~W., {Tofflemire}, B.~M., {et~al.} 2019, \apjl, 880,
  L17

\bibitem[{{Ohta} {et~al.}(2005){Ohta}, {Taruya}, \&
  {Suto}}]{2005ApJ...622.1118O}
{Ohta}, Y., {Taruya}, A., \& {Suto}, Y. 2005, \apj, 622, 1118

\bibitem[{{Oshagh} {et~al.}(2018){Oshagh}, {Triaud}, {Burdanov}, {Figueira},
  {Reiners}, {Santos}, {Faria}, {Boue}, {D{\'\i}az}, {Dreizler}, {Boldt},
  {Delrez}, {Ducrot}, {Gillon}, {Guzman Mesa}, {Jehin}, {Khalafinejad}, {Kohl},
  {Serrano}, \& {Udry}}]{2018A&A...619A.150O}
{Oshagh}, M., {Triaud}, A.~H.~M.~J., {Burdanov}, A., {et~al.} 2018, \aap, 619,
  A150

\bibitem[{{Owen}(2020)}]{2020MNRAS.498.5030O}
{Owen}, J.~E. 2020, \mnras, 498, 5030

\bibitem[{{Owen} \& {Adams}(2014)}]{2014MNRAS.444.3761O}
{Owen}, J.~E. \& {Adams}, F.~C. 2014, \mnras, 444, 3761

\bibitem[{{Owen} \& {Wu}(2013)}]{2013ApJ...775..105O}
{Owen}, J.~E. \& {Wu}, Y. 2013, \apj, 775, 105

\bibitem[{{Palle} {et~al.}(2020){Palle}, {Oshagh}, {Casasayas-Barris},
  {Hirano}, {Stangret}, {Luque}, {Strachan}, {Gaidos}, {Anglada-Escude},
  {Plavchan}, \& {Addison}}]{PalleAUMic}
{Palle}, E., {Oshagh}, M., {Casasayas-Barris}, N., {et~al.} 2020, \aap, 643,
  A25

\bibitem[{{Parviainen} \& {Aigrain}(2015)}]{ldtk}
{Parviainen}, H. \& {Aigrain}, S. 2015, \mnras, 453, 3821

\bibitem[{{Penz} {et~al.}(2008){Penz}, {Micela}, \& {Lammer}}]{Penz08a}
{Penz}, T., {Micela}, G., \& {Lammer}, H. 2008, \aap, 477, 309

\bibitem[{{Pepe} {et~al.}(2021){Pepe}, {Cristiani}, {Rebolo}, {Santos},
  {Dekker}, {Cabral}, {Di Marcantonio}, {Figueira}, {Lo Curto}, {Lovis},
  {Mayor}, {M{\'e}gevand}, {Molaro}, {Riva}, {Zapatero Osorio}, {Amate},
  {Manescau}, {Pasquini}, {Zerbi}, {Adibekyan}, {Abreu}, {Affolter}, {Alibert},
  {Aliverti}, {Allart}, {Allende Prieto}, {{\'A}lvarez}, {Alves}, {Avila},
  {Baldini}, {Bandy}, {Barros}, {Benz}, {Bianco}, {Borsa}, {Bourrier},
  {Bouchy}, {Broeg}, {Calderone}, {Cirami}, {Coelho}, {Conconi}, {Coretti},
  {Cumani}, {Cupani}, {D'Odorico}, {Damasso}, {Deiries}, {Delabre},
  {Demangeon}, {Dumusque}, {Ehrenreich}, {Faria}, {Fragoso}, {Genolet},
  {Genoni}, {G{\'e}nova Santos}, {Gonz{\'a}lez Hern{\'a}ndez}, {Hughes},
  {Iwert}, {Kerber}, {Knudstrup}, {Landoni}, {Lavie}, {Lillo-Box}, {Lizon},
  {Maire}, {Martins}, {Mehner}, {Micela}, {Modigliani}, {Monteiro}, {Monteiro},
  {Moschetti}, {Murphy}, {Nunes}, {Oggioni}, {Oliveira}, {Oshagh}, {Pall{\'e}},
  {Pariani}, {Poretti}, {Rasilla}, {Rebord{\~a}o}, {Redaelli}, {Santana
  Tschudi}, {Santin}, {Santos}, {S{\'e}gransan}, {Schmidt}, {Segovia},
  {Sosnowska}, {Sozzetti}, {Sousa}, {Span{\`o}}, {Su{\'a}rez Mascare{\~n}o},
  {Tabernero}, {Tenegi}, {Udry}, \& {Zanutta}}]{2021A&A...645A..96P}
{Pepe}, F., {Cristiani}, S., {Rebolo}, R., {et~al.} 2021, \aap, 645, A96

\bibitem[{{Pepe} {et~al.}(2002){Pepe}, {Mayor}, {Galland}, {Naef}, {Queloz},
  {Santos}, {Udry}, \& {Burnet}}]{2002A&A...388..632P}
{Pepe}, F., {Mayor}, M., {Galland}, F., {et~al.} 2002, \aap, 388, 632

\bibitem[{{Pillitteri} {et~al.}(2013){Pillitteri}, {Remage Evans}, {Wolk}, \&
  {Bruck Syal}}]{2013AJ....145..143P}
{Pillitteri}, I., {Remage Evans}, N., {Wolk}, S.~J., \& {Bruck Syal}, M. 2013,
  \aj, 145, 143

\bibitem[{{Pizzolato} {et~al.}(2003){Pizzolato}, {Maggio}, {Micela},
  {Sciortino}, \& {Ventura}}]{Pizz03}
{Pizzolato}, N., {Maggio}, A., {Micela}, G., {Sciortino}, S., \& {Ventura}, P.
  2003, \aap, 397, 147

\bibitem[{{Plavchan} {et~al.}(2020){Plavchan}, {Barclay}, {Gagn{\'e}}, {Gao},
  {Cale}, {Matzko}, {Dragomir}, {Quinn}, {Feliz}, {Stassun}, {Crossfield},
  {Berardo}, {Latham}, {Tieu}, {Anglada-Escud{\'e}}, {Ricker}, {Vanderspek},
  {Seager}, {Winn}, {Jenkins}, {Rinehart}, {Krishnamurthy}, {Dynes}, {Doty},
  {Adams}, {Afanasev}, {Beichman}, {Bottom}, {Bowler}, {Brinkworth}, {Brown},
  {Cancino}, {Ciardi}, {Clampin}, {Clark}, {Collins}, {Davison},
  {Foreman-Mackey}, {Furlan}, {Gaidos}, {Geneser}, {Giddens}, {Gilbert},
  {Hall}, {Hellier}, {Henry}, {Horner}, {Howard}, {Huang}, {Huber}, {Kane},
  {Kenworthy}, {Kielkopf}, {Kipping}, {Klenke}, {Kruse}, {Latouf}, {Lowrance},
  {Mennesson}, {Mengel}, {Mills}, {Morton}, {Narita}, {Newton}, {Nishimoto},
  {Okumura}, {Palle}, {Pepper}, {Quintana}, {Roberge}, {Roccatagliata},
  {Schlieder}, {Tanner}, {Teske}, {Tinney}, {Vanderburg}, {von Braun}, {Walp},
  {Wang}, {Wang}, {Weigand }, {White}, {Wittenmyer}, {Wright}, {Youngblood},
  {Zhang}, \& {Zilberman}}]{2020Natur.582..497P}
{Plavchan}, P., {Barclay}, T., {Gagn{\'e}}, J., {et~al.} 2020, \nat, 582, 497

\bibitem[{{Poppenhaeger} {et~al.}(2020){Poppenhaeger}, {Ketzer}, \&
  {Mallonn}}]{Pop20}
{Poppenhaeger}, K., {Ketzer}, L., \& {Mallonn}, M. 2020, \mnras

\bibitem[{{Ricker} {et~al.}(2015){Ricker}, {Winn}, {Vanderspek}, {Latham},
  {Bakos}, {Bean}, {Berta-Thompson}, {Brown}, {Buchhave}, {Butler}, {Butler},
  {Chaplin}, {Charbonneau}, {Christensen-Dalsgaard}, {Clampin}, {Deming},
  {Doty}, {De Lee}, {Dressing}, {Dunham}, {Endl}, {Fressin}, {Ge}, {Henning},
  {Holman}, {Howard}, {Ida}, {Jenkins}, {Jernigan}, {Johnson}, {Kaltenegger},
  {Kawai}, {Kjeldsen}, {Laughlin}, {Levine}, {Lin}, {Lissauer}, {MacQueen},
  {Marcy}, {McCullough}, {Morton}, {Narita}, {Paegert}, {Palle}, {Pepe},
  {Pepper}, {Quirrenbach}, {Rinehart}, {Sasselov}, {Sato}, {Seager},
  {Sozzetti}, {Stassun}, {Sullivan}, {Szentgyorgyi}, {Torres}, {Udry}, \&
  {Villasenor}}]{2015JATIS...1a4003R}
{Ricker}, G.~R., {Winn}, J.~N., {Vanderspek}, R., {et~al.} 2015, Journal of
  Astronomical Telescopes, Instruments, and Systems, 1, 014003

\bibitem[{{Rizzuto} {et~al.}(2017){Rizzuto}, {Mann}, {Vanderburg}, {Kraus}, \&
  {Covey}}]{2017AJ....154..224R}
{Rizzuto}, A.~C., {Mann}, A.~W., {Vanderburg}, A., {Kraus}, A.~L., \& {Covey},
  K.~R. 2017, \aj, 154, 224

\bibitem[{{Rizzuto} {et~al.}(2020){Rizzuto}, {Newton}, {Mann}, {Tofflemire},
  {Vanderburg}, {Kraus}, {Wood}, {Quinn}, {Zhou}, {Thao}, {Law}, {Ziegler}, \&
  {Brice{\~n}o}}]{rizzuto2020}
{Rizzuto}, A.~C., {Newton}, E.~R., {Mann}, A.~W., {et~al.} 2020, \aj, 160, 33

\bibitem[{{Rossiter}(1924)}]{1924ApJ....60...15R}
{Rossiter}, R.~A. 1924, \apj, 60, 15

\bibitem[{{Sanz-Forcada} {et~al.}(2011){Sanz-Forcada}, {Micela}, {Ribas},
  {Pollock}, {Eiroa}, {Velasco}, {Solano}, \& {Garc{\'\i}a-{\'A}lvarez}}]{SF11}
{Sanz-Forcada}, J., {Micela}, G., {Ribas}, I., {et~al.} 2011, \aap, 532, A6

\bibitem[{{Shaikhislamov} {et~al.}(2020){Shaikhislamov}, {Fossati},
  {Khodachenko}, {Lammer}, {Garc{\'\i}a Mu{\~n}oz}, {Youngblood}, {Dwivedi}, \&
  {Rumenskikh}}]{2020A&A...639A.109S}
{Shaikhislamov}, I.~F., {Fossati}, L., {Khodachenko}, M.~L., {et~al.} 2020,
  \aap, 639, A109

\bibitem[{{Snellen} {et~al.}(2008){Snellen}, {Albrecht}, {de Mooij}, \& {Le
  Poole}}]{2008A&A...487..357S}
{Snellen}, I.~A.~G., {Albrecht}, S., {de Mooij}, E.~J.~W., \& {Le Poole}, R.~S.
  2008, \aap, 487, 357

\bibitem[{{Tabernero} {et~al.}(2020){Tabernero}, {Zapatero Osorio}, {Allart},
  {Borsa}, {Casasayas-Barris}, {Demangeon}, {Ehrenreich}, {Lillo-Box}, {Lovis},
  {Pall{\'e}}, {Sousa}, {Rebolo}, {Santos}, {Pepe}, {Cristiani}, {Adibekyan},
  {Allende Prieto}, {Alibert}, {Barros}, {Bouchy}, {Bourrier}, {D'Odorico},
  {Dumusque}, {Faria}, {Figueira}, {G{\'e}nova Santos}, {Gonz{\'a}lez
  Hern{\'a}ndez}, {Hojjatpanah}, {Lo Curto}, {Lavie}, {Martins}, {Martins},
  {Mehner}, {Micela}, {Molaro}, {Nunes}, {Poretti}, {Seidel}, {Sozzetti},
  {Su{\'a}rez Mascare{\~n}o}, {Udry}, {Aliverti}, {Affolter}, {Alves}, {Amate},
  {Avila}, {Bandy}, {Benz}, {Bianco}, {Broeg}, {Cabral}, {Conconi}, {Coelho},
  {Cumani}, {Deiries}, {Dekker}, {Delabre}, {Fragoso}, {Genoni}, {Genolet},
  {Hughes}, {Knudstrup}, {Kerber}, {Landoni}, {Lizon}, {Maire}, {Manescau}, {Di
  Marcantonio}, {M{\'e}gevand}, {Monteiro}, {Monteiro}, {Moschetti}, {Mueller},
  {Modigliani}, {Oggioni}, {Oliveira}, {Pariani}, {Pasquini}, {Rasilla},
  {Redaelli}, {Riva}, {Santana-Tschudi}, {Santin}, {Santos}, {Segovia},
  {Sosnowska}, {Span{\`o}}, {Tenegi}, {Iwert}, {Zanutta}, \&
  {Zerbi}}]{tabernero_esp}
{Tabernero}, H.~M., {Zapatero Osorio}, M.~R., {Allart}, R., {et~al.} 2020,
  arXiv e-prints, arXiv:2011.12197

\bibitem[{{Ter Braak}(2006)}]{TerBraak2006}
{Ter Braak}, C. J.~F. 2006, Statistics and Computing, 16, 239

\bibitem[{{Thao} {et~al.}(2020){Thao}, {Mann}, {Johnson}, {Newton}, {Guo},
  {Kain}, {Rizzuto}, {Charbonneau}, {Dalba}, {Gaidos}, {Irwin}, \&
  {Kraus}}]{2020AJ....159...32T}
{Thao}, P.~C., {Mann}, A.~W., {Johnson}, M.~C., {et~al.} 2020, \aj, 159, 32

\bibitem[{{Wang} \& {Dai}(2019)}]{2019ApJ...873L...1W}
{Wang}, L. \& {Dai}, F. 2019, \apjl, 873, L1

\bibitem[{{Wyttenbach} {et~al.}(2015){Wyttenbach}, {Ehrenreich}, {Lovis},
  {Udry}, \& {Pepe}}]{wyttenbach}
{Wyttenbach}, A., {Ehrenreich}, D., {Lovis}, C., {Udry}, S., \& {Pepe}, F.
  2015, \aap, 577, A62

\bibitem[{{Yu} {et~al.}(2017){Yu}, {Donati}, {H{\'e}brard}, {Moutou}, {Malo},
  {Grankin}, {Hussain}, {Collier Cameron}, {Vidotto}, {Baruteau}, {Alencar},
  {Bouvier}, {Petit}, {Takami}, {Herczeg}, {Gregory}, {Jardine}, {Morin},
  {M{\'e}nard}, \& {Matysse Collaboration}}]{yu2017}
{Yu}, L., {Donati}, J.-F., {H{\'e}brard}, E.~M., {et~al.} 2017, \mnras, 467,
  1342

\bibitem[{{Zechmeister} {et~al.}(2009){Zechmeister}, {K{\"u}rster}, \&
  {Endl}}]{2009A&A...505..859Z}
{Zechmeister}, M., {K{\"u}rster}, M., \& {Endl}, M. 2009, \aap, 505, 859

\bibitem[{{Zeng} \& {Sasselov}(2013)}]{2013PASP..125..227Z}
{Zeng}, L. \& {Sasselov}, D. 2013, \pasp, 125, 227

\bibitem[{{Zhou} {et~al.}(2021){Zhou}, {Quinn}, {Irwin}, {Huang}, {Collins},
  {Bouma}, {Khan}, {Landrigan}, {Vanderburg}, {Rodriguez}, {Latham}, {Torres},
  {Douglas}, {Bieryla}, {Esquerdo}, {Berlind}, {Calkins}, {Buchhave},
  {Charbonneau}, {Collins}, {Kielkopf}, {Jensen}, {Tan}, {Hart}, {Carter},
  {Stockdale}, {Ziegler}, {Law}, {Mann}, {Howell}, {Matson}, {Scott}, {Furlan},
  {White}, {Hellier}, {Anderson}, {West}, {Ricker}, {Vanderspek}, {Seager},
  {Jenkins}, {Winn}, {Mireles}, {Rowden}, {Yahalomi}, {Wohler}, {Brasseur},
  {Daylan}, \& {Col{\'o}n}}]{2021AJ....161....2Z}
{Zhou}, G., {Quinn}, S.~N., {Irwin}, J., {et~al.} 2021, \aj, 161, 2

\bibitem[{{Zhou} {et~al.}(2020){Zhou}, {Winn}, {Newton}, {Quinn}, {Rodriguez},
  {Mann}, {Rizzuto}, {Vand erburg}, {Huang}, {Latham}, {Teske}, {Wang},
  {Shectman}, {Butler}, {Crane}, {Thompson}, {Henry}, {Paredes}, {Jao},
  {James}, \& {Hinojosa}}]{2020ApJ...892L..21Z}
{Zhou}, G., {Winn}, J.~N., {Newton}, E.~R., {et~al.} 2020, \apjl, 892, L21

\bibitem[{{Zuckerman} {et~al.}(2011){Zuckerman}, {Rhee}, {Song}, \&
  {Bessell}}]{2011ApJ...732...61Z}
{Zuckerman}, B., {Rhee}, J.~H., {Song}, I., \& {Bessell}, M.~S. 2011, \apj,
  732, 61

\end{thebibliography}

\end{document}